\documentclass{article}

\pdfoutput=1
\usepackage{graphicx,amsmath,amssymb,epsf} \usepackage{latexsym,bm,
  slashed} \usepackage{xcolor} \definecolor{dark}{rgb}{0.10,0.2,0.3}
\definecolor{magenta}{rgb}{0.7,0.1,0.3}
\definecolor{purpure}{rgb}{0.5,0.15,0.3}
\usepackage[font=small,format=plain,labelfont=bf,up,textfont=it,up]{caption}
\usepackage{hyperref, cite} \hypersetup{colorlinks, citecolor=blue,
  filecolor=blue, linkcolor=magenta,
  urlcolor=purpure,hyperfootnotes=true,pdftex}

\usepackage{subcaption}

\title{Average minijet rapidity ratios in Mueller-Navelet jets} 
\author{N. Bethencourt de Le{\'o}n$^{1}$, G. Chachamis$^2$, A. Sabio Vera$^{1,3}$\\ 
\\
\small $^1$ Instituto de F{\'\i}sica Te{\'o}rica UAM/CSIC, Nicol{\'a}s Cabrera 15, E-28049 Madrid, Spain.\\
\small $^2$ Laborat{\' o}rio de Instrumenta\c{c}{\~ a}o e F{\' \i}sica Experimental de Part{\' \i}culas (LIP),\\
\small Av. Prof. Gama Pinto, 2, P-1649-003 Lisboa, Portugal.\\
\small $^3$ Theoretical Physics Department, Universidad Aut{\' o}noma de Madrid, E-28049 Madrid, Spain.\\} 

\begin{document} 

\maketitle 

\begin{abstract}
We investigate different final state features in Mueller-Navelet jets events at hadron colliders. The focus lies on the average rapidity ratio between subsequent minijet emissions which has been investigated in previous works but now is modified to also incorporate the transverse momenta together with the rapidities of the emitted jets. We study the dependence of this observable on a lower transverse momentum veto which does affect the typical minijet multiplicity of the events under scrutiny. We find that this observable is stable when including higher order quantum  corrections, also when collinear terms are resummed to all orders. 
\end{abstract}

\section{Introduction}

An active area of research in QCD phenomenology at high energies 
is to pin down novel observables where the dominant contributions 
stem from the Balitsky-Fadin-Kuraev-Lipatov (BFKL) domain~\cite{Kuraev:1977fs,Kuraev:1976ge,Fadin:1975cb,Lipatov:1976zz,Balitsky:1978ic,Lipatov:1985uk,Fadin:1998py,Ciafaloni:1998gs}. This is a challenging task since for typical observables calculations based on matrix elements computed at fixed order 
along with the Dokshitzer-Gribov-Lipatov-Altarelli-Parisi (DGLAP) evolution~\cite{Gribov:1972ri,Gribov:1972rt,Lipatov:1974qm,Altarelli:1977zs,Dokshitzer:1977sg} to account for the PDFs tend to describe the bulk of the data adequately. It is then needed to move towards corners of the 
phase space to isolate BFKL effects. This can be done by studying the structure of final states in Mueller-Navelet (MN) jets events~\cite{Mueller:1986ey}  currently produced at the LHC.

In a nutshell, MN events have two jets with similar and large enough $p_T$ such that they serve as a hard scale, $\Lambda^2_{QCD} \ll p_T^2 \ll s$, where $s$ is the c.o.m. energy squared. The two tagged jets should also be separated by a large rapidity interval Y while there is a rich mini-jet activity in between. Considering two tagged jets of similar $p_T$ ensures that we lie within multi-Regge kinematics, this being one of the pillars for the BFKL program. Requiring a large rapidity separation between the two MN jets sets one of them in the very forward and the other in the very backward regions. Numerous studies took place on MN jets both at leading-order (LO) BFKL as well as at next-to-leading order (NLO). The main quantity of interest in most studies was the azimuthal decorrelation between the two outermost jets, for a non exhaustive list of theoretical works see Refs.~\cite{DelDuca:1993mn,Stirling:1994zs, DelDuca:1994ng,Orr:1997im,Kwiecinski:2001nh,Vera:2006un, Vera:2007kn,Marquet:2007xx,Colferai:2010wu,Bartels:2001ge,Bartels:2002yj,Caporale:2011cc,Caporale:2014gpa} while relevant experimental analyses by ATLAS and CMS can be found in Refs.~\cite{Aad:2011jz,Chatrchyan:2012pb,Aad:2014pua,Khachatryan:2016udy}.

One of the conclusions of the comparison to experimental data from different approaches is that a more precise theoretical work is needed (e.g. see~\cite{Khachatryan:2016udy}). This is more pressing recently since experimental uncertainties appear to be smaller than the theoretical ones and extra emphasis needs to be given in the theory side in order to have more accurate predictions. It is in this context that we present new observables which could be useful for  this quest. 

As mentioned above, the rapidity ratios put forward in~\cite{Chachamis:2015ico} aim at probing novel multi-Regge kinematics signatures. The proposal was to examine the average per event jet $p_T$, the average jet azimuthal angle $\phi$ and the average ratio of rapidities between neighboring jets. It was shown that the NLO corrections could modify the  predictions when compared to the LO ones. From a theoretical stand point it would be desirable to propose infrared finite observables which are robust under higher order corrections. We will show that the 
quantities discussed in the following will lie within this category.  

Besides the issues just described, there appears another complication one needs to deal with. For our proposed observables
we see that events where the minijets have relatively low $p_T$,  contrary to what one would naively expect, give a very significant contribution to the gluon Green's function (Fig. 1 in~\cite{Chachamis:2015ico}) and consequently to the cross-section. The experimental analyses however  (mainly to deal with jet energy reconstruction uncertainties) impose a veto on the $p_T$ of any resolved minijet\footnote{`minijets' correspond to any of the jets in between  the two outermost MN jets.}. Usually the $p_T$ veto value for ATLAS and CMS is $Q_0 = 20$ GeV which is rather large if we compare it to the 
$Q_0 = 1$ GeV value which was the jet $p_T$ infrared cutoff for the plots in~\cite{Chachamis:2015ico}.

With the present work we want to address the two issues just described in order to pave the path for a proper comparison between a full BFKL phenomenological analysis and experimental data for our proposed observables. The aim is not a full fledged study but rather to isolate the intricacies and deal with them. A full phenomenological study including PDFs and integration over the $p_T$  of the MN jets and over different rapidity ranges is well beyond the scope of this paper. However, we think that the issues we discuss here, namely, the difficulty of ascertaining the NLO uncertainties and the role of the presence of a large $p_T$ veto for any resolved minijet, will be of help to whoever proceed to such a study.

From the list with the average jet azimuthal angle $\phi$,  $p_T$ and rapidities, here we will  address the last two.  Since in the experimental analyses there is a large $p_T$ veto for resolved minijets we would miss the largest part of any multi-Regge kinematics effects because these trimmed sectors build up the bulk of the cross section. We will show that this problem can be diminished when mixing $p_T$ and rapidities in our newly proposed quantities. 

In the next section we will introduce our notation, set up the basics for the two MN observables we will be studying and the numerical framework to analyze them. In Section 3 we will present new results and draw our conclusions in Section 4.

\section{The observables}

We focus on events where two jets with rapidities $y_a$ in the forward direction and $y_b$  in the backward direction can be clearly identified. If the difference $Y=y_a-y_b$ is large enough then terms of the form $\alpha_s^n Y^n$ are important order-by-order to get a good description of measured cross sections which can be written in the factorized form 
\begin{eqnarray}
\sigma (Q_1,Q_2,Y) = \int d^2 \vec{k}_A d^2 \vec{k}_B \, {\phi_A(Q_1,\vec{k}_a) \, 
\phi_B(Q_2,\vec{k}_b)} \, {f (\vec{k}_a,\vec{k}_b,Y)}.
\end{eqnarray}
In this expression $\phi_{A,B}$ are impact factors depending on external scales, $Q_{1,2}$, and the off-shell reggeized gluon momenta, $\vec{k}_{a,b}$.  The gluon Green function $f$ depends on $\vec{k}_{a,b}$ and the center-of-mass energy in the scattering $\sim e^{Y/2}$.
For simplicity, we will consider that the impact factors are $\alpha_s$ and  keep $|\vec{k}_{a}|$ and $|\vec{k}_{b}|$ constant such that the gluon Green's function can be also referred to as the cross-section.

For BFKL phenomenology at the LHC it is mandatory to work within the NLO approximation at least for the gluon Green's function which introduces the dependence on physical scales such as the one associated to the running of the coupling and the one related to the choice of energy scale in the resummed logarithms~\cite{Forshaw:2000hv,Chachamis:2004ab,Forshaw:1999xm,Schmidt:1999mz}. It is possible to write the gluon Green function in an iterative way in transverse momentum and rapidity space at LO~\cite{Schmidt:1996fg} and NLO~ \cite{Andersen:2003an,Andersen:2003wy}.  The iterative solution has the form (at LO, for the NLO expressions see Refs.~ \cite{Andersen:2003an,Andersen:2003wy})
\begin{eqnarray}
f &=& e^{\omega \left(\vec{k}_A\right) Y}  \Bigg\{\delta^{(2)} \left(\vec{k}_A-\vec{k}_B\right) + \sum_{N=1}^\infty \prod_{i=1}^N \frac{\alpha_s N_c}{\pi}  \int d^2 \vec{k}_i  
\frac{\theta\left(k_i^2-\lambda^2\right)}{\pi k_i^2} \nonumber\\
&&\hspace{-.6cm}\times \int_0^{y_{i-1}} \hspace{-.3cm}d y_i e^{\left(\omega \left(\vec{k}_A+\sum_{l=1}^i \vec{k}_l\right) -\omega \left(\vec{k}_A+\sum_{l=1}^{i-1} \vec{k}_l\right)\right) y_i} \delta^{(2)} \hspace{-.16cm}
\left(\vec{k}_A+ \sum_{l=1}^n \vec{k}_l - \vec{k}_B\right) \hspace{-.2cm}\Bigg\} \, , 
\label{BFKL_iter}
 \end{eqnarray}
where 
\begin{eqnarray}
\omega \left(\vec{q}\right) &=& - \frac{\alpha_s N_c}{\pi} \ln{\frac{q^2}{\lambda^2}} 
\end{eqnarray}
corresponds to the gluon Regge trajectory which carries a regulator, $\lambda$, of infrared divergences. All these expressions have been implemented  in the Monte Carlo code {\tt BFKLex} which we have already used for different applications ranging from collider phenomenology to more formal studies in the calculation of scattering amplitudes in supersymmetric theories~\cite{Chachamis:2011rw,Chachamis:2011nz,Chachamis:2012fk,Chachamis:2012qw,Caporale:2013bva,Chachamis:2015zzp}. 
The right-hand-side of Eq.~(\ref{BFKL_iter}) states that the gluon Green's function can be obtained by an infinite sum of terms, however the sum is
in reality convergent to a numerically accepted value after a finite number of terms, hence in actuality the upper limit of the sum is not infinity but rather some
properly chosen value for the sought  numerical accuracy $N_{\text{final}}$. Let us assume that we truncate the sum at some $ N = N_{\text{trunc}}$.
Since all the terms in the sum corresponding to $ N = 1, 2, ...,  N_{\text{trunc}}$ contain phase space integrations, there is an inherited
statistical error in the result accumulated thus far and if the contribution from $ N = N_{\text{trunc + 1}}$ is smaller than the statistical error, one can 
omit that and set $N_{\text{final}} = N_{\text{trunc }}$\footnote{The actual
$N_{\text{final}}$ depends mainly on the rapidity difference between the two tagged jets and to a lesser degree on their transverse momenta. The larger the rapidity difference, the larger
$N_{\text{final}}$ is needed.}.

\begin{eqnarray}
 f (\vec{k}_a,\vec{k}_b,Y)&=& \sum_{N=1}^{N_{\text{final}}} f_N (\vec{k}_a,\vec{k}_b,Y) \,\,\,\text{or equivalently for the rest of our discussion} \nonumber\\
 \sigma &=& \sum_{N=1}^{N_{\text{final}}} \sigma_N \,\,\,\text{with} \,\,\sigma_N = f_N .
\end{eqnarray}

It turns out that the BFKL formalism can  be quite sensitive to collinear regions of phase space, in particular when the process-dependent impact factors are broad and allow for the external scales $Q_i$ to significantly deviate from the internal reggeized gluon transverse momenta $k_i$. In this case there exists a dominant double-log term in the NLO BFKL kernel in the collinear regions which takes the form
\begin{eqnarray}
\theta \left(k_i^2-\lambda^2\right) \to \theta \left(k_i^2-\lambda^2\right)  
-{\frac{\bar{\alpha}_s}{4} \ln^2{\left(\frac{\vec{k}_A^2}{\left(\vec{k}_A+\vec{k}_i\right)^2}\right)}},
\end{eqnarray}
which needs to be resummed to all orders to stabilize the behavior of the BFKL cross sections and to apply the formalism beyond the original multi-Regge kinematics. These issues have been investigated in~\cite{Salam:1998tj,Ciafaloni:2003ek}. In particular,  in~\cite{Vera:2005jt}, it was shown that the collinear corrections can be resummed to all-orders using the prescription
\begin{eqnarray}
\theta \left(k_i^2-\lambda^2\right) \to \theta \left(k_i^2-\lambda^2\right)  + \sum_{n=1}^\infty 
\frac{\left(-\bar{\alpha}_s\right)^n}{2^n n! (n+1)!} \ln^{2n}{\left(\frac{\vec{k}_A^2}{\left(\vec{k}_A+\vec{k}_i\right)^2}\right)}. 
\label{SumBessel}
\end{eqnarray}
The double logarithms in the expression above resum to a Bessel function of the first kind~\cite{Vera:2005jt}  (similar results have recently been obtained in coordinate representation in~\cite{Iancu:2015vea}). Phenomenological applications of this resummation, not using a Monte Carlo approach, show  good perturbative convergence~\cite{Vera:2006un,Vera:2007kn,Caporale:2007vs,Vera:2007dr,Caporale:2008fj,Hentschinski:2012kr,Hentschinski:2013id,Caporale:2013uva,Chachamis:2015ona} whereas in~\cite{Chachamis:2015zzp} we implemented this collinear resummation in {\tt BFKLex} and investigated what is its effect in the behavior of the gluon Green's  function. In the following, whenever we use the term NLO in figures, we will refer to the NLO kernel with the collinear contributions from the double logarithms resummed by using the Bessel function (NLO+DLs). The exact expression is given below:
\begin{eqnarray}
\widetilde{{\cal K}}_{r}\left({\bf q}, {\bf q}^{\prime}\right)
&=& \frac{1}{\pi \left({\bf q}-{\bf q}^{\prime}\right)^{2}}
\Bigg\{
-\bar{\alpha}_{s} + a \, \bar{\alpha}_{s}^{2} - b \, \bar{\alpha}_{s}^{2} \frac{|{\bf q}^{\prime}-{\bf q}|}{{\bf q}^{\prime}-{\bf q}} \ln \left(\frac{{\bf q}^{2}}{{\bf q}^{\prime 2}}\right) \nonumber\\
&&  +\left(\frac{{\bf q}^{2}}{{\bf q}^{\prime 2}}\right)^{b \, \bar{\alpha}_s \frac{|{\bf q}^{\prime}-{\bf q}|}{{\bf q}^{\prime}-{\bf q}}}  \sqrt{\frac{2\left(\bar{\alpha}_{s}-a \bar{\alpha}_{s}^{2}\right)}{\ln ^{2}\left(\frac{{\bf q}^{2}}{{\bf q}^{\prime 2}}\right)}} J_{1}\left(\sqrt{2\left(\bar{\alpha}_{s}-a \, \bar{\alpha}_{s}^{2}\right) 
\ln ^{2}\left(\frac{{\bf q}^{2}}{{\bf q}^{\prime 2}}\right)}\right)\Bigg\} 
\nonumber\\
&+&
\frac{\bar{\alpha}_{s}^{2}}{4 \pi}\left\{\left(1+\frac{n_{f}}{N_{c}^{3}}\right)\left(\frac{3\left({\bf q} \cdot {\bf q}^{\prime}\right)^{2}-2 {\bf q}^{2} {\bf q}^{\prime 2}}{16 {\bf q}^{2} {\bf q}^{\prime 2}}\right)\right.  \left(\frac{2}{{\bf q}^{2}}+\frac{2}{{\bf q}^{\prime 2}}+\left(\frac{1}{{\bf q}^{\prime 2}}-\frac{1}{{\bf q}^{2}}\right) \ln \frac{{\bf q}^{2}}{{\bf q}^{\prime 2}}\right) \nonumber\\
&-& \left(3+\left(1+\frac{n_{f}}{N_{c}^{3}}\right)\left(1-\frac{\left({\bf q}^{2}+{\bf q}^{\prime 2}\right)^{2}}{8 {\bf q}^{2} {\bf q}^{\prime 2}}-\frac{\left(2 {\bf q}^{2} {\bf q}^{\prime 2}-3 {\bf q}^{4}-3 {\bf q}^{\prime 4}\right)}{16 {\bf q}^{4} {\bf q}^{\prime 4}}\left({\bf q} \cdot {\bf q}^{\prime}\right)^{2}\right)\right) \nonumber\\
&\times& \int_{0}^{\infty} d x \frac{1}{{\bf q}^{2}+x^{2} {\bf q}^{\prime 2}} \ln \left|\frac{1+x}{1-x}\right| \nonumber\\
&+&\frac{2\left({\bf q}^{2}-{\bf q}^{\prime 2}\right)}{\left({\bf q}-{\bf q}^{\prime}\right)^{2}\left({\bf q}+{\bf q}^{\prime}\right)^{2}}\left(\frac{1}{2} \ln \frac{{\bf q}^{2}}{{\bf q}^{\prime 2}} \ln \frac{{\bf q}^{2} {\bf q}^{\prime 2}\left({\bf q}-{\bf q}^{\prime}\right)^{4}}{\left({\bf q}^{2}+{\bf q}^{\prime 2}\right)^{4}}+\left(\int_{0}^{-\frac{{\bf q}^{2}}{{\bf q}^{\prime 2}}}-\int_{0}^{-\frac{{\bf q}^{\prime 2}}{{\bf q}^{2}}}\right) d t \frac{\ln (1-t)}{t}\right) \nonumber\\
&-&\left.\left(1-\frac{\left({\bf q}^{2}-{\bf q}^{\prime 2}\right)^{2}}{\left({\bf q}-{\bf q}^{\prime}\right)^{2}\left({\bf q}+{\bf q}^{\prime}\right)^{2}}\right)\left(\left(\int_{0}^{1}-\int_{1}^{\infty}\right) d z \frac{1}{\left({\bf q}^{\prime}-z {\bf q}\right)^{2}} \ln \frac{(z {\bf q})^{2}}{{\bf q}^{\prime 2}}\right)\right\}\,,
\end{eqnarray}
where,
\begin{eqnarray}
a = \frac{13}{36} \frac{n_{f}}{N_{c}^{3}} + \frac{55}{36} \, \, , \, \, 
b = \frac{1}{6} \frac{n_{f}}{N_{c}^{3}} + \frac{11}{12} 
\end{eqnarray}
The BFKL equation at NLO is
\begin{eqnarray}
\left(\omega-\omega_{0}\left({\bf k}_{a}^{2}, \lambda^{2}\right)\right) f_{\omega}\left({\bf k}_{a}, {\bf k}_{b}\right) &=& \delta^{(2)}\left({\bf k}_{a}-{\bf k}_{b}\right) \nonumber\\
&&\hspace{-5cm}+\int d^{2} {\bf k}\left(\frac{\Gamma^{\rm cusp}_0}{\pi {\bf k}^{2}}  \theta\left({\bf k}^{2}-\lambda^{2}\right)+\widetilde{{\cal K}}_{r}\left({\bf k}_{a}, {\bf k}_{a}+{\bf k}\right)\right) f_{\omega}\left({\bf k}_{a}+{\bf k}, {\bf k}_{b}\right)
\end{eqnarray}
with
\begin{eqnarray}
\omega_{0}\left(q^{2}, \lambda\right) &=& 
-\int_{\lambda^{2}}^{q^{2}} \frac{d k^{2}}{k^{2}} \Gamma_{\rm cusp}^0  +\bar{\alpha}_{s}^{2} \frac{3}{2} \zeta (3)  \\ 
\Gamma_{\rm cusp}0 &\equiv& \bar{\alpha}_{s}+ \bar{\alpha}_{s}^{2}
\left(\frac{1}{3}-  \frac{\zeta(2)}{2}\right)
\end{eqnarray}

As mentioned previously, have two tagged jets such that one is in the forward region at rapidity $y_a$ and the other in the backward direction at rapidity $y_b$ while in between we have emissions of minijets with transverse momenta $k_i$ and rapidities $y_i$ with $1 \le i \le N$ and $y_b < y_i < y_a$. Due to the additive nature of rapidity under boosts, we shift the rapidities of the tagged jets and of the minijets such that $y_b=0$. We also set $y_{i-1} > y_i$, $y_0 = y_a$ and
$y_{N+1} = y_b = 0$. Finally, we define as jet multiplicity of an event, $N$, the number of emitted minijets. We will not be using any jet clustering algorithm in this study not to distort effects that stem from BFKL dynamics. Nevertheless, we do need to introduce a minimum $p_T$ cutoff for the minijets in order to ensure IR safety. Therefore, we consider $J_{min}$ which is an infrared cutoff simulating the resolution power of the collider experiments or in other words the minijet veto: $J_{min}$ should be interpreted as the minimum transverse momentum resolution limit, any minijets above $J_{min}$ contribute to the multiplicity of an event whereas minijets below $J_{min}$ are unresolved. 

To be more precise, let us picture a MN event where the minijets are having transverse momenta $k_1 = 15$ GeV, $k_2 = 3$ GeV, $k_3 = 7$ GeV, $k_4 = 30$ GeV, $k_5 = 22$ GeV, then its multiplicity will be $N = 5$ for $J_{min} = 2$ GeV while for $J_{min} = 10$ GeV its multiplicity will be $N = 3$.
Schematically, a MN event is characterized by:
\begin{eqnarray}
 ka, \,kb&:  & \text{transverse momenta of the MN jets}   \nonumber
 \\
  y_0 = y_a = Y,\, y_{N+1} = y_b = 0&:  & \text{rapidities of the MN jets}  \nonumber
 \\
 k_1, k_2, ... , k_N&:  &  \text{transverse momenta of the minijets}   \nonumber
 \\
 y_1, y_2, ... , y_N&: &   \text{rapidities of the minijets with } y_{i-1} > y_i
\label{eq:MN}
\end{eqnarray}

Now we focus on the study of the average rapidity ratio of a MN event of multiplicity $N$, {\it i.e.}
\begin{eqnarray}
\langle {\mathcal R}_y \rangle &=& \frac{1}{N-1}  \sum_{i=1}^{N-1} \frac{y_i}{y_{i-1}}\,,
\label{eq:observableRy}
\end{eqnarray}
and the newly defined average rapidity ratio scaled with the $p_T$ of the minijets $\langle {\mathcal R}_{p_T,y} \rangle$, defined as
\begin{eqnarray}
\langle {\mathcal R}_{p_T, y} \rangle &=& \frac{1}{N-1}  \sum_{i=1}^{N-1} \frac{k_i e^{y_i}}{k_{i-1} e^{y_{i-1}}} \,.
\label{eq:observableRpTy}
\end{eqnarray}
Eq.~(\ref{eq:observableRy}) differs from the original definition in~\cite{Chachamis:2015ico} since now $i$ runs over the minijets and excludes the leading MN jets.  $\langle {\mathcal R}_{p_T, y} \rangle$ incorporates a $p_T$ dependence which  carries information related to the decoupling between transverse and longitudinal components of the emitted gluons. As we will see in the next section, this quantity  is more sensitive to the increase of the total rapidity $Y$ than $\langle {\mathcal R}_y \rangle$.

\begin{figure}
\begin{subfigure}{.5\textwidth}
  \centering
  \includegraphics[width=.8\linewidth]{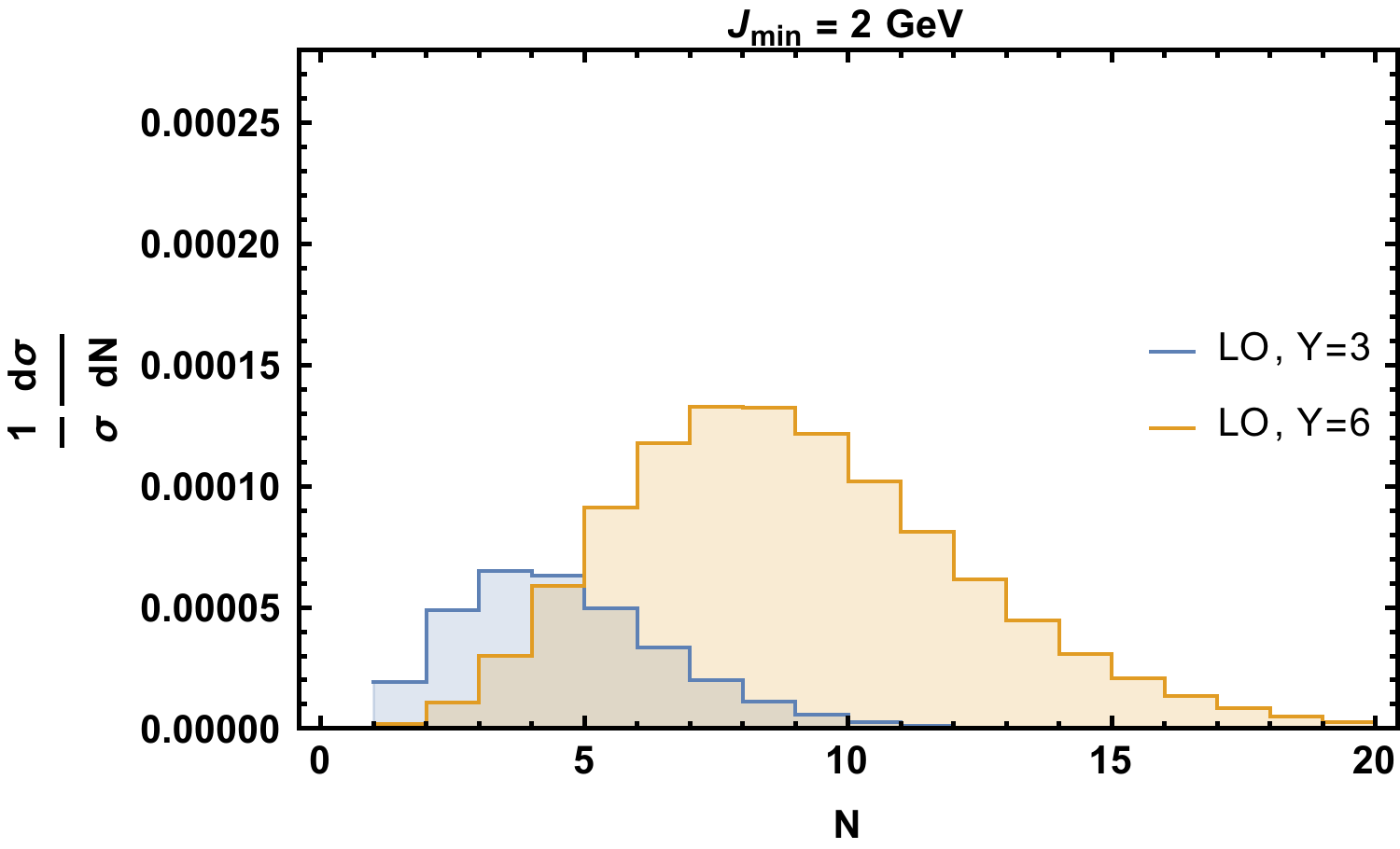}
  \caption{}
  \label{fig:sfig1}
\end{subfigure}%
\begin{subfigure}{.5\textwidth}
  \centering
  \includegraphics[width=.8\linewidth]{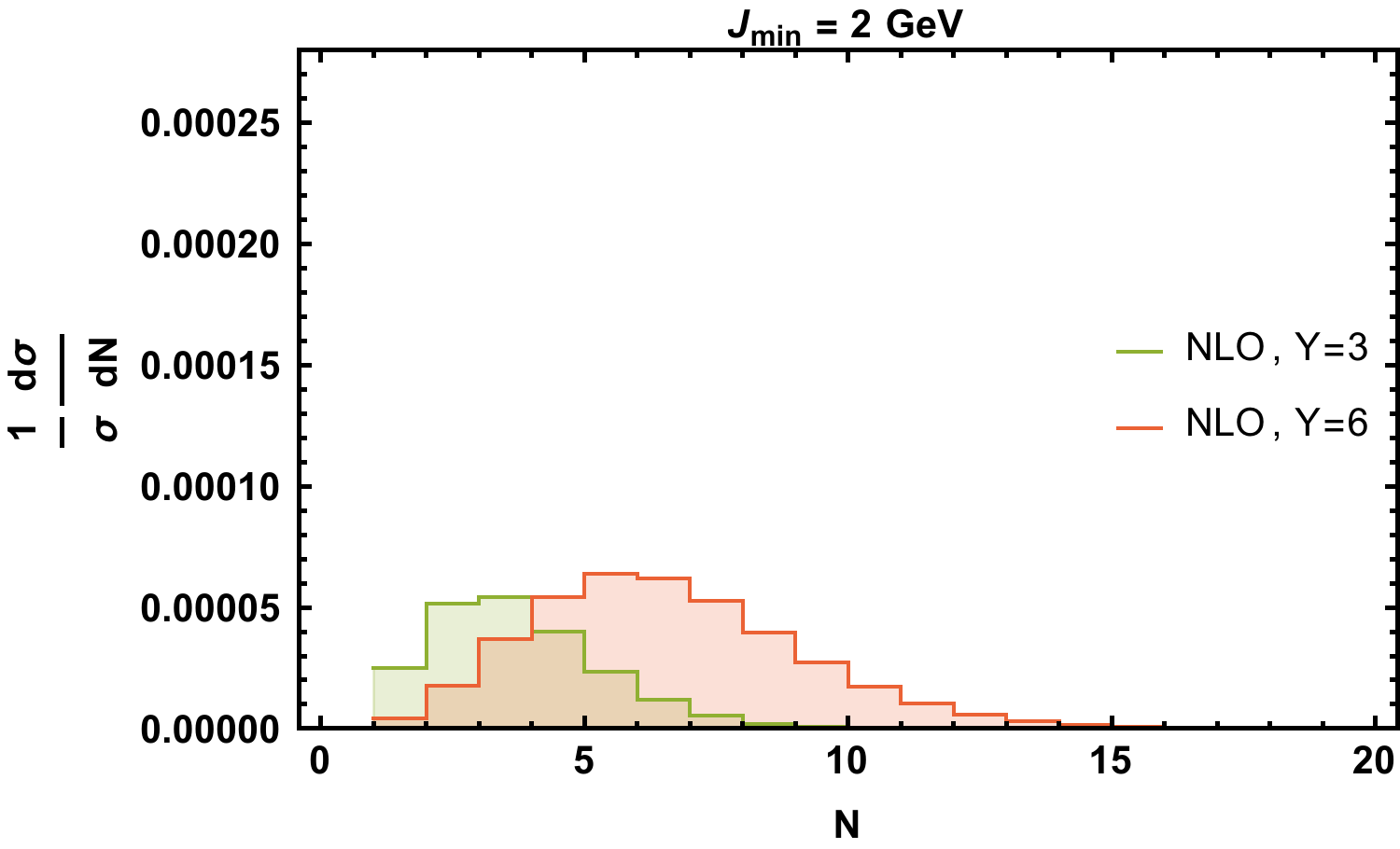}
  \caption{}
  \label{fig:sfig2}
\end{subfigure}
\\
\begin{subfigure}{.5\textwidth}
  \centering
  \includegraphics[width=.8\linewidth]{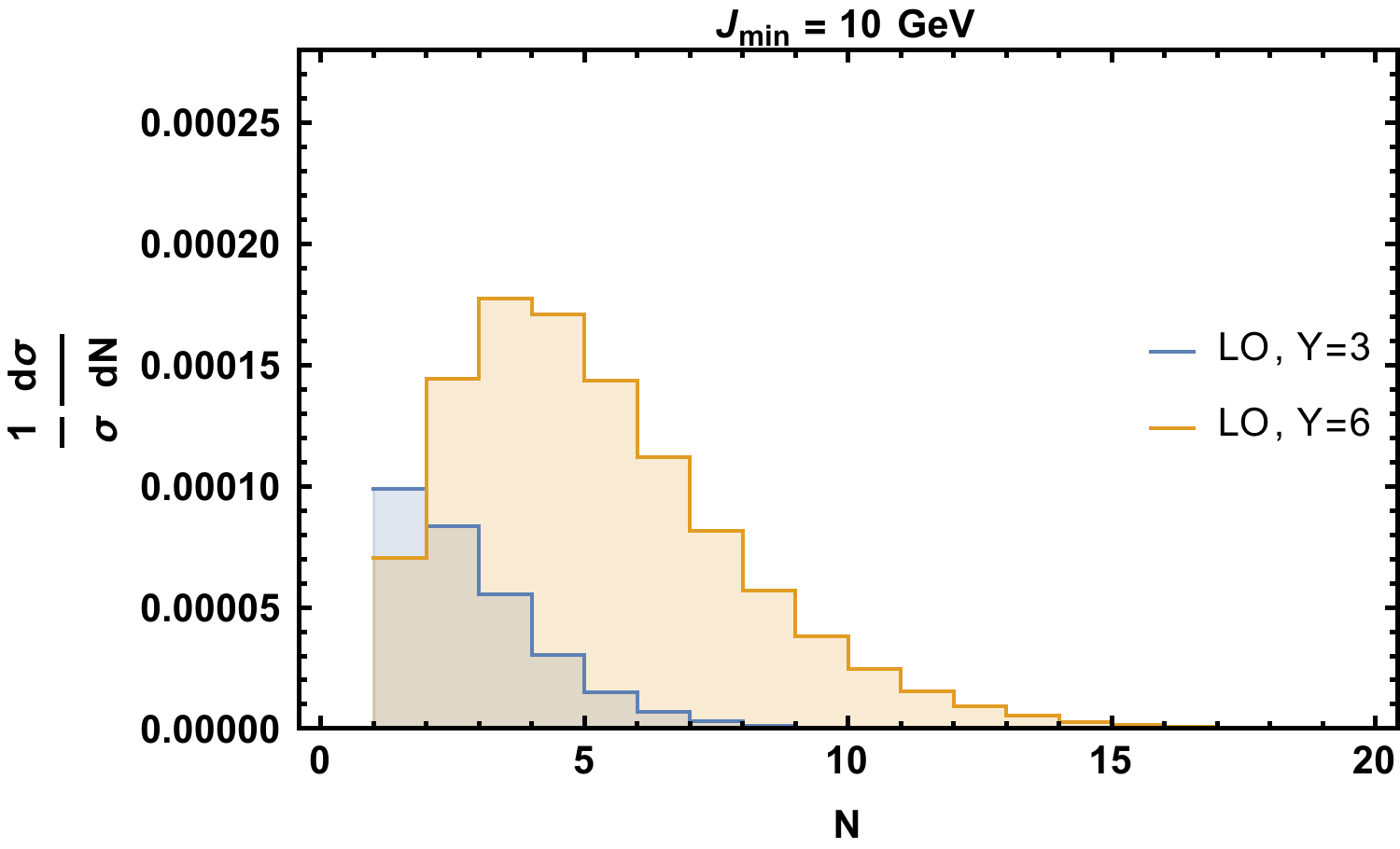}
  \caption{}
  \label{fig:sfig1}
\end{subfigure}%
\begin{subfigure}{.5\textwidth}
  \centering
  \includegraphics[width=.8\linewidth]{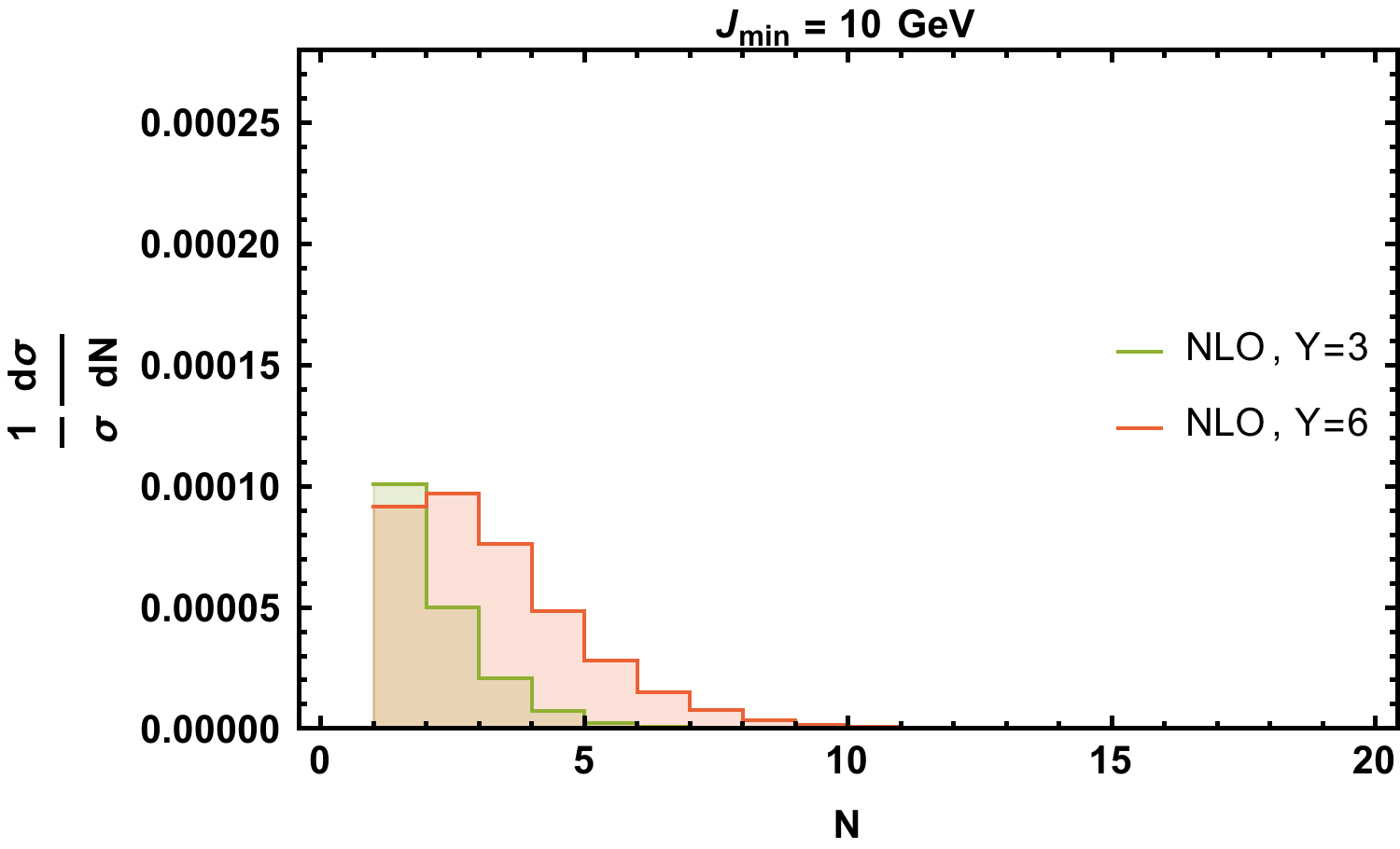}
  \caption{}
  \label{fig:sfig2}
\end{subfigure}
\\
\begin{subfigure}{.5\textwidth}
  \centering
  \includegraphics[width=.8\linewidth]{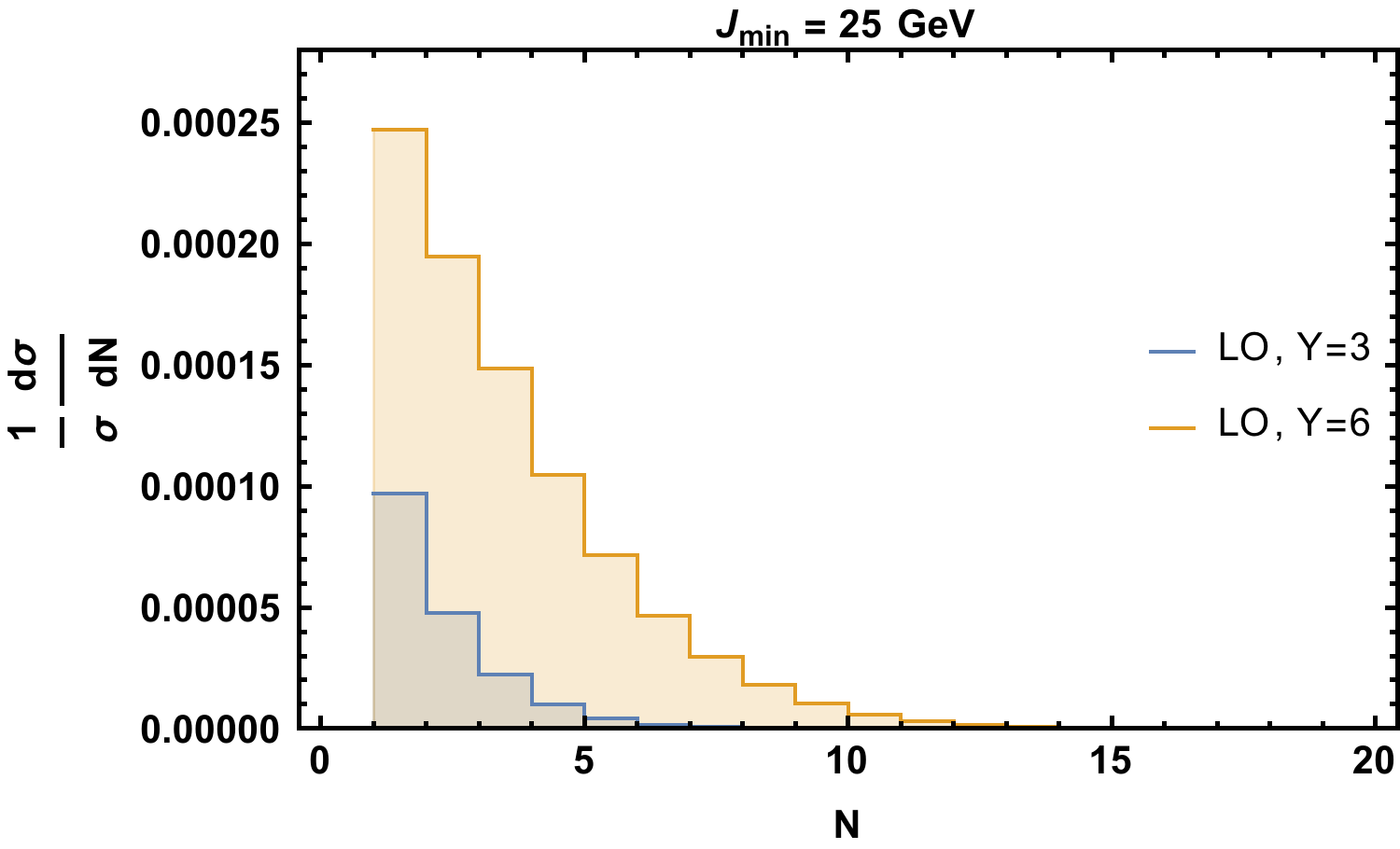}
  \caption{}
  \label{fig:sfig1}
\end{subfigure}%
\begin{subfigure}{.5\textwidth}
  \centering
  \includegraphics[width=.8\linewidth]{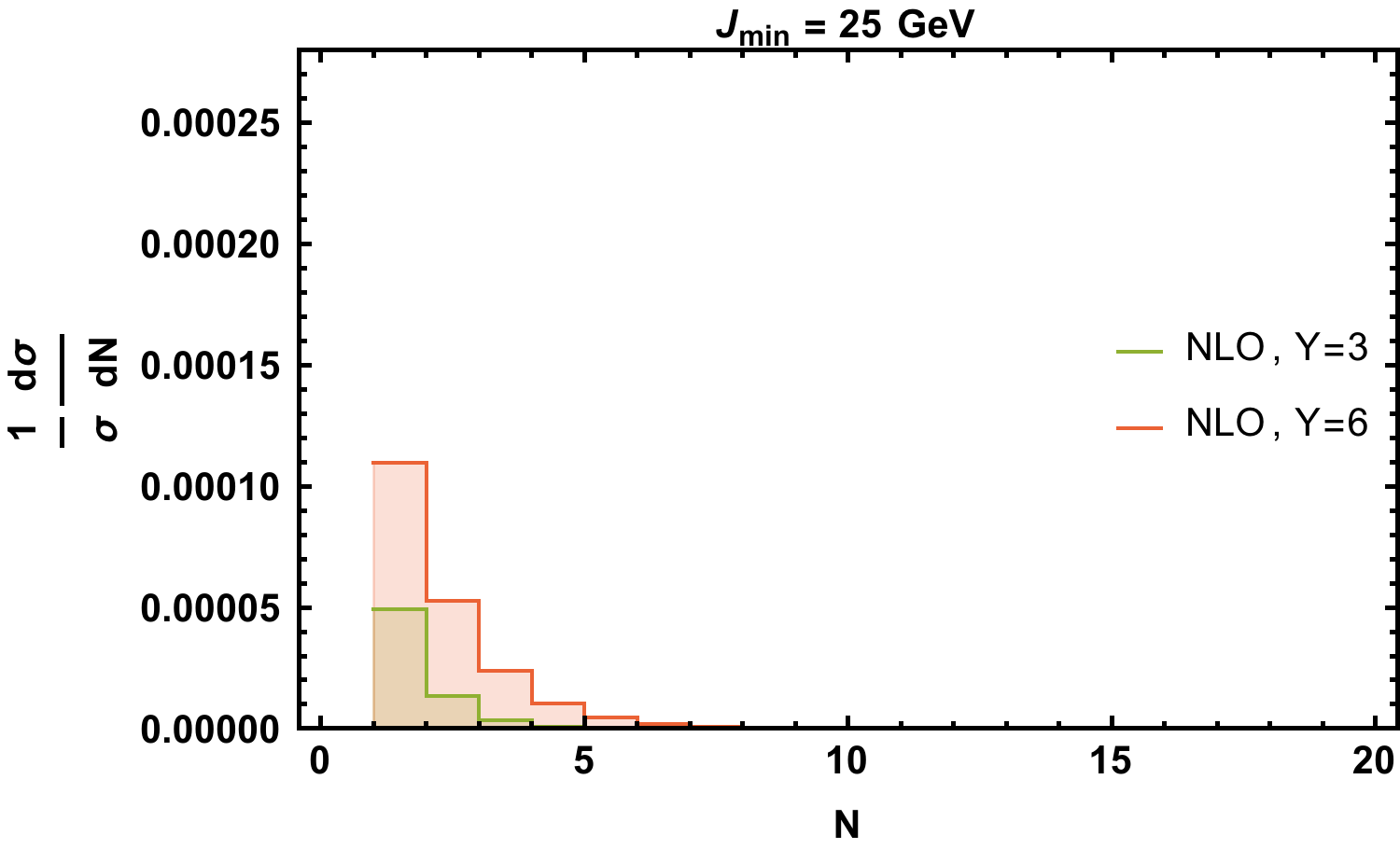}
  \caption{}
  \label{fig:sfig2}
\end{subfigure}
\caption{Multiplicity distributions for $Y = 3, 6$ and $J_{\text{min}} = 2, 10$ and $25$ GeV (top, middle, bottom).}
\label{fig:mult}
\end{figure}

In Ref.~\cite{Chachamis:2015ico}, we offered numerical results for $\langle {\mathcal R}_y \rangle$ considering the whole
gluon Green's function which includes the  sum over all $ N = 1, 2, ...,  N_{\text{final}}$ multiplicities.  We now consider a less academic situation where an experimental resolution scale is present. We start presenting  Fig.~\ref{fig:mult} where we plot the contribution from each multiplicity for rapidities $Y=3$ and $Y=6$, for $J_{min} = 2$ (top) and $J_{min}=10$ (middle) and $J_{min}=25$ (bottom) both at LO (left) and NLO+DLs (right) with $k_a = 20$ GeV and $k_b = 30$ GeV. 

Let us first focus on Fig.~\ref{fig:mult} (a) where $J_{min}=2$ GeV. When $Y=6$  we see that the largest contributions to the cross-section come from $N=7, 8$ multiplicities whereas in Fig.~\ref{fig:mult} (e) where $J_{min}=25$ GeV, the important multiplicities are $N=1, 2$. However, in our conventions here, a MN event with $N=1,2$ is essentially a final state with 3, 4 hard jets since the experimental veto on the resolved minijets is 20 GeV and there is no reason at present energies at the LHC to believe that a NLO fixed order calculation will fail to describe the hard scattering part of such a process. One does not expect BFKL dynamics to add much to fixed order predictions when the multiplicity is that small.
 BFKL dynamics is more relevant when we have events compatible with the multi-Regge kinematics and that implies {\it de facto} larger multiplicities.
One important complication is that the subset of events with small multiplicities when $J_{min}$ is large is the dominant one and therefore it will be this part of events that will drive the behavior of the observables we defined above. Any BFKL related effects present in the subset of events with larger multiplicities will be washed away since it provides a much smaller contribution than that of the low  multiplicities subset.

To improve this situation we can set the multiplicity of the final state to some fixed $N$ and study $\langle {\mathcal R}_y \rangle$ and $\langle {\mathcal R}_{p_T,y} \rangle$ for that specific multiplicity with the additional constraint that $N$ should be rather large. This implies of course that the available events in the experimental data sets with fixed multiplicity will be much fewer that the whole MN jets events. Setting the multiplicity to $N=3, 4, 5$ should give enough number of events for good statistics in the experimental analyses and be large enough for BFKL effects not to be overshadowed in future phenomenological analysis of LHC data.

Let us now provide a more extensive presentation of our results.

\section{Results}

In this section we will present the results or our numerical analysis performed at LO and NLO+DLs, first 
 for $\langle {\mathcal R}_{y} \rangle$ and then for 
$\langle {\mathcal R}_{p_T, y} \rangle$. In the first two subsections we mainly compare the differences found for the two ratios after increasing the rapidity span $Y$ while in the last subsection we present plots with a comparison between LO and NLO+DLs
distributions. Since this  is not a phenomenological study against data, we show normalized distributions as the best way to compare shapes and qualitative
characteristics in the plots. In any case, even in a future phenomenological analysis, if one imposes additional selection criteria
for the events in the experimental data ({\it e.g.} fixed multiplicity) a good practice to deal with systematic uncertainties should be the usage of normalized distributions. Therefore, we will present plots for $\frac{1}{\sigma_N} \frac{d \sigma_N}{d \langle {\mathcal R}_{y} \rangle}$ and $\frac{1}{\sigma_N} \frac{d \sigma_N}{d \langle {\mathcal R}_{p_T, y} \rangle}$.

We fix the minijet
multiplicity to $N=3, 4, 5$ and take a range of values of $J_{min} = 2, 10, 25$ GeV. We use the two reference points at rapidity $Y=3$ and $Y=6$ for the MN jets setting $k_a = 20$ GeV and $k_b = 30$ GeV
for their transverse momenta. This  is similar to the lower end of the kinematical cuts that can be used
in MN experimental analyses at the LHC.  Larger values of rapidity separation, {\it e.g.} $Y=8, 9$, are indeed covered by the LHC experiments and in principle would be more favourable for BFKL related effects to be manifest. However, the number of MN events in that case is really small and statistical uncertainties are dominant, especially so, when one imposes the additional restriction of a fixed jet (minijet) multiplicity. With $Y=3$ and $Y=6$ we study two distinct regions in $Y$, the former is to cover the lower end of rapidities where one
should not expect large contributions from BFKL dynamics and the latter is large enough for the effects to kick in but not too large in order to have good statistics.

\subsection{$\langle {\mathcal R}_{y} \rangle$}

\begin{figure}
\begin{subfigure}{.5\textwidth}
  \centering
  \includegraphics[width=.8\linewidth]{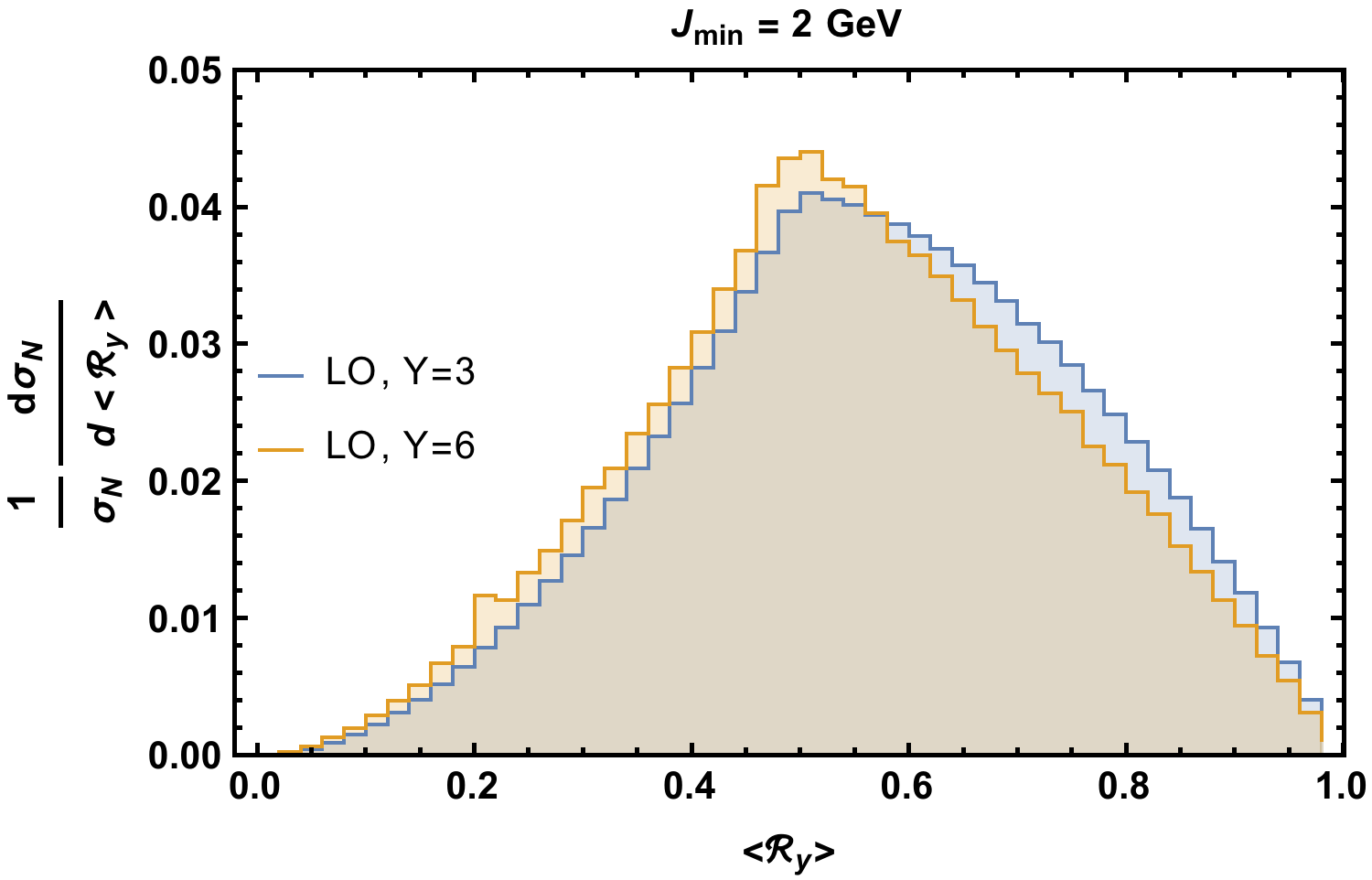}
  \caption{}
  \label{fig:sfig1}
\end{subfigure}%
\begin{subfigure}{.5\textwidth}
  \centering
  \includegraphics[width=.8\linewidth]{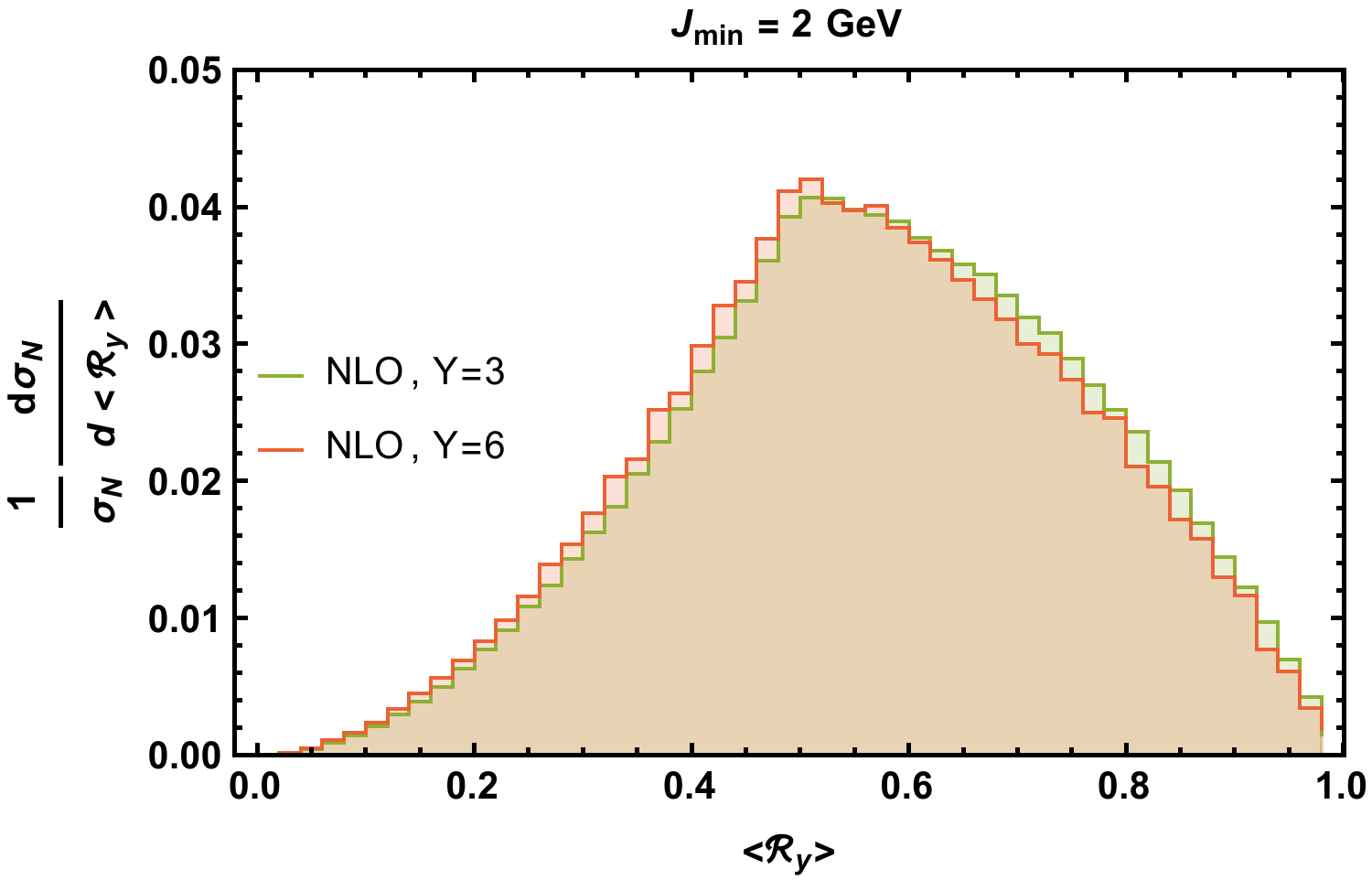}
  \caption{}
  \label{fig:sfig2}
\end{subfigure}
\\
\begin{subfigure}{.5\textwidth}
  \centering
  \includegraphics[width=.8\linewidth]{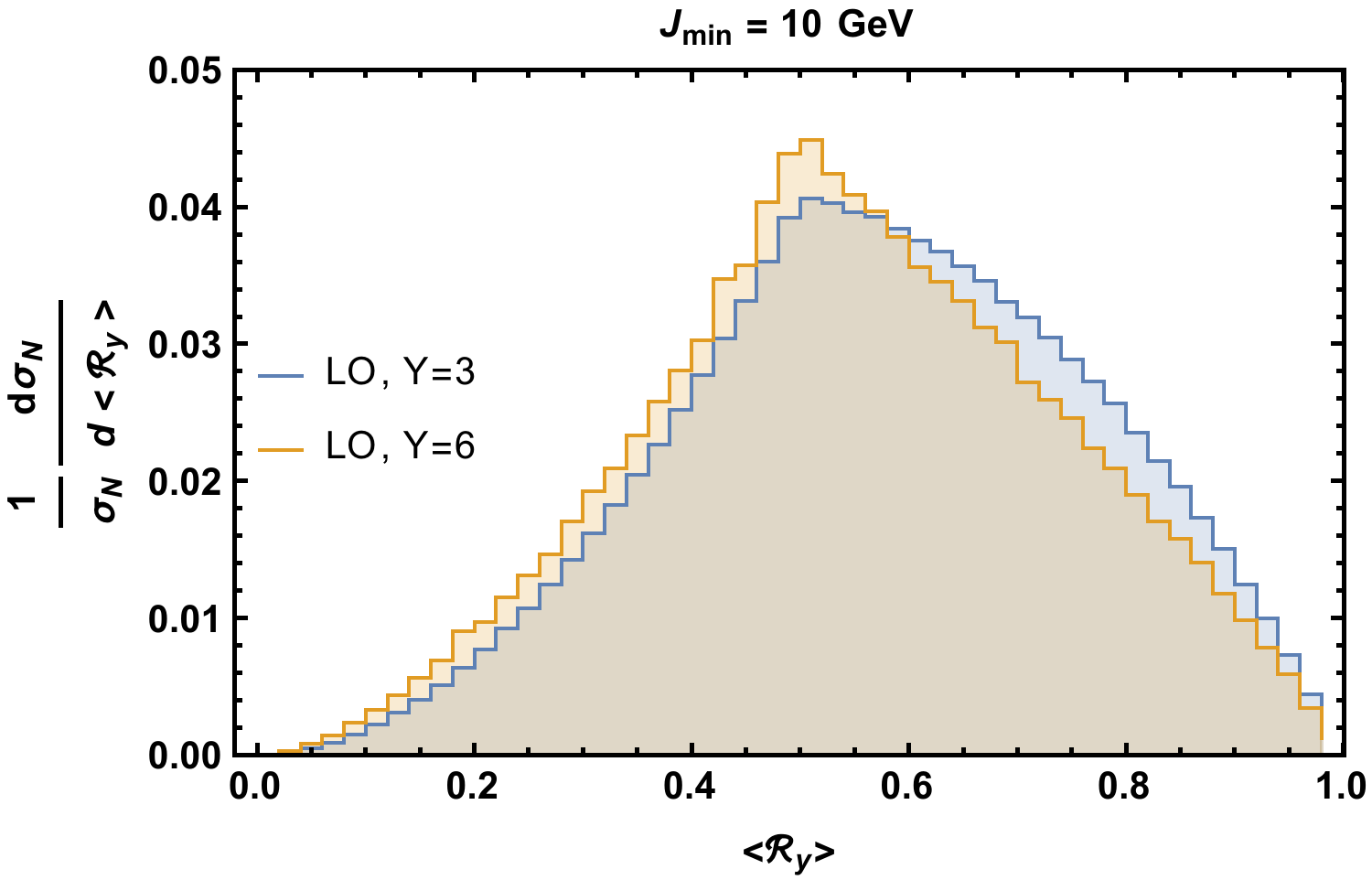}
  \caption{}
  \label{fig:sfig1}
\end{subfigure}%
\begin{subfigure}{.5\textwidth}
  \centering
  \includegraphics[width=.8\linewidth]{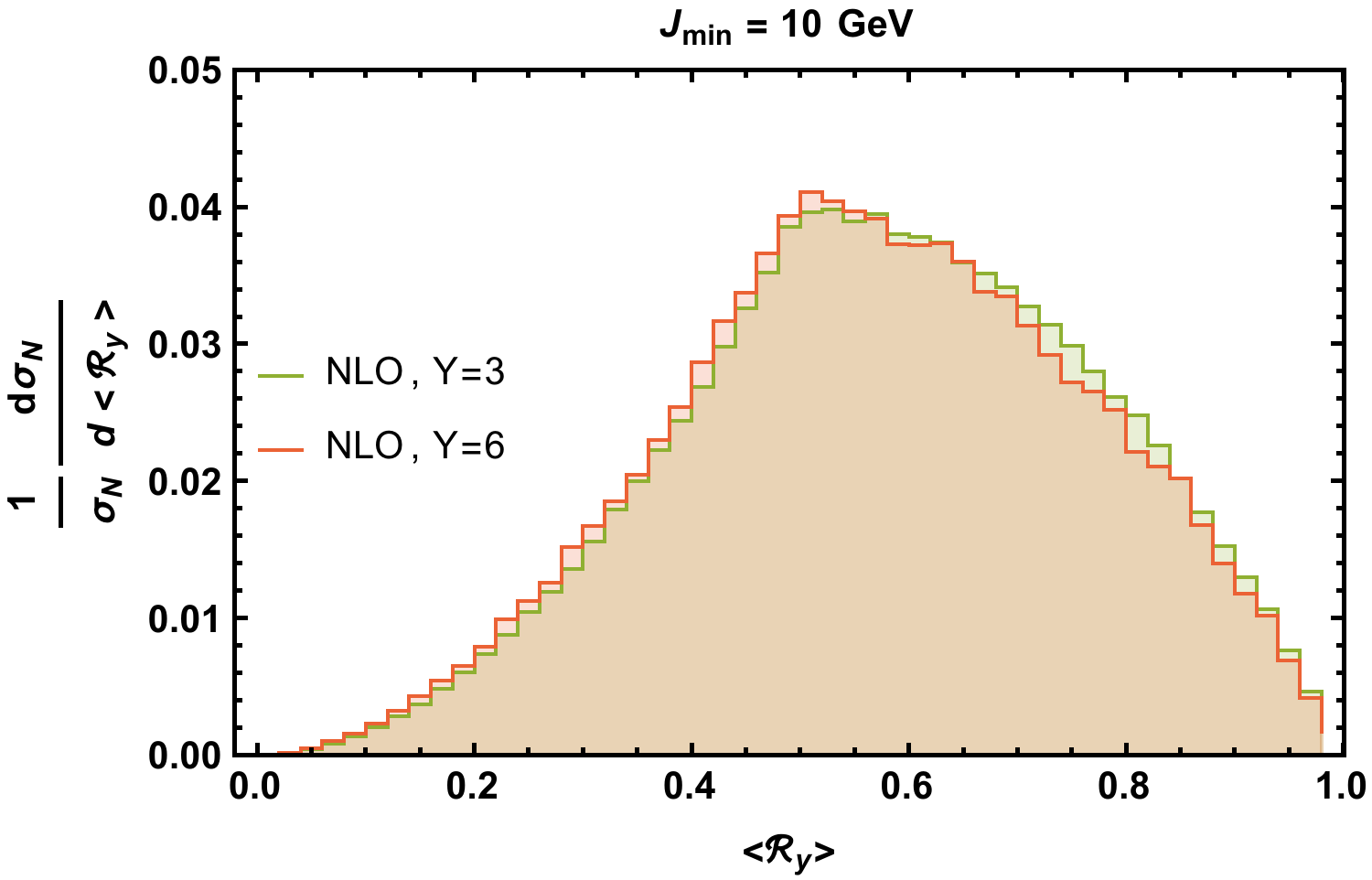}
  \caption{}
  \label{fig:sfig2}
\end{subfigure}
\\
\begin{subfigure}{.5\textwidth}
  \centering
  \includegraphics[width=.8\linewidth]{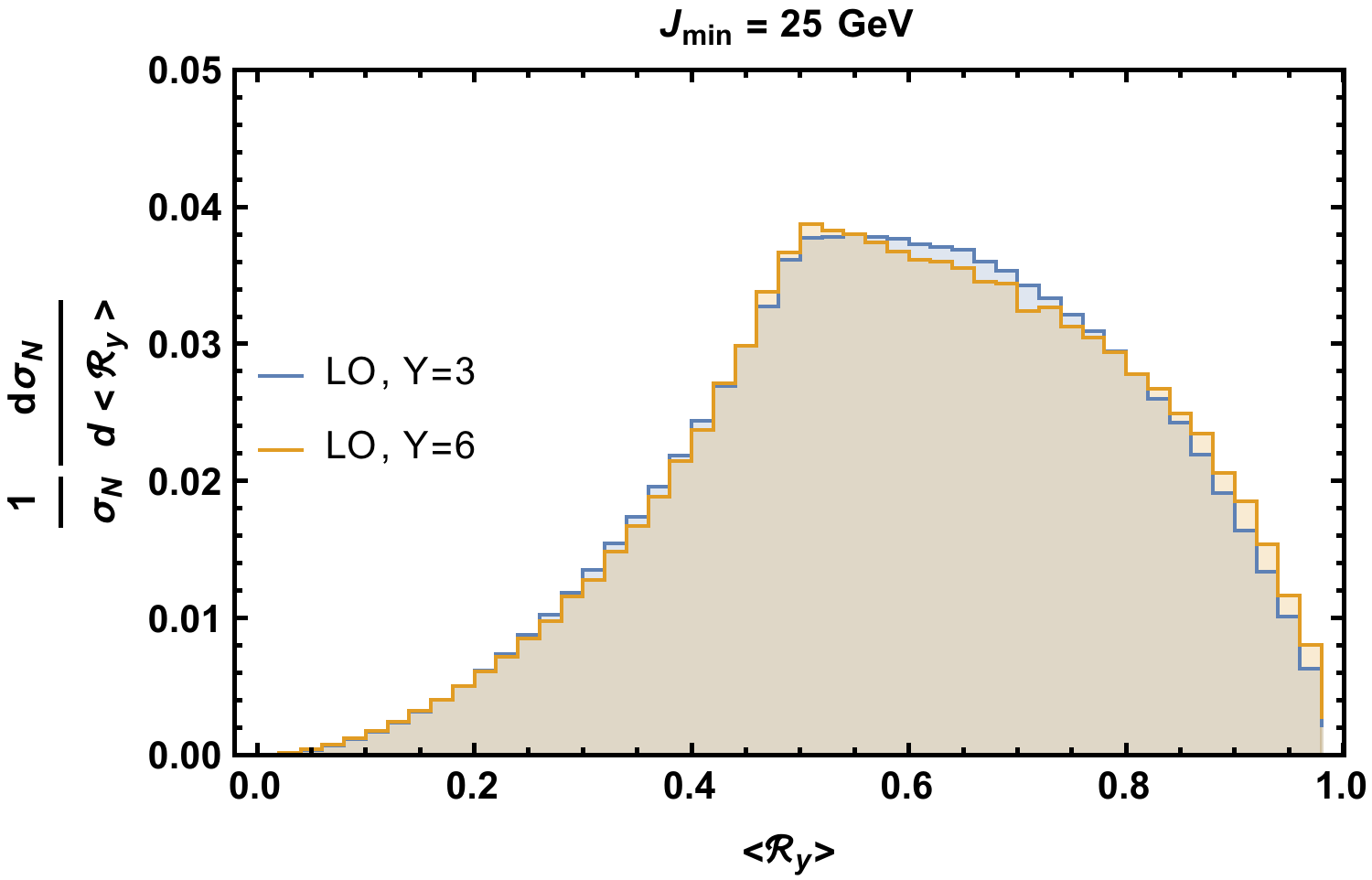}
  \caption{}
  \label{fig:sfig1}
\end{subfigure}%
\begin{subfigure}{.5\textwidth}
  \centering
  \includegraphics[width=.8\linewidth]{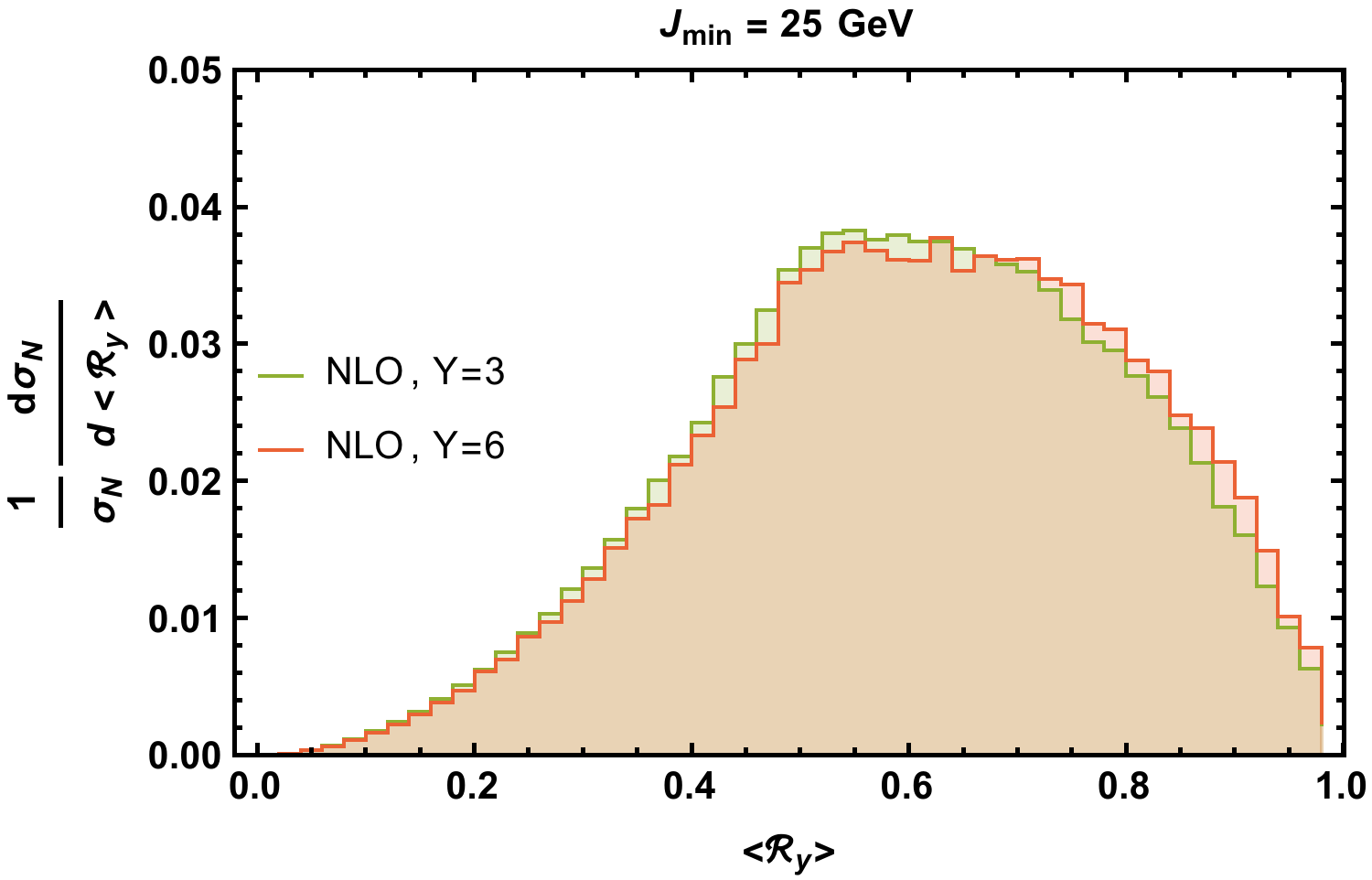}
  \caption{}
  \label{fig:sfig2}
\end{subfigure}
\caption{Normalized $\langle {\mathcal R}_{y} \rangle$ at LO (left) and NLO+DLs (right) for multiplicity $N = 3$, rapidity difference $Y = 3, 6$ and $J_{\text{min}} = 2$ (top), $J_{\text{min}} = 10$ (middle), $J_{\text{min}} = 25$ (bottom).}
\label{fig:ry3}
\end{figure}

\begin{figure}
\begin{subfigure}{.5\textwidth}
  \centering
  \includegraphics[width=.8\linewidth]{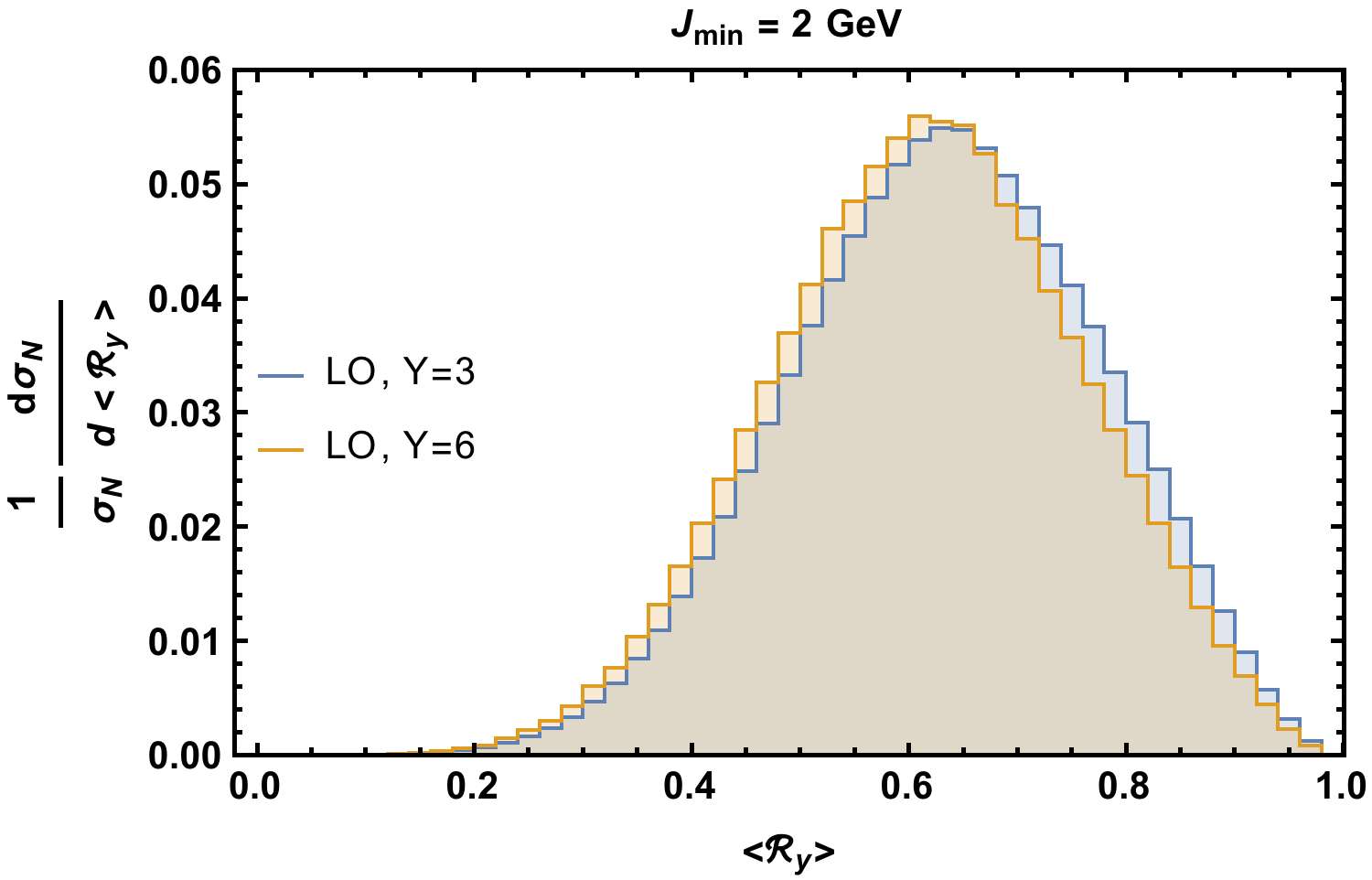}
  \caption{}
  \label{fig:sfig1}
\end{subfigure}%
\begin{subfigure}{.5\textwidth}
  \centering
  \includegraphics[width=.8\linewidth]{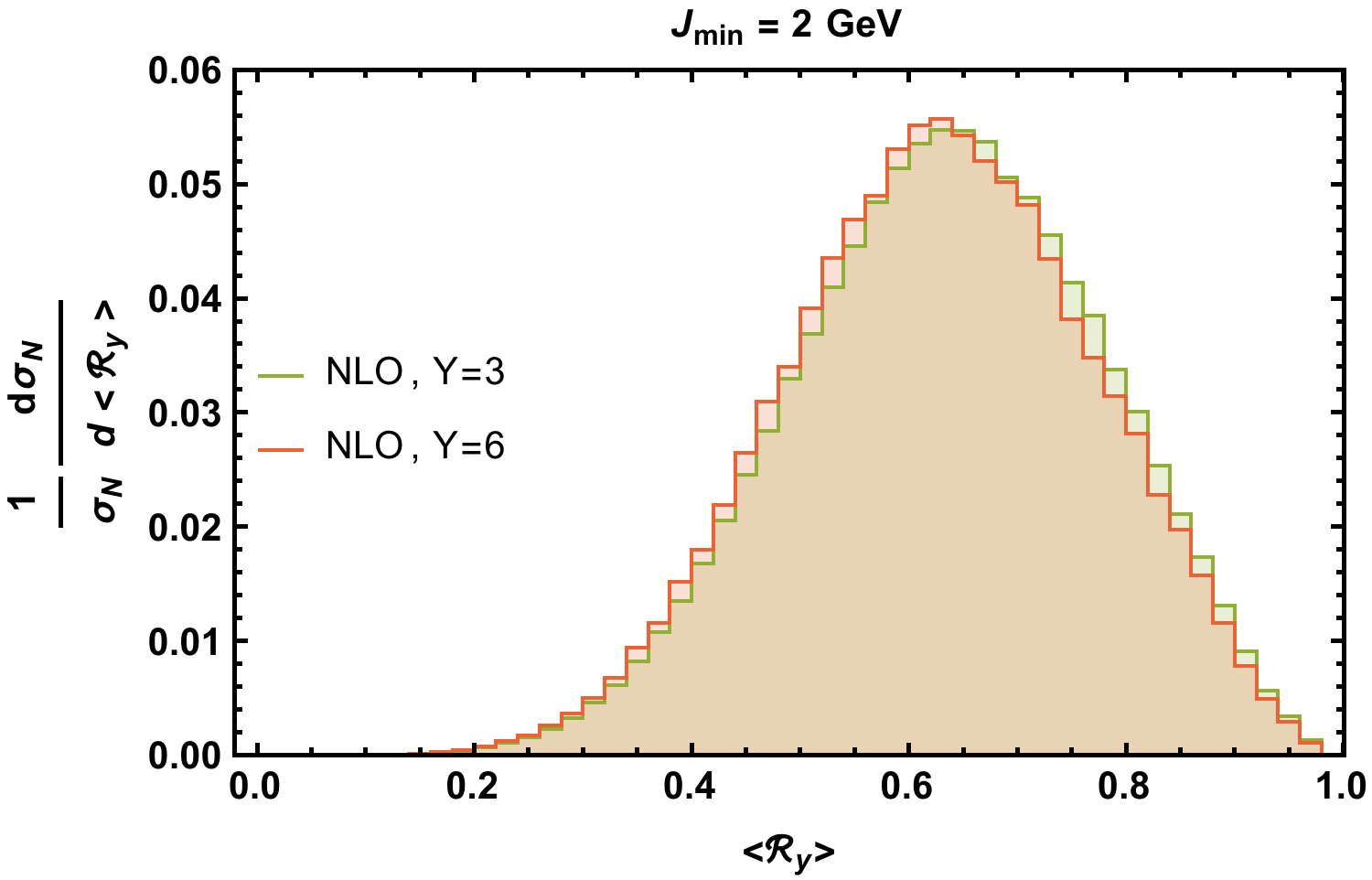}
  \caption{}
  \label{fig:sfig2}
\end{subfigure}
\\
\begin{subfigure}{.5\textwidth}
  \centering
  \includegraphics[width=.8\linewidth]{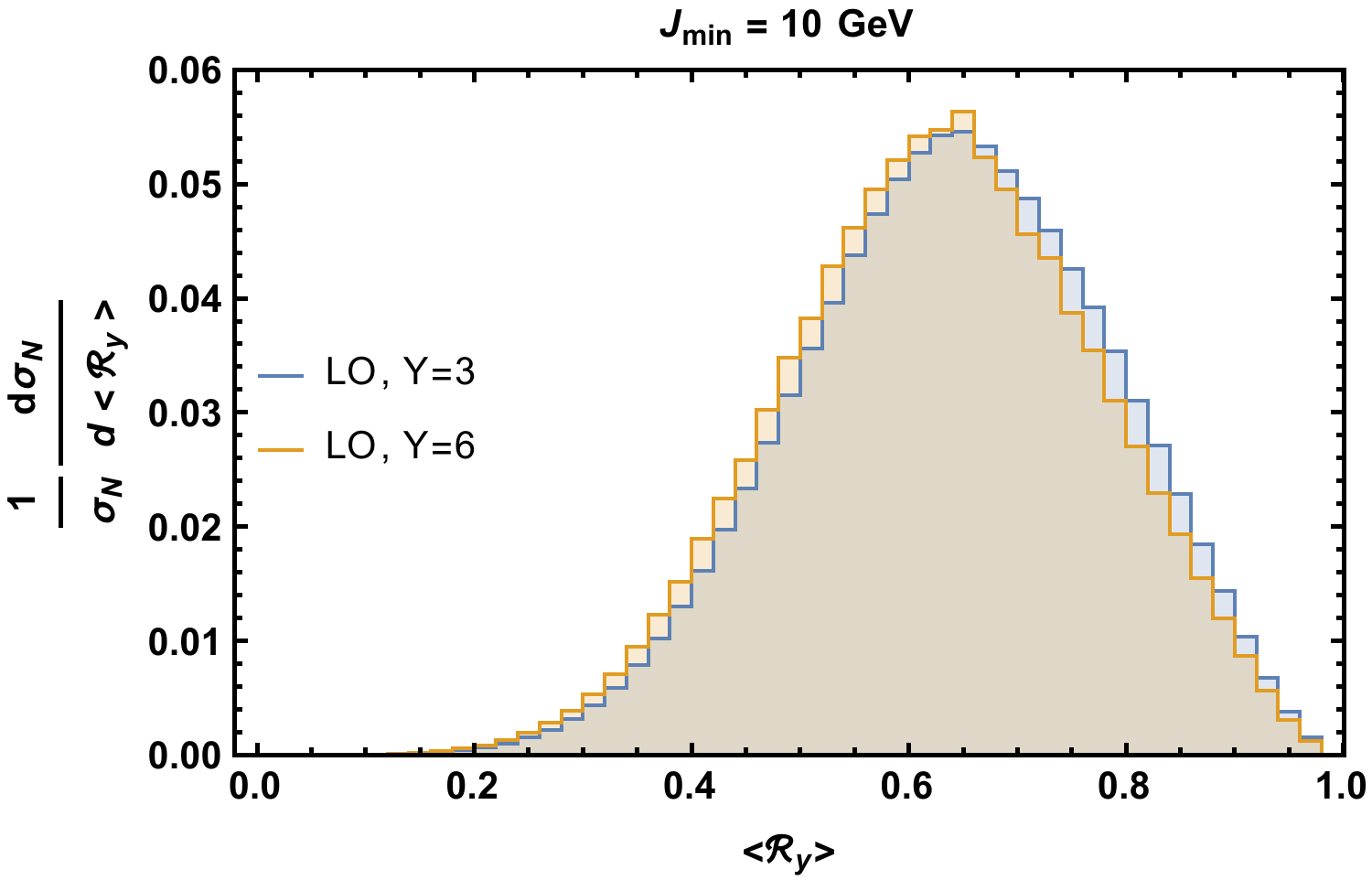}
  \caption{}
  \label{fig:sfig1}
\end{subfigure}%
\begin{subfigure}{.5\textwidth}
  \centering
  \includegraphics[width=.8\linewidth]{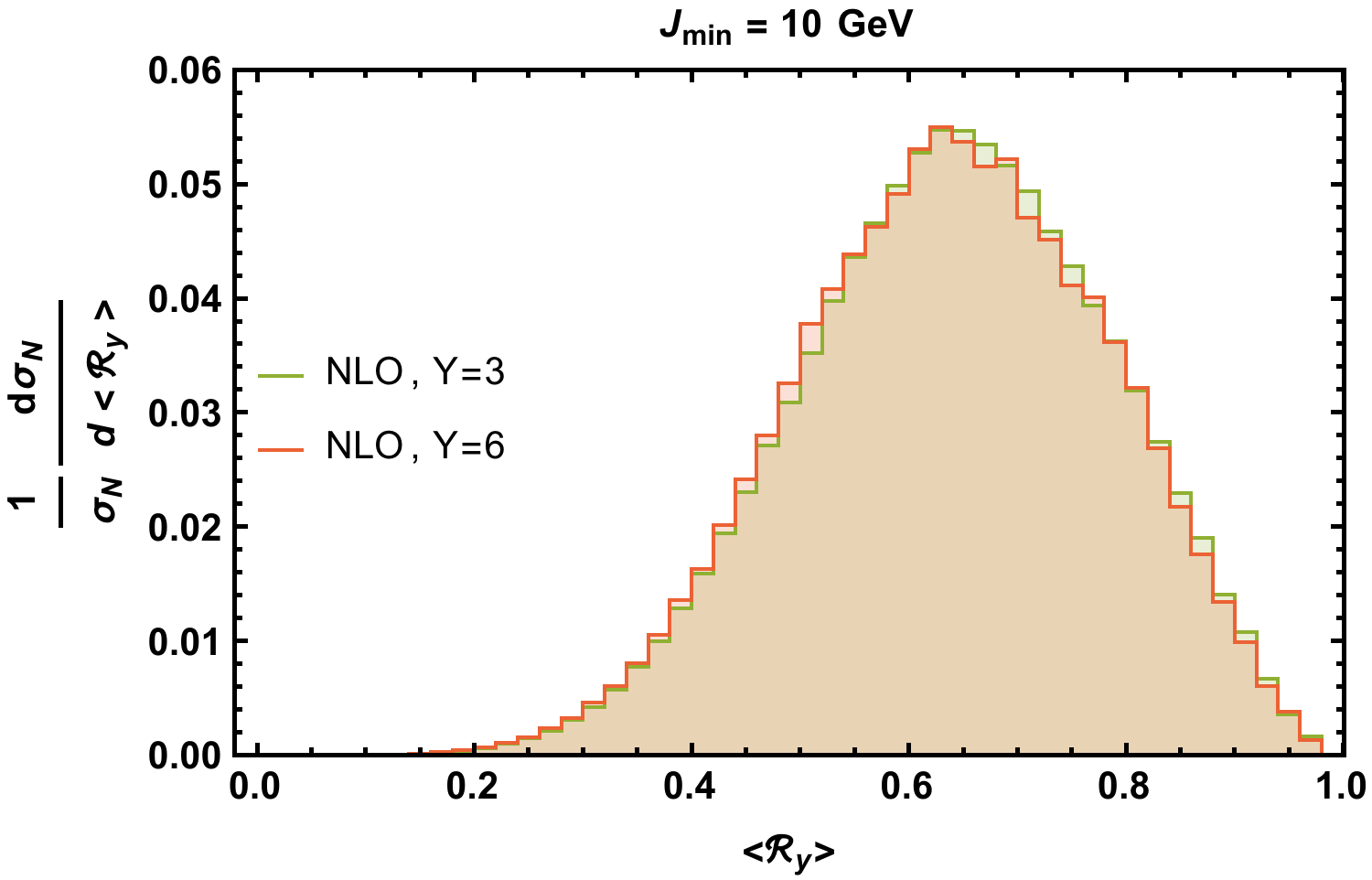}
  \caption{}
  \label{fig:sfig2}
\end{subfigure}
\\
\begin{subfigure}{.5\textwidth}
  \centering
  \includegraphics[width=.8\linewidth]{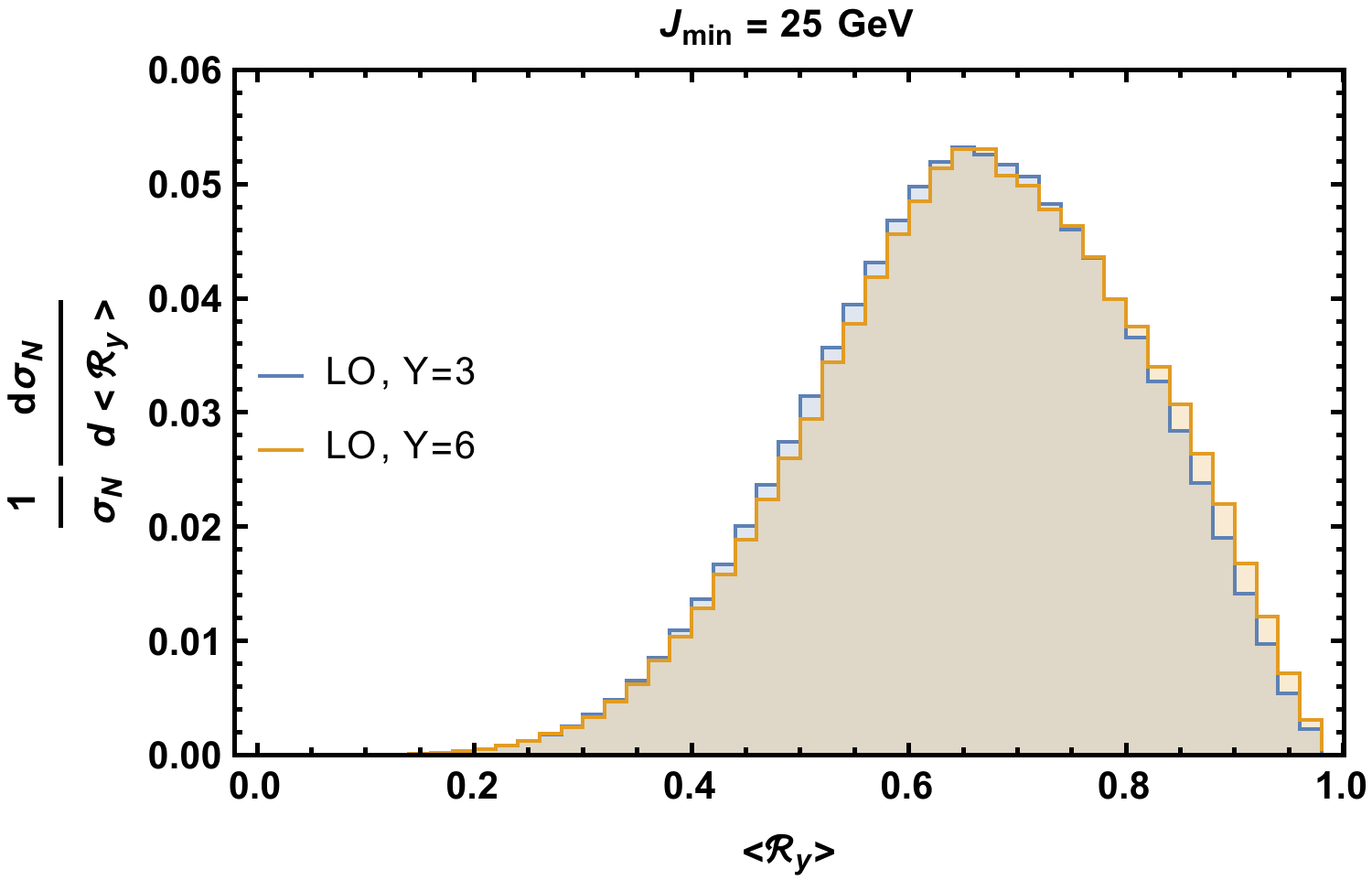}
  \caption{}
  \label{fig:sfig1}
\end{subfigure}%
\begin{subfigure}{.5\textwidth}
  \centering
  \includegraphics[width=.8\linewidth]{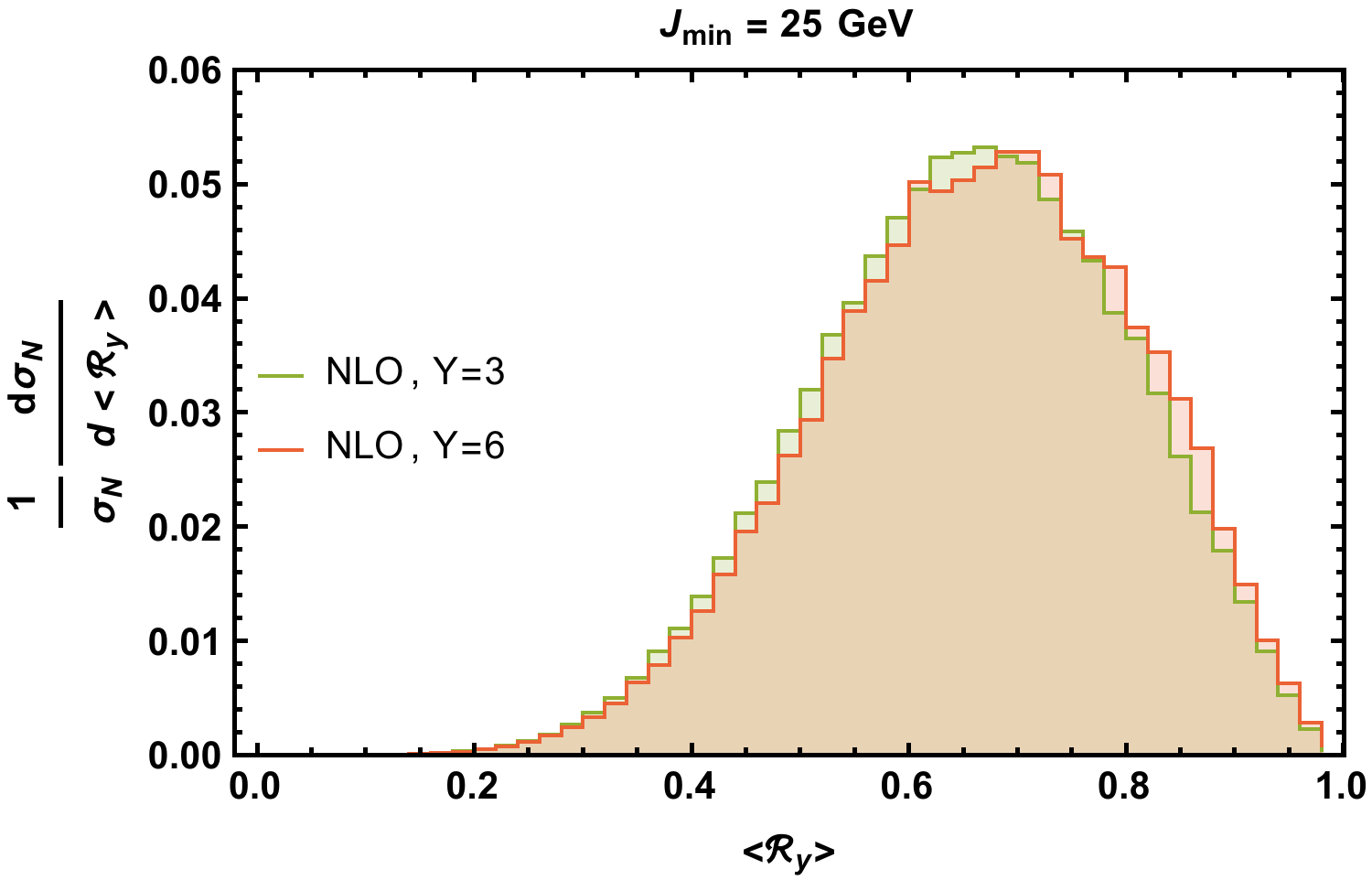}
  \caption{}
  \label{fig:sfig2}
\end{subfigure}
\caption{Normalized $\langle {\mathcal R}_{y} \rangle$ at LO (left) and NLO+DLs (right) for multiplicity $N = 4$, rapidity difference $Y = 3, 6$ and $J_{\text{min}} = 2$ (top), $J_{\text{min}} = 10$ (middle), $J_{\text{min}} = 25$ (bottom).}
\label{fig:ry4}
\end{figure}

\begin{figure}
\begin{subfigure}{.5\textwidth}
  \centering
  \includegraphics[width=.8\linewidth]{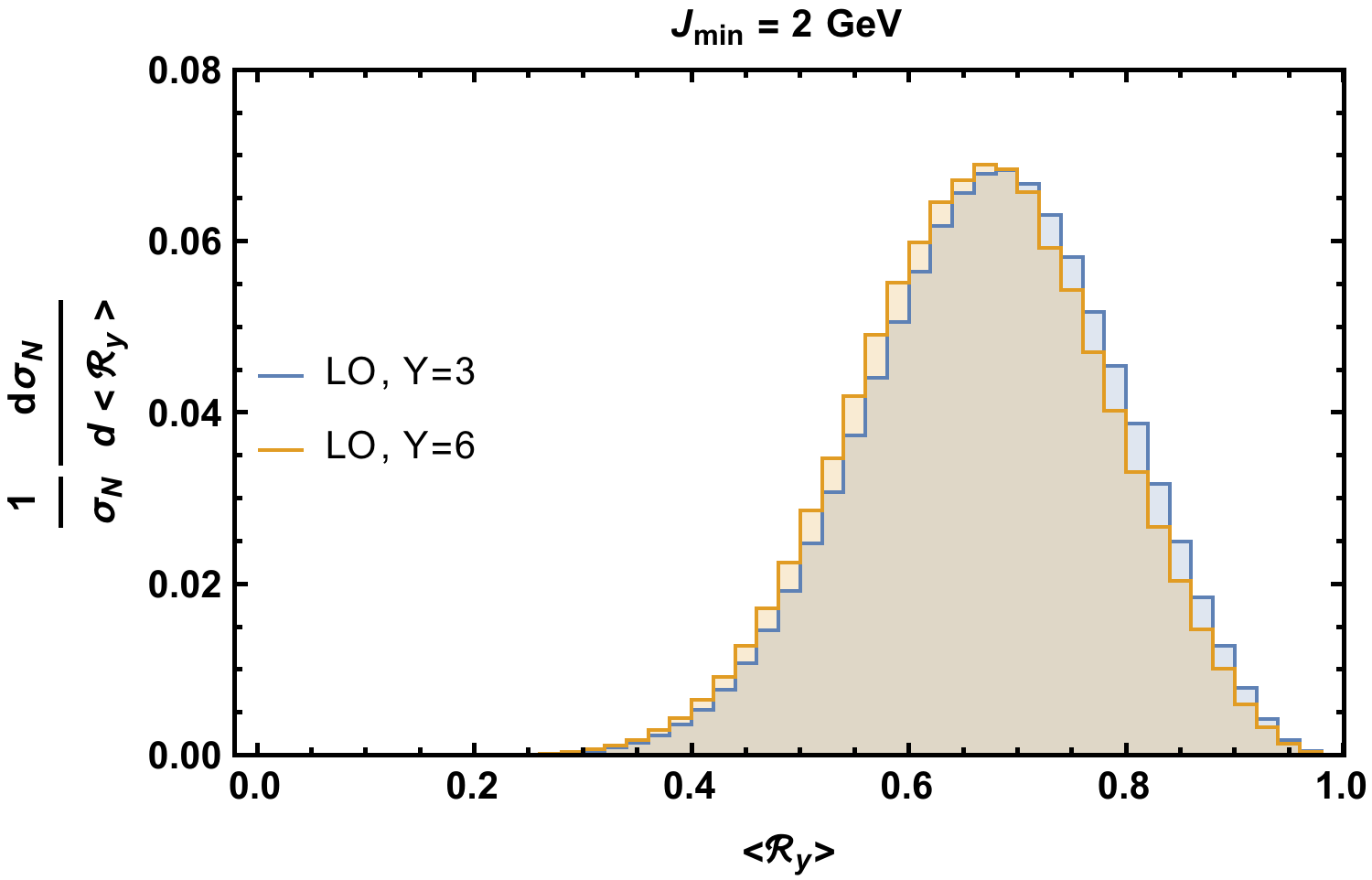}
  \caption{}
  \label{fig:sfig1}
\end{subfigure}%
\begin{subfigure}{.5\textwidth}
  \centering
  \includegraphics[width=.8\linewidth]{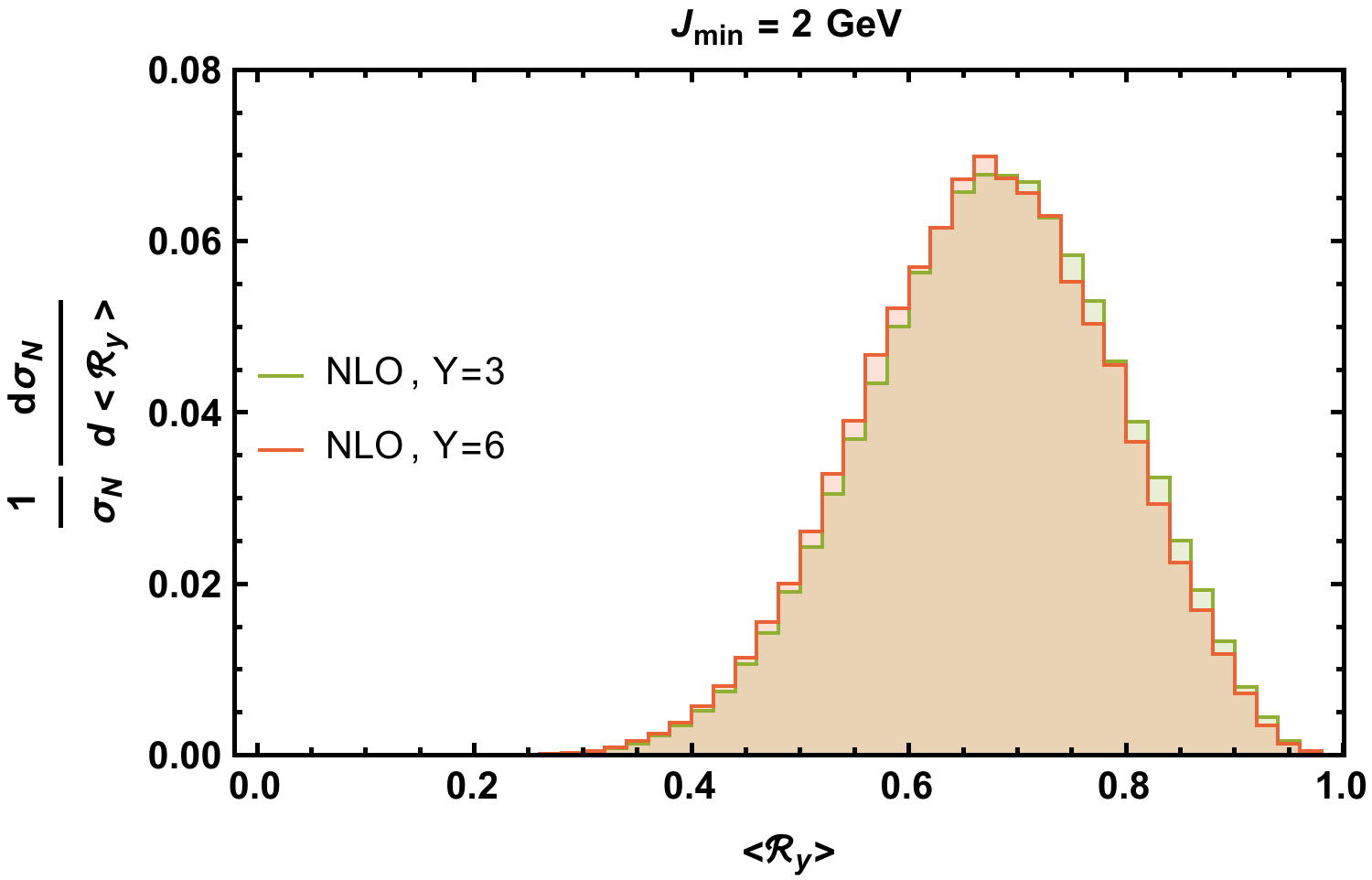}
  \caption{}
  \label{fig:sfig2}
\end{subfigure}
\\
\begin{subfigure}{.5\textwidth}
  \centering
  \includegraphics[width=.8\linewidth]{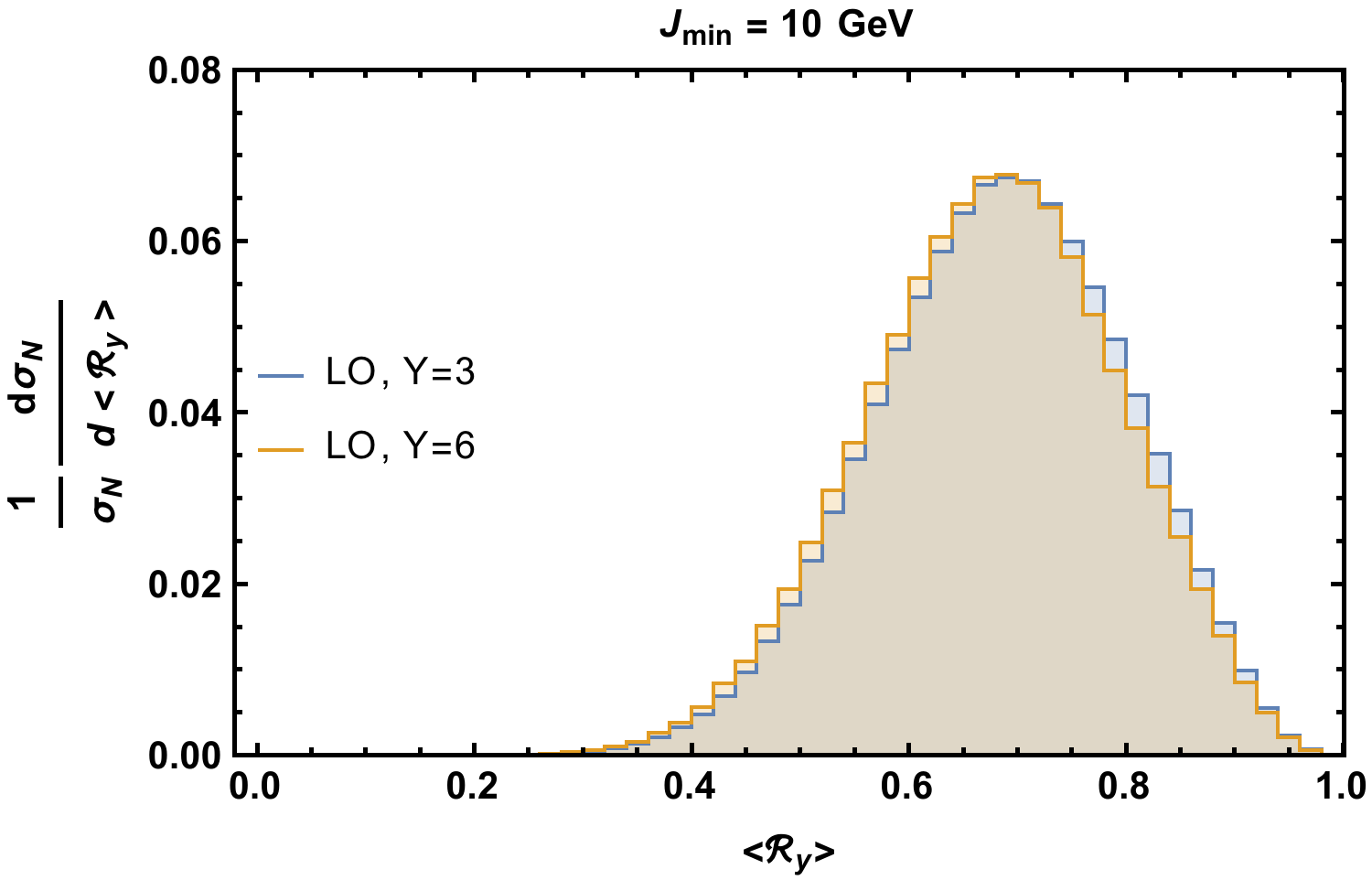}
  \caption{}
  \label{fig:sfig1}
\end{subfigure}%
\begin{subfigure}{.5\textwidth}
  \centering
  \includegraphics[width=.8\linewidth]{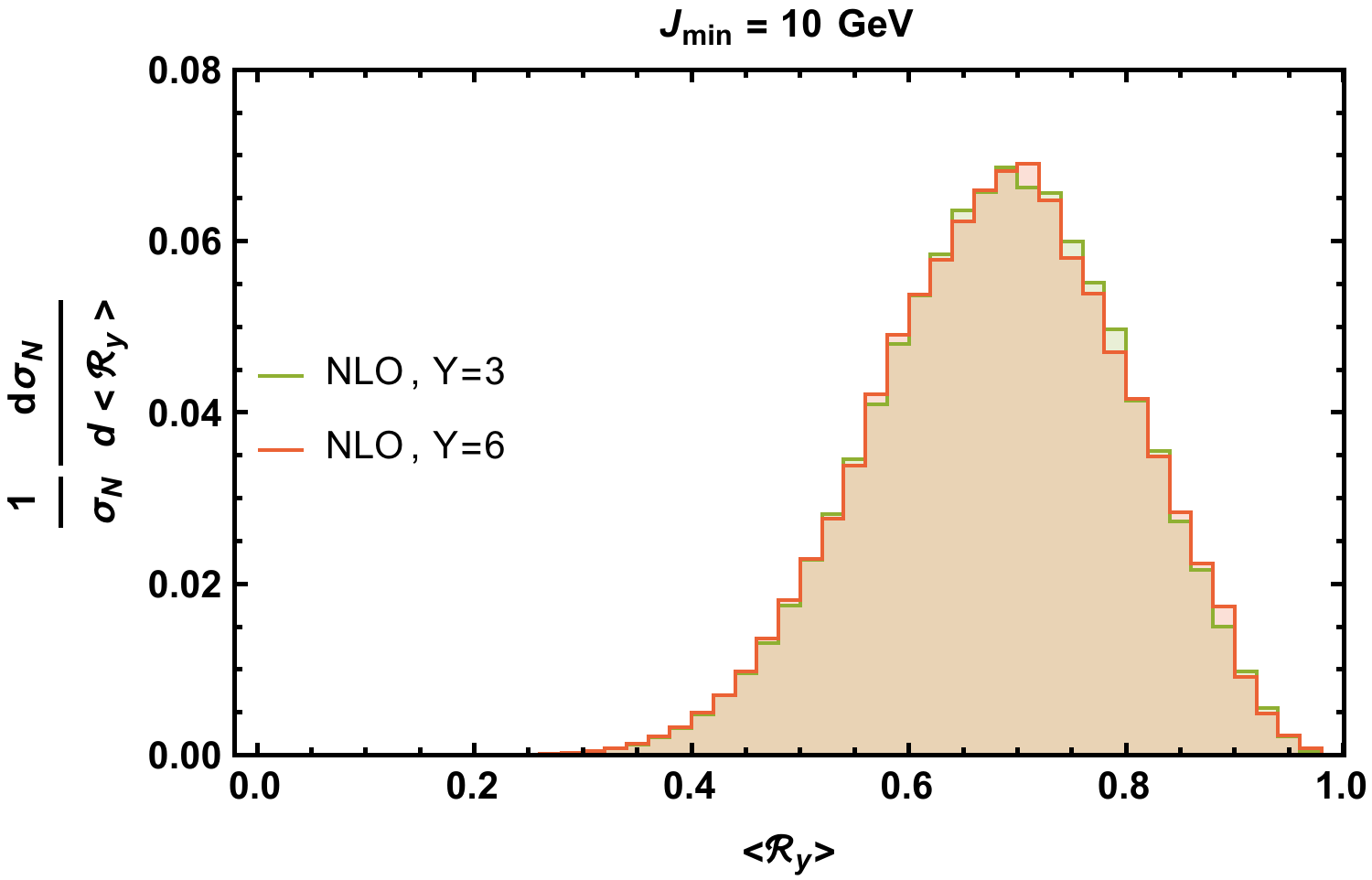}
  \caption{}
  \label{fig:sfig2}
\end{subfigure}
\\
\begin{subfigure}{.5\textwidth}
  \centering
  \includegraphics[width=.8\linewidth]{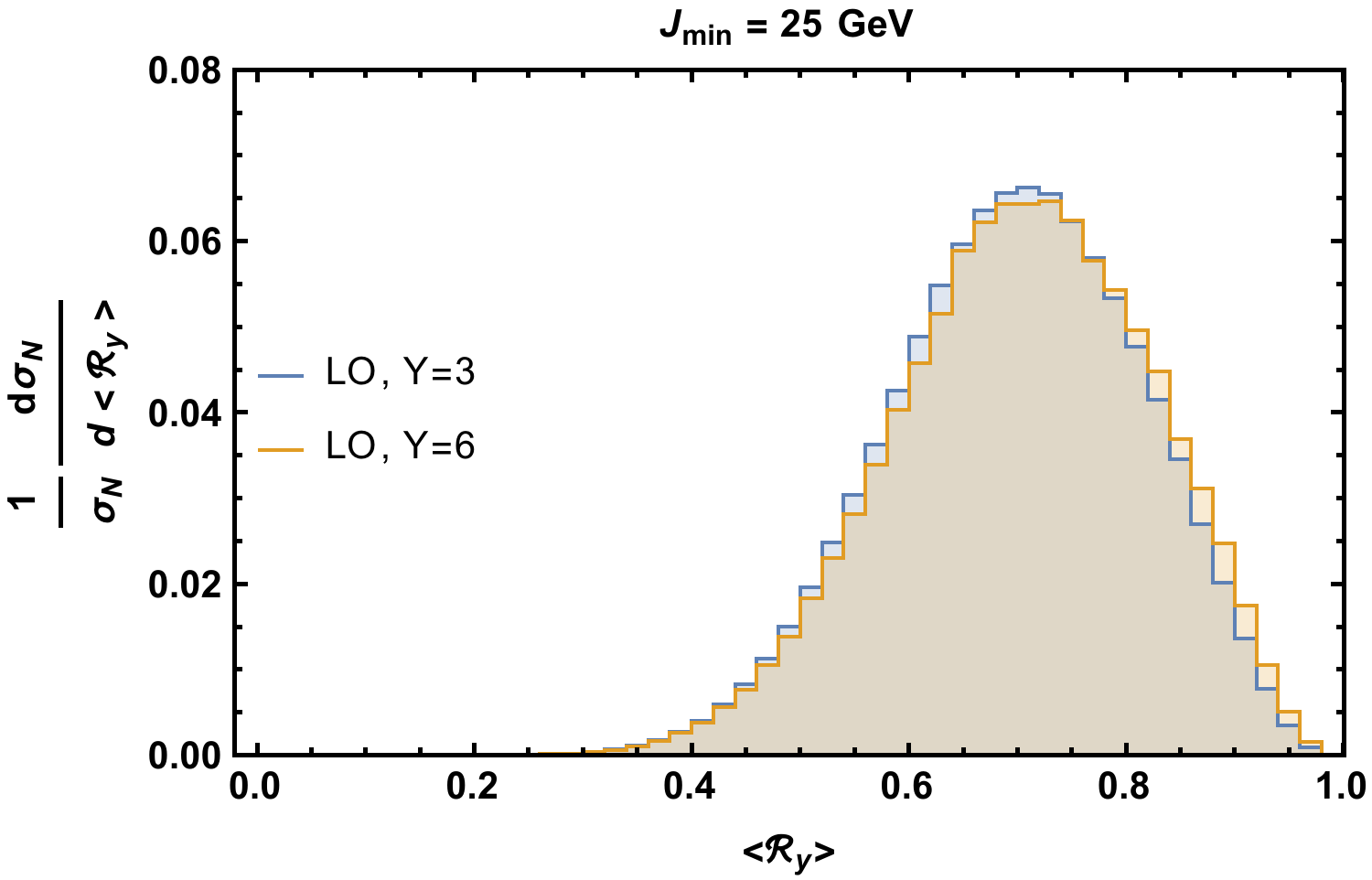}
  \caption{}
  \label{fig:sfig1}
\end{subfigure}%
\begin{subfigure}{.5\textwidth}
  \centering
  \includegraphics[width=.8\linewidth]{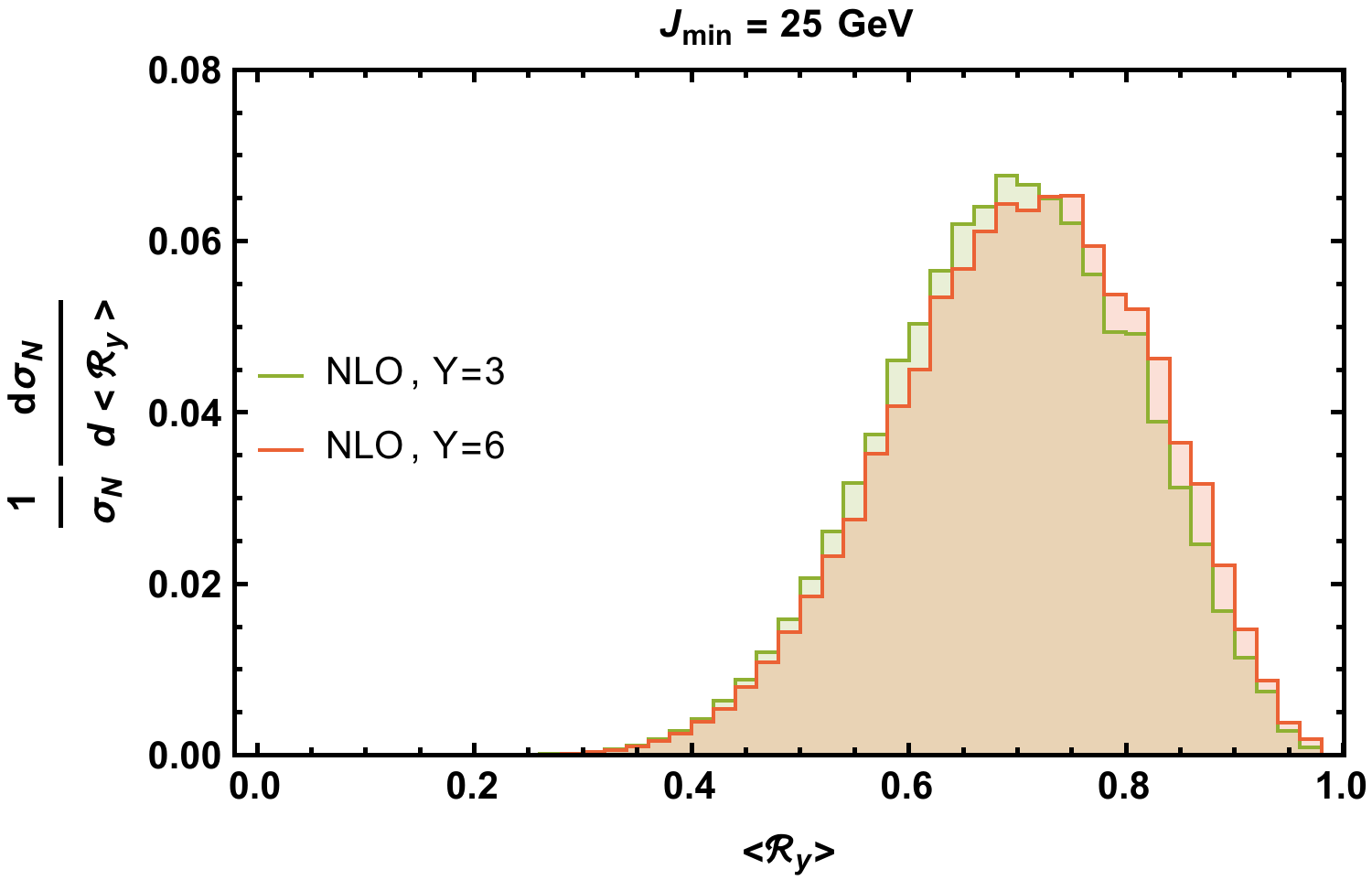}
  \caption{}
  \label{fig:sfig2}
\end{subfigure}
\caption{Normalized $\langle {\mathcal R}_{y} \rangle$ at LO (left) and NLO+DLs (right) for multiplicity $N = 5$, rapidity difference $Y = 3, 6$ and $J_{\text{min}} = 2$ (top), $J_{\text{min}} = 10$ (middle), $J_{\text{min}} = 25$ (bottom).}
\label{fig:ry5}
\end{figure}

In Figs.~\ref{fig:ry3},~\ref{fig:ry4} and ~\ref{fig:ry5} we plot $\frac{1}{\sigma_N} \frac{d \sigma_N}{d \langle {\mathcal R}_{y} \rangle}$ for $N=3, 4, 5$ multiplicities respectively. In each subplot of all three figures, we show two distributions, one
curve is reserved for rapidity $Y=3$ and the other for $Y=6$ whereas, subplots $(a), (c)$ and $(e)$ are at LO and $(b), (d)$ 
and $(f)$ are at NLO+DLs. The two subplots on the top row have $J_{min} = 2$ GeV, the ones in the middle have $J_{min} = 10$ GeV
and the two subplots at the bottom have $J_{min} = 25$ GeV. In all cases, the values of $\langle {\mathcal R}_{y} \rangle$ are bounded by 0 and 1.

For multiplicity $N=3$, in Fig.~\ref{fig:ry3} we see that there is a peak at around $\langle {\mathcal R}_{y} \rangle = 0.5$ and the distributions are quite broad. There is a clear asymmetry between the $\langle {\mathcal R}_{y} \rangle < 0.5$ and
$\langle {\mathcal R}_{y} \rangle > 0.5$ regions that has to do with the asymmetry in $k_a$ and $k_b$. Since we need to have
momentum conservation on the transverse plane, $\vec{k}_b - \vec{k}_a - \vec{k}_1 - \vec{k}_2 = \vec{k}_3$ which really means that only two out of three minijets are generated freely and hence the actual values of the two outermost jets do have an impact. That effect is more pronounced as $J_{min}$ becomes larger.
While the distributions for $Y=3,6$ at LO and for smaller values of $J_{min}$ present some differences, they are very similar at NLO+DLs regardless of $J_{min}$ and the same holds at LO but for $J_{min} = 25$ GeV.

For multiplicity $N=4$, in Fig.~\ref{fig:ry4} we see that the peak is shifted to larger values and in particular a bit beyond
 $\langle {\mathcal R}_{y} \rangle = 0.65$. The distributions are considerably less broad
 and  more symmetric to the left and right of the peak although there is still some noticeable skewness. 
We again see noticeable differences between the distributions for $Y=3,6$ at LO and for smaller values of $J_{min}$ (Fig.~\ref{fig:ry4}$(a),(b)$) while in the remaining subplots the distributions for the two values of rapidity are very similar. The peak seems to shift very slightly to larger values as $J_{min}$ becomes larger.

Lastly, in Fig.~\ref{fig:ry5} for $N=5$ the same trend continues. We see that the peak is shifted to even larger values, closer now to
 $\langle {\mathcal R}_{y} \rangle \sim 0.7$. The distributions are more narrow
 and more symmetric than for $N=4$, nevertheless, the tail of the distributions at small values drops more gradually than the tail near $\langle {\mathcal R}_{y} \rangle \sim 1$. They are also more similar for the two rapidity values $Y=3, 6$ in all six subplots than in smaller multiplicities. The peak again shifts very slightly to larger values as $J_{min}$ increases.

It is remarkable that for $J_{min} = 2$ and especially for $J_{min} = 10$, in all Figs.~\ref{fig:ry3},~\ref{fig:ry4},~\ref{fig:ry5}, the distributions  at NLO+DLs are very similar for $Y=3$ and $Y=6$.

Last but not least, these observables are invariant under the introduction of higher order corrections. This provides a very robust set of predictions from the theoretical view point.

\subsection{$\langle {\mathcal R}_{p_T, y} \rangle$}

\begin{figure}
\begin{subfigure}{.5\textwidth}
  \centering
  \includegraphics[width=.8\linewidth]{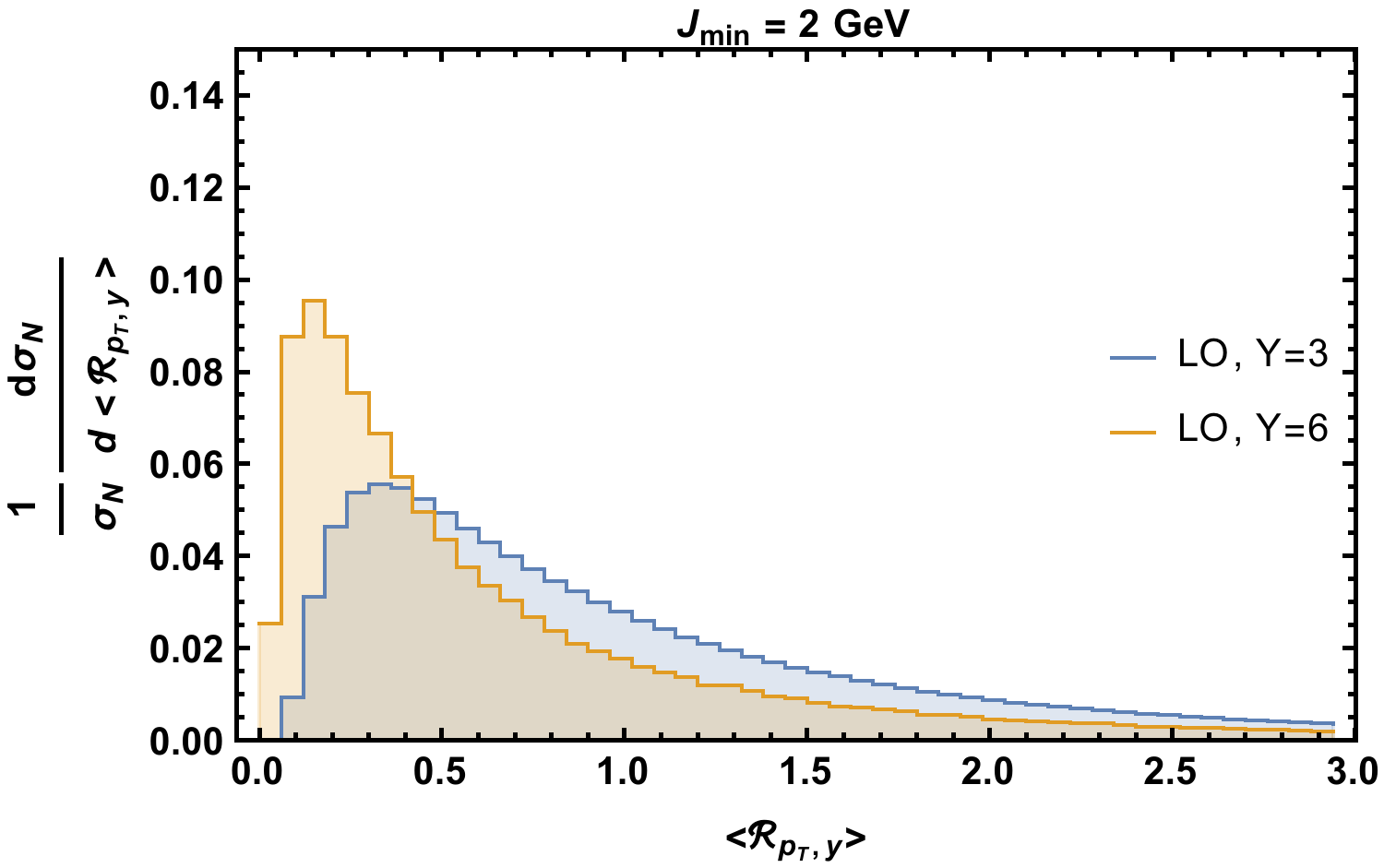}
  \caption{}
  \label{fig:sfig1}
\end{subfigure}%
\begin{subfigure}{.5\textwidth}
  \centering
  \includegraphics[width=.8\linewidth]{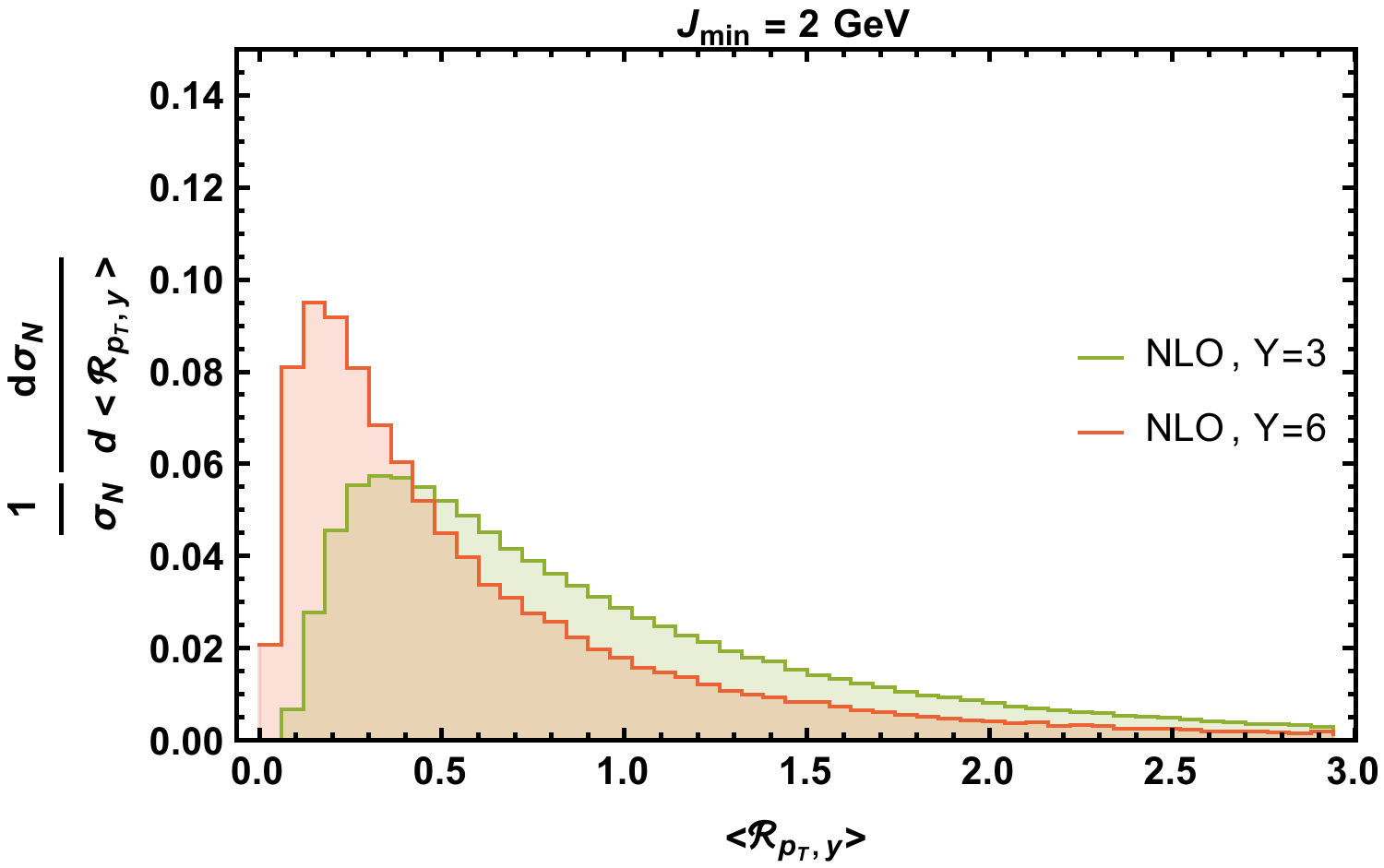}
  \caption{}
  \label{fig:sfig2}
\end{subfigure}
\\
\begin{subfigure}{.5\textwidth}
  \centering
  \includegraphics[width=.8\linewidth]{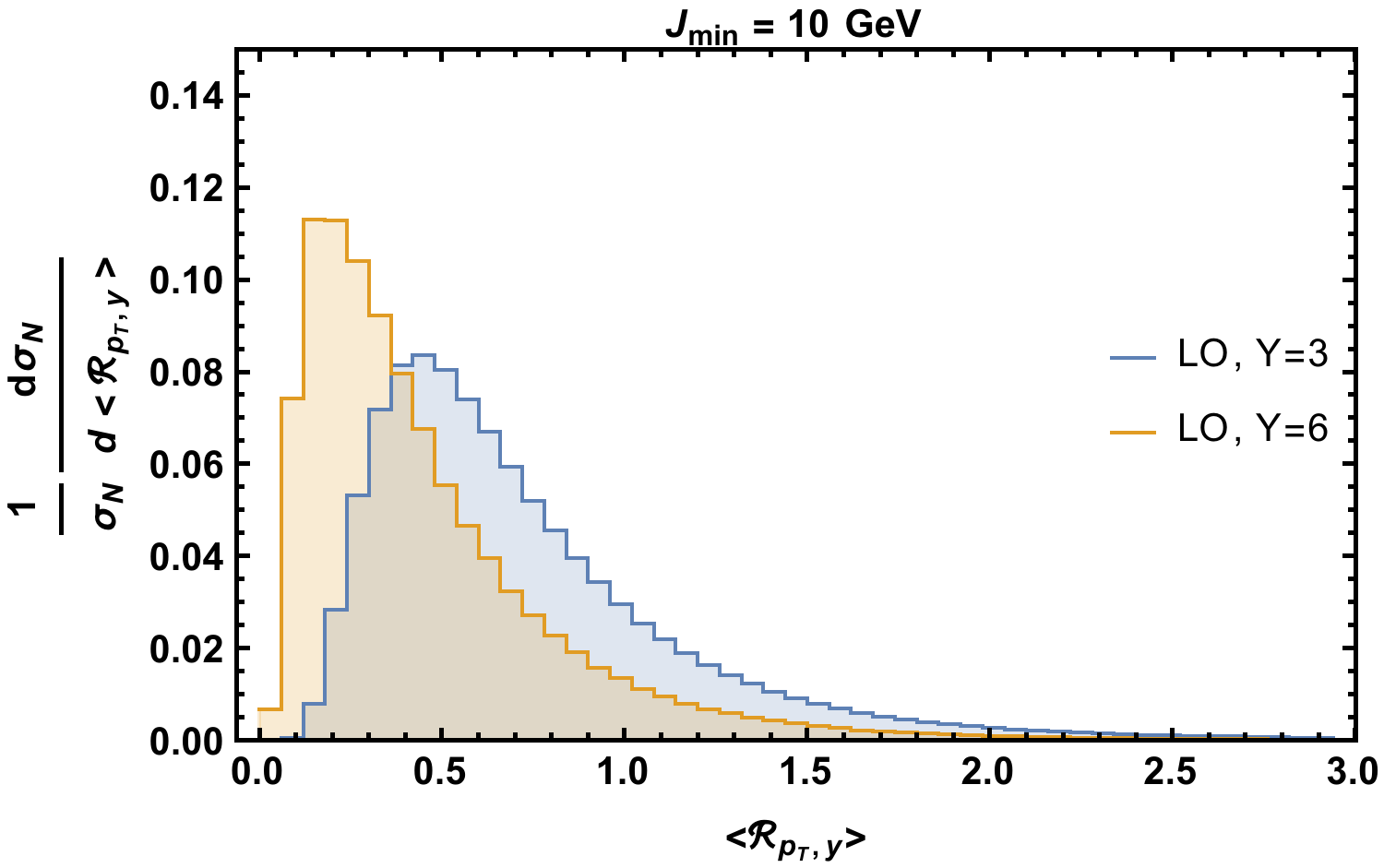}
  \caption{}
  \label{fig:sfig1}
\end{subfigure}%
\begin{subfigure}{.5\textwidth}
  \centering
  \includegraphics[width=.8\linewidth]{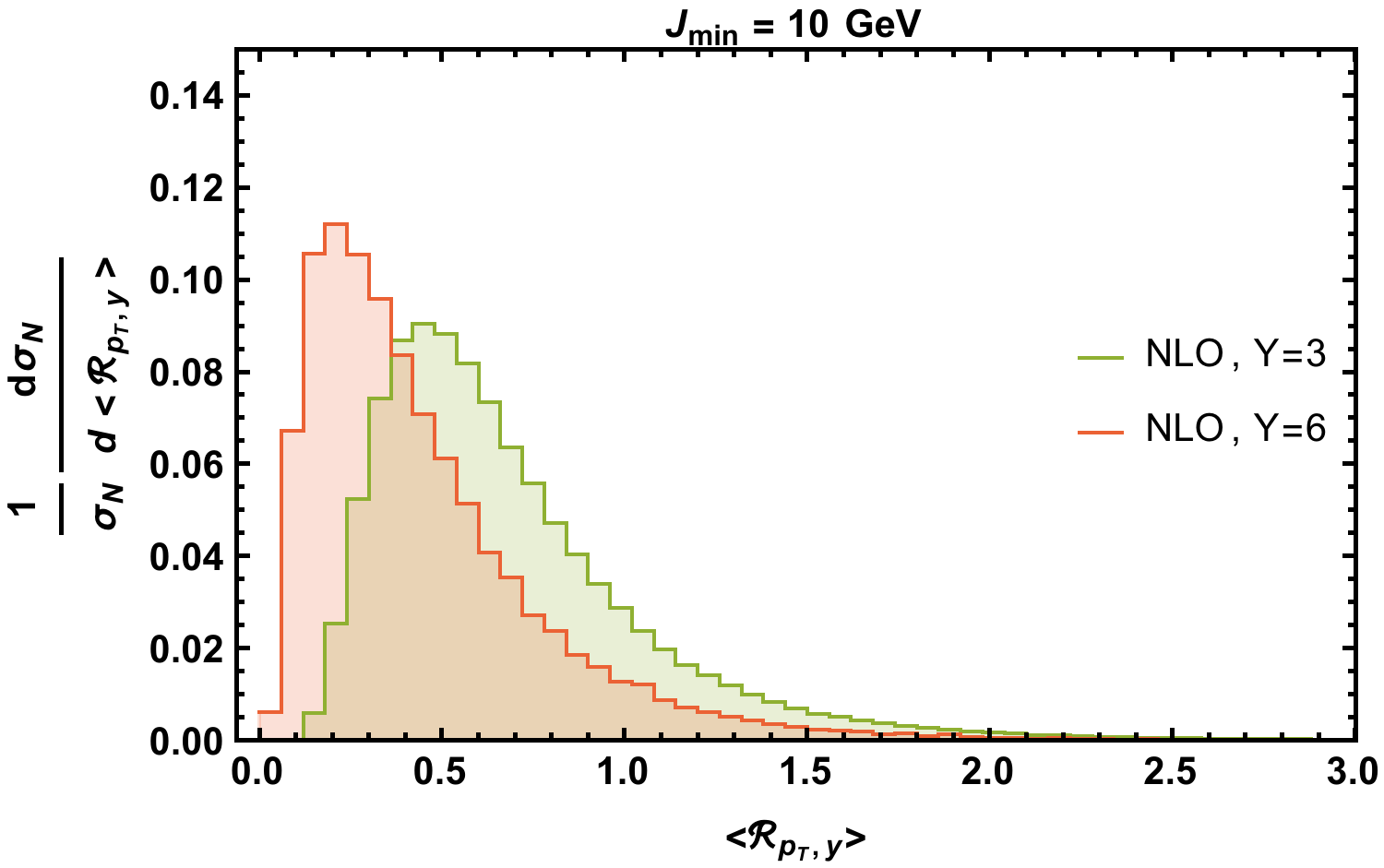}
  \caption{}
  \label{fig:sfig2}
\end{subfigure}
\\
\begin{subfigure}{.5\textwidth}
  \centering
  \includegraphics[width=.8\linewidth]{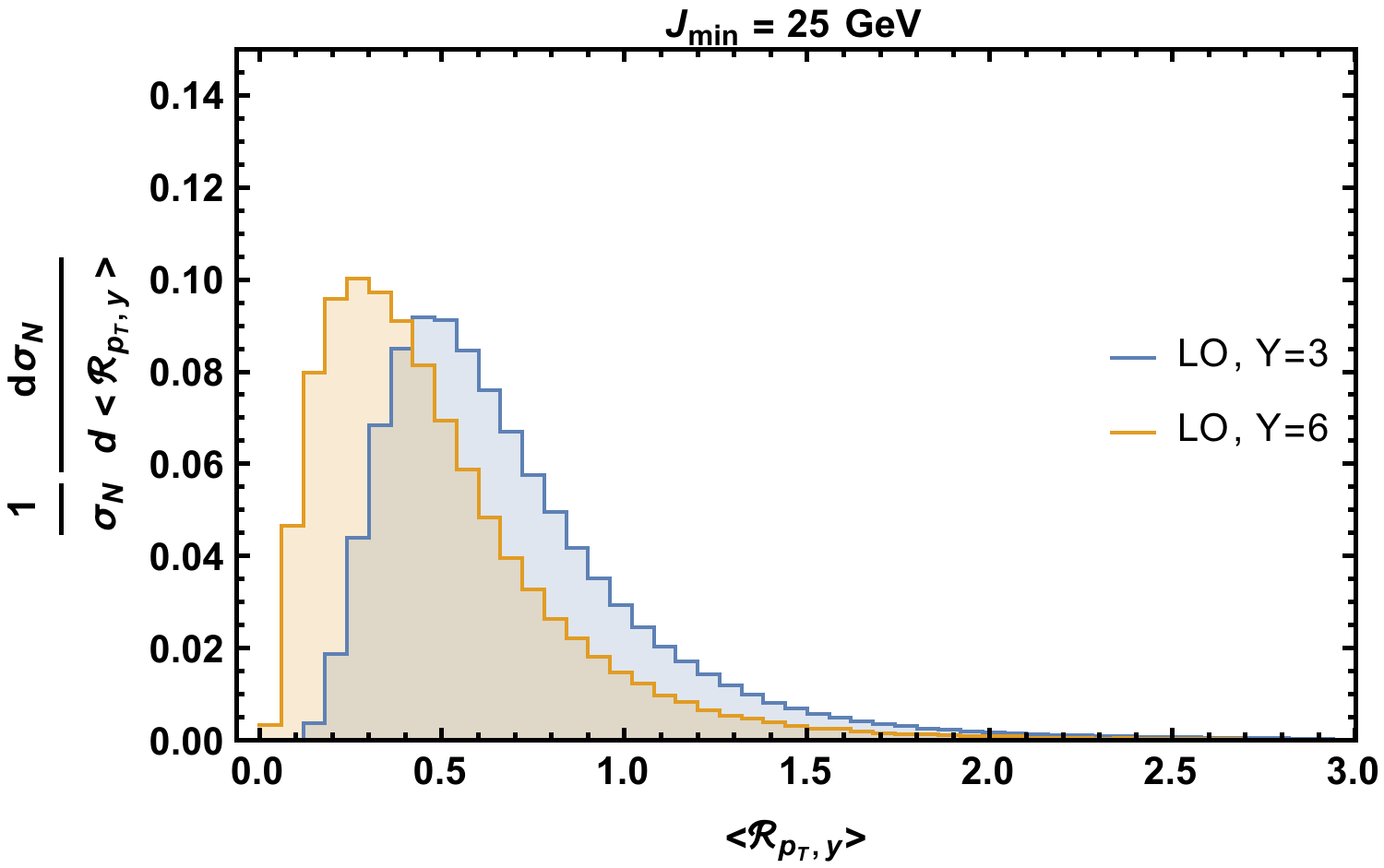}
  \caption{}
  \label{fig:sfig1}
\end{subfigure}%
\begin{subfigure}{.5\textwidth}
  \centering
  \includegraphics[width=.8\linewidth]{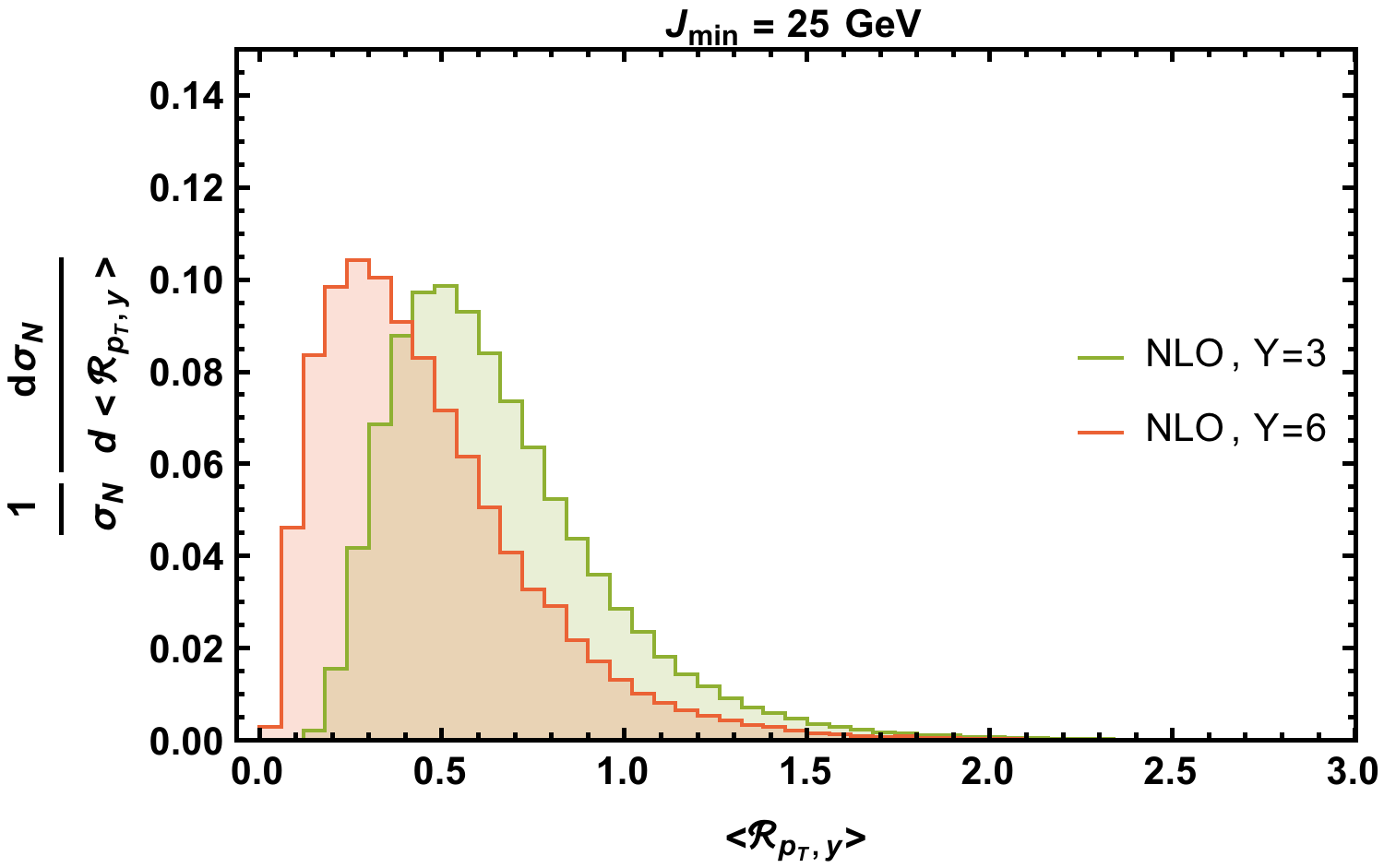}
  \caption{}
  \label{fig:sfig2}
\end{subfigure}
\caption{Normalized $\langle {\mathcal R}_{p_T, y} \rangle$ at LO (left) and NLO+DLs (right) for multiplicity $N = 3$, rapidity difference $Y = 3, 6$ and $J_{\text{min}} = 2$ (top), $J_{\text{min}} = 10$ (middle), $J_{\text{min}} = 25$ (bottom).}
\label{fig:rpty3}
\end{figure}

\begin{figure}
\begin{subfigure}{.5\textwidth}
  \centering
  \includegraphics[width=.8\linewidth]{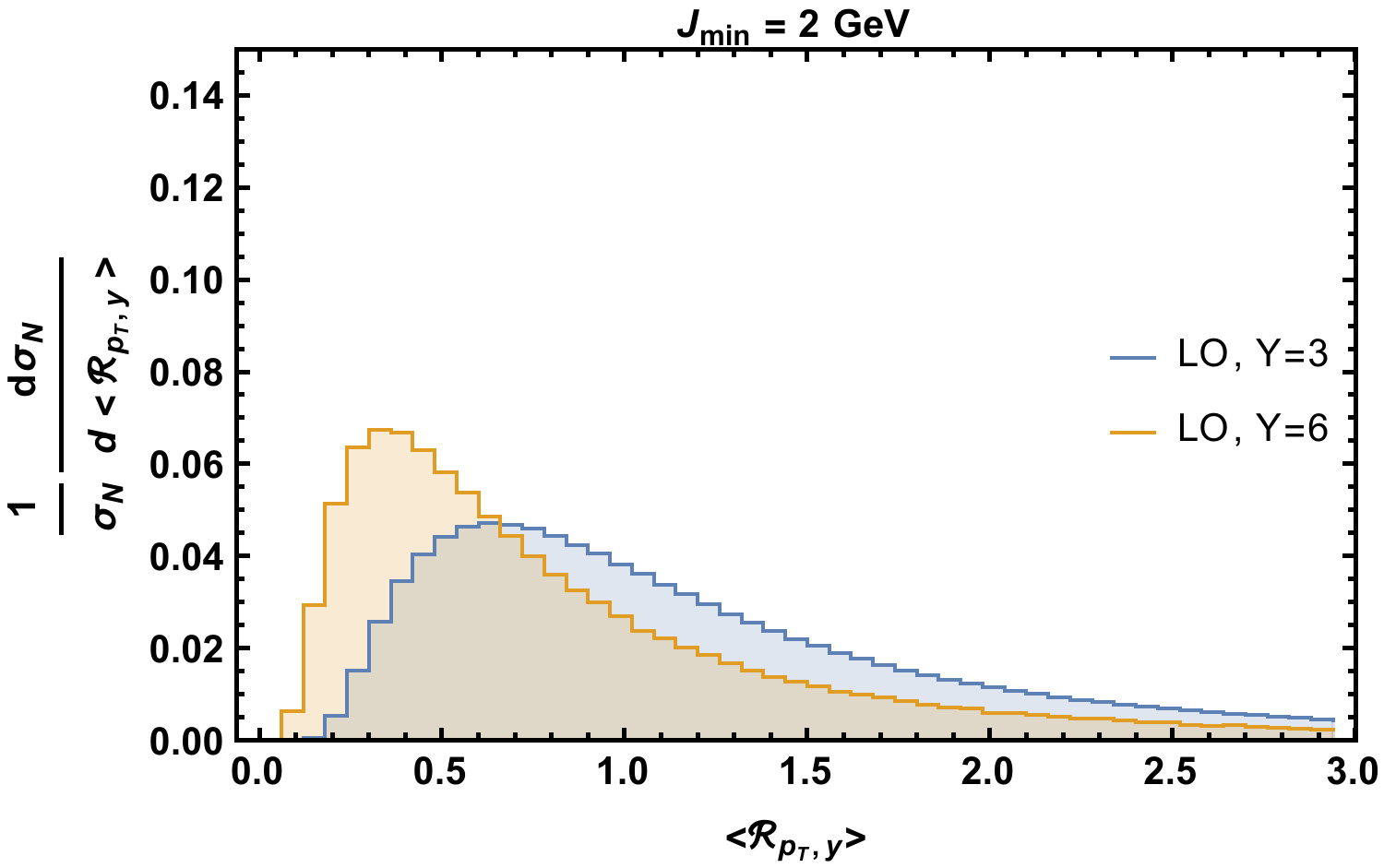}
  \caption{}
  \label{fig:sfig1}
\end{subfigure}%
\begin{subfigure}{.5\textwidth}
  \centering
  \includegraphics[width=.8\linewidth]{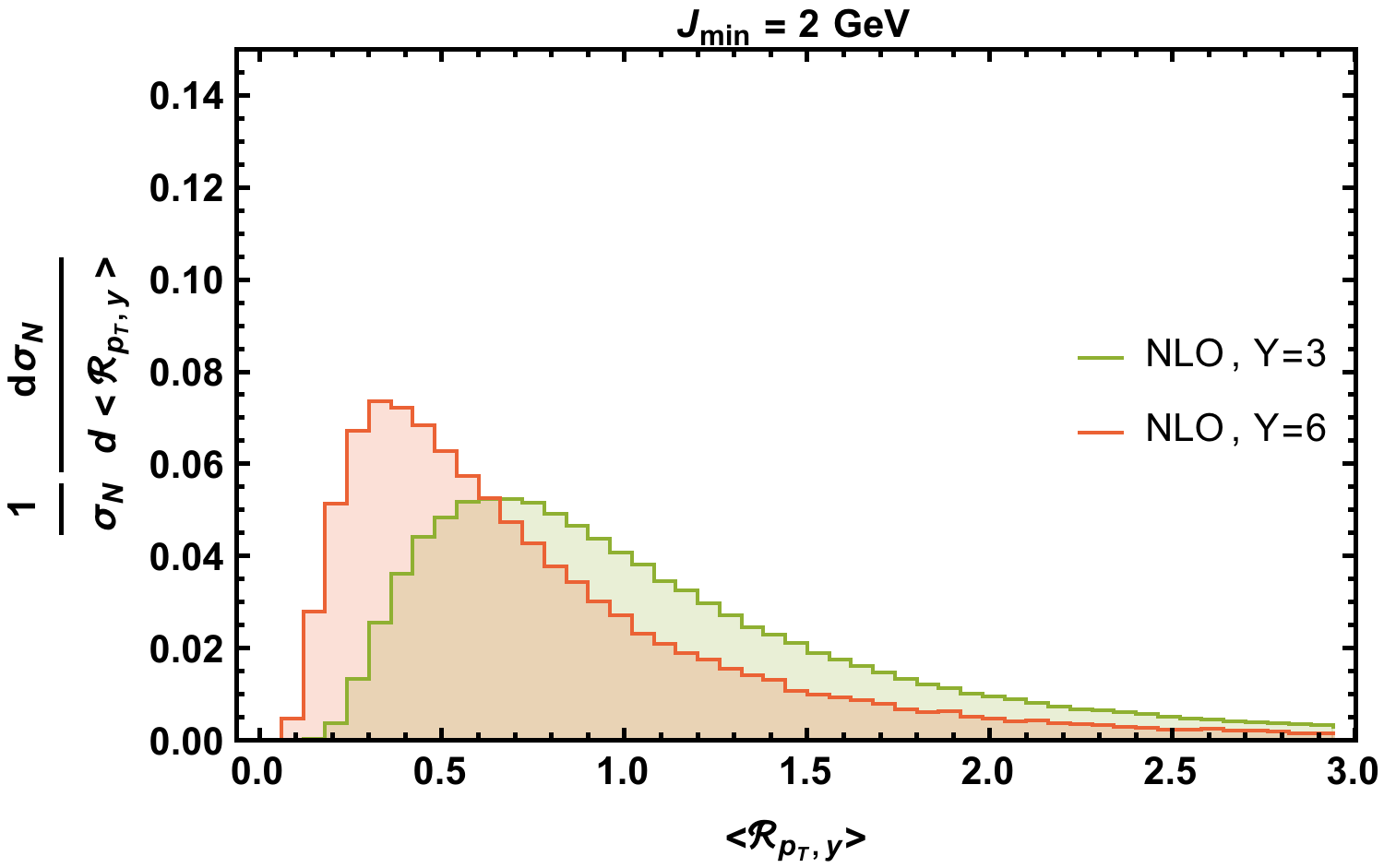}
  \caption{}
  \label{fig:sfig2}
\end{subfigure}
\\
\begin{subfigure}{.5\textwidth}
  \centering
  \includegraphics[width=.8\linewidth]{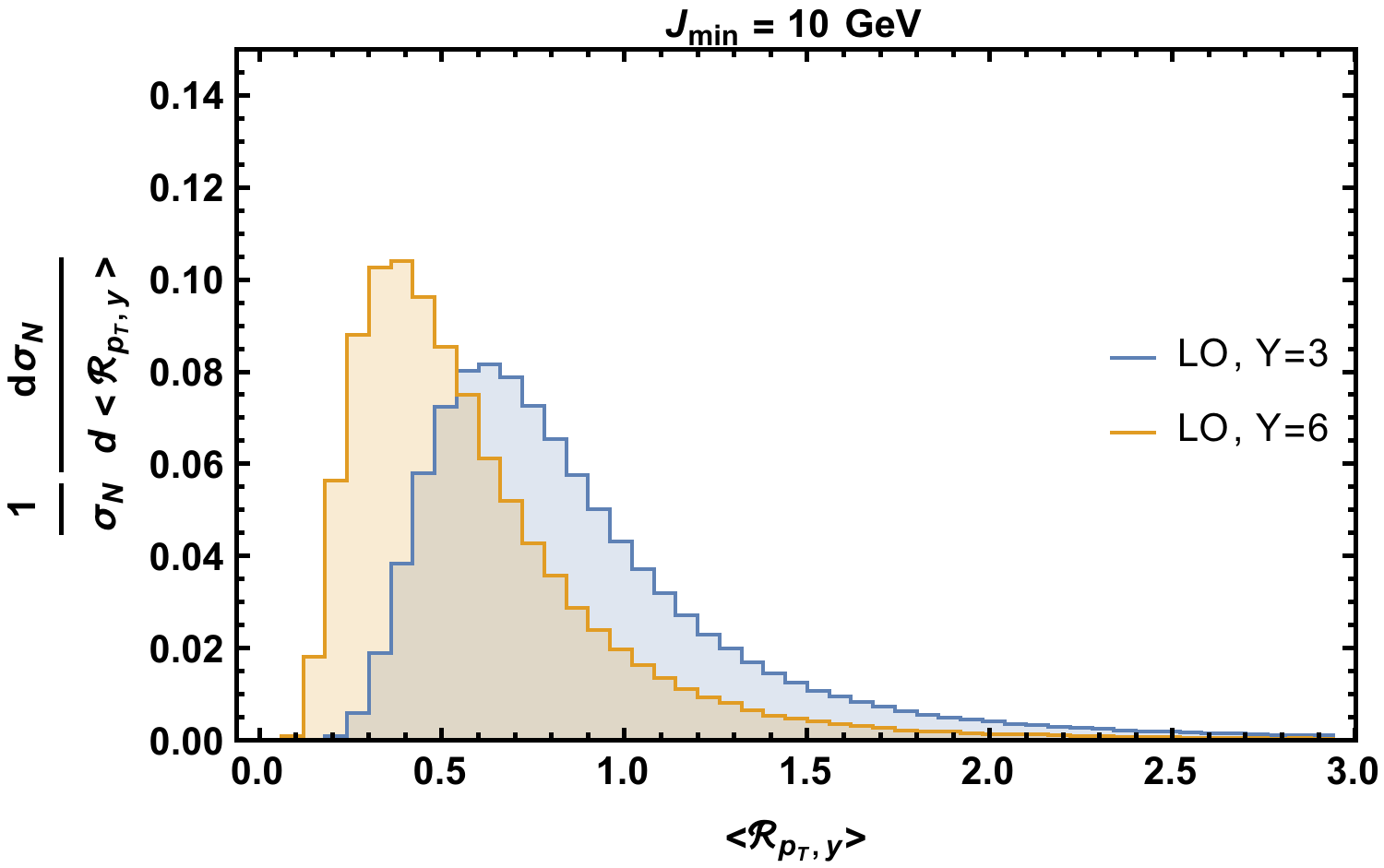}
  \caption{}
  \label{fig:sfig1}
\end{subfigure}%
\begin{subfigure}{.5\textwidth}
  \centering
  \includegraphics[width=.8\linewidth]{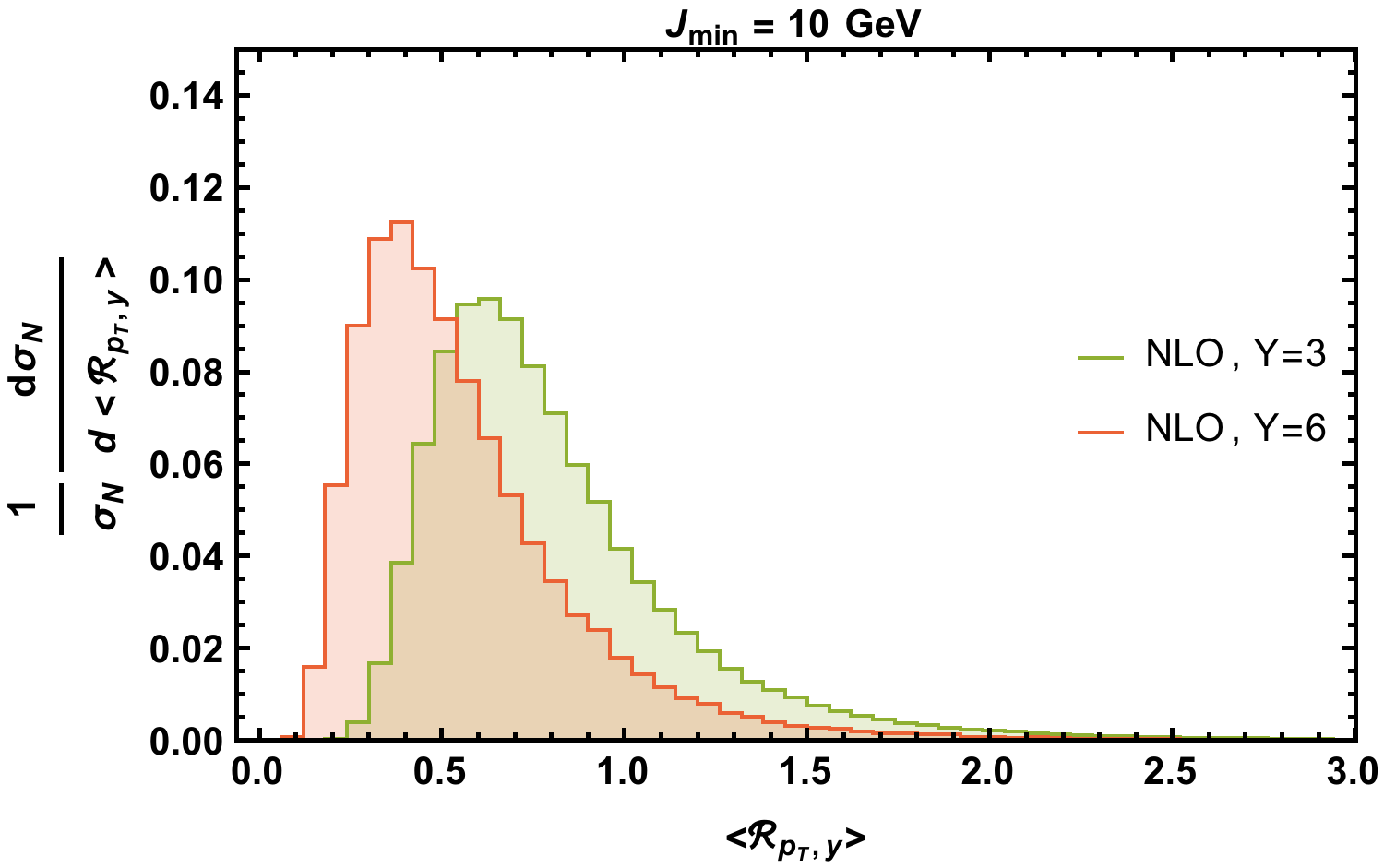}
  \caption{}
  \label{fig:sfig2}
\end{subfigure}
\\
\begin{subfigure}{.5\textwidth}
  \centering
  \includegraphics[width=.8\linewidth]{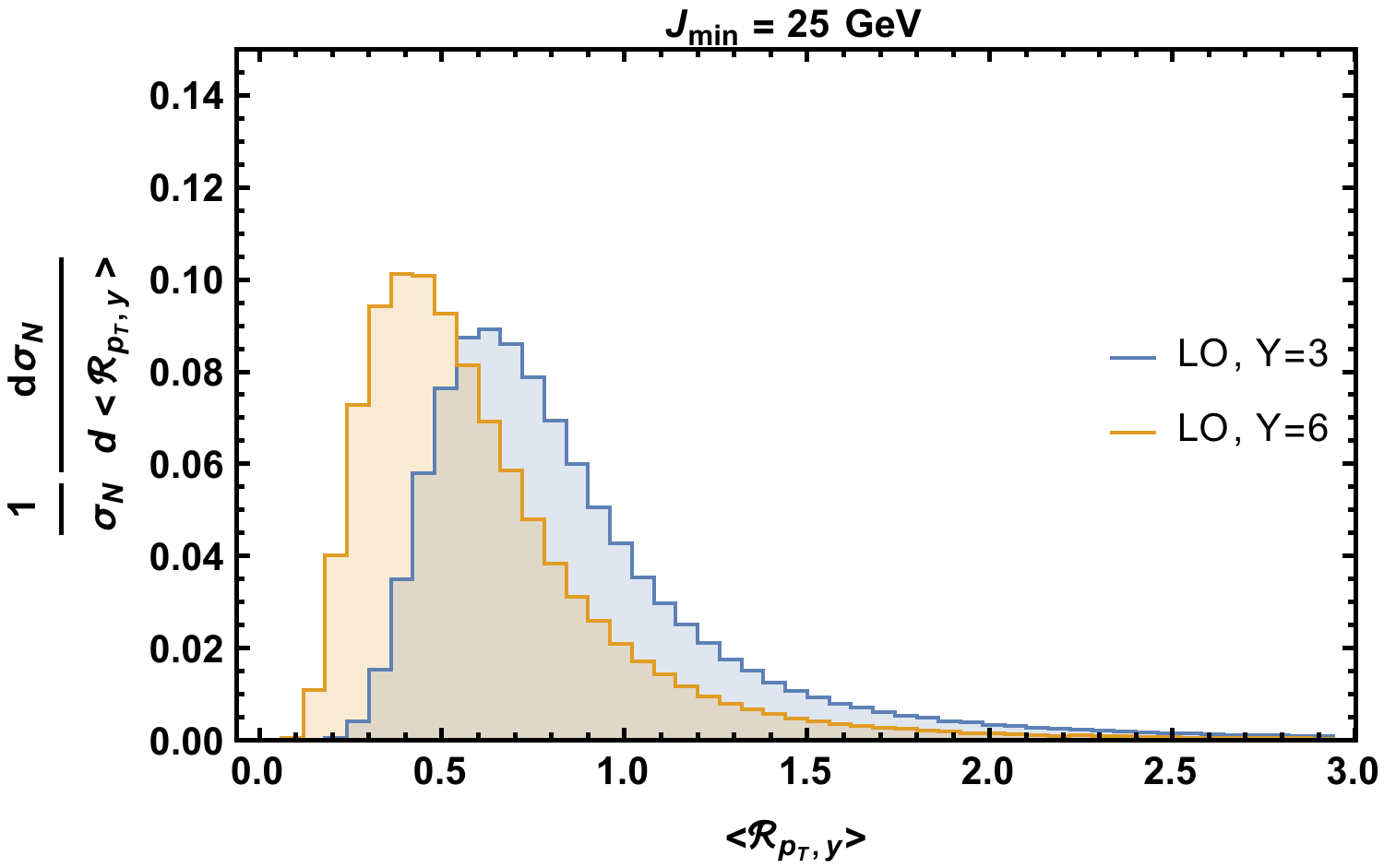}
  \caption{}
  \label{fig:sfig1}
\end{subfigure}%
\begin{subfigure}{.5\textwidth}
  \centering
  \includegraphics[width=.8\linewidth]{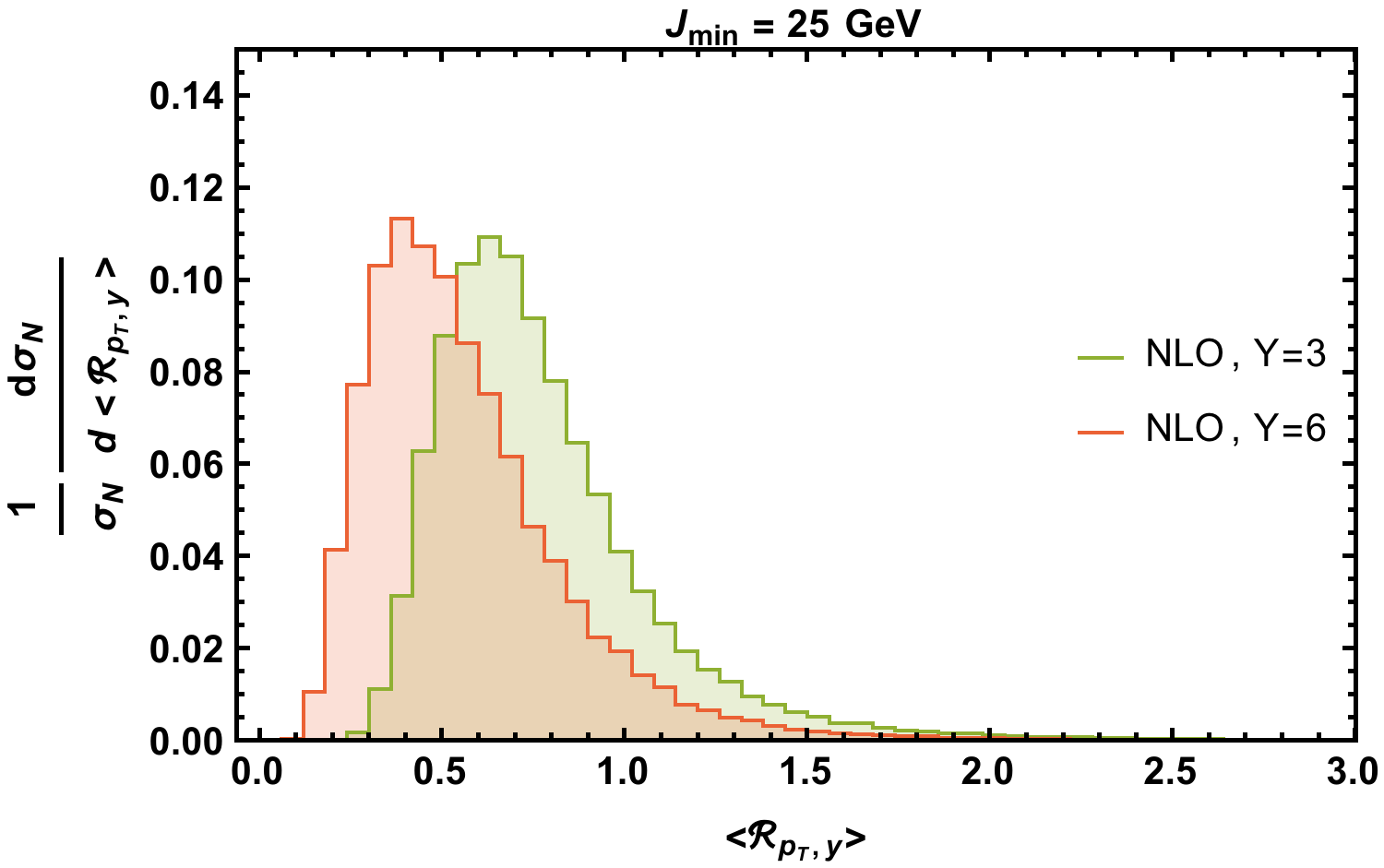}
  \caption{}
  \label{fig:sfig2}
\end{subfigure}
\caption{Normalized $\langle {\mathcal R}_{p_T, y} \rangle$ at LO (left) and NLO+DLs (right) for multiplicity $N = 4$, rapidity difference $Y = 3, 6$ and $J_{\text{min}} = 2$ (top), $J_{\text{min}} = 10$ (middle), $J_{\text{min}} = 25$ (bottom).}
\label{fig:rpty4}
\end{figure}

\begin{figure}
\begin{subfigure}{.5\textwidth}
  \centering
  \includegraphics[width=.8\linewidth]{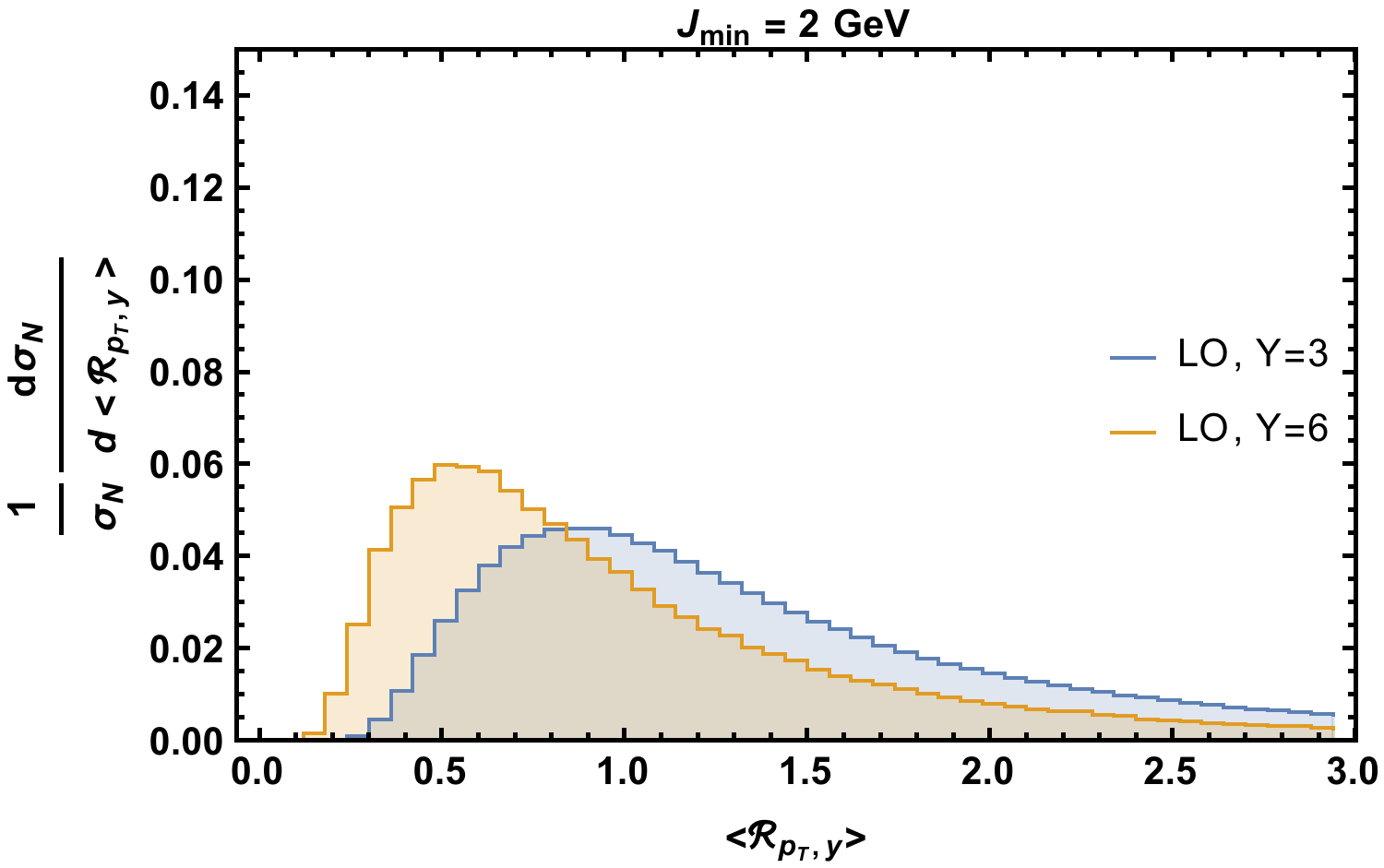}
  \caption{}
  \label{fig:sfig1}
\end{subfigure}%
\begin{subfigure}{.5\textwidth}
  \centering
  \includegraphics[width=.8\linewidth]{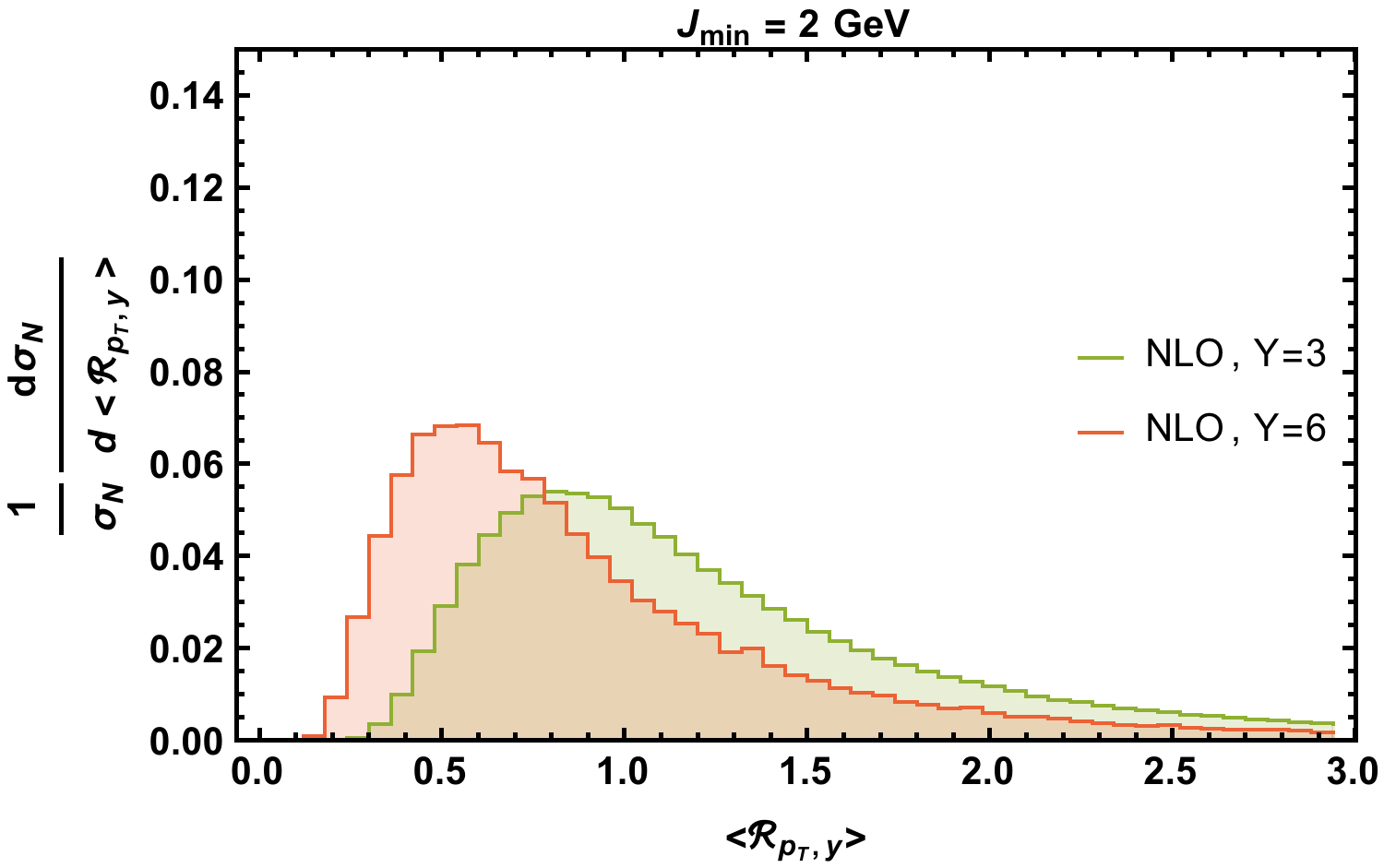}
  \caption{}
  \label{fig:sfig2}
\end{subfigure}
\\
\begin{subfigure}{.5\textwidth}
  \centering
  \includegraphics[width=.8\linewidth]{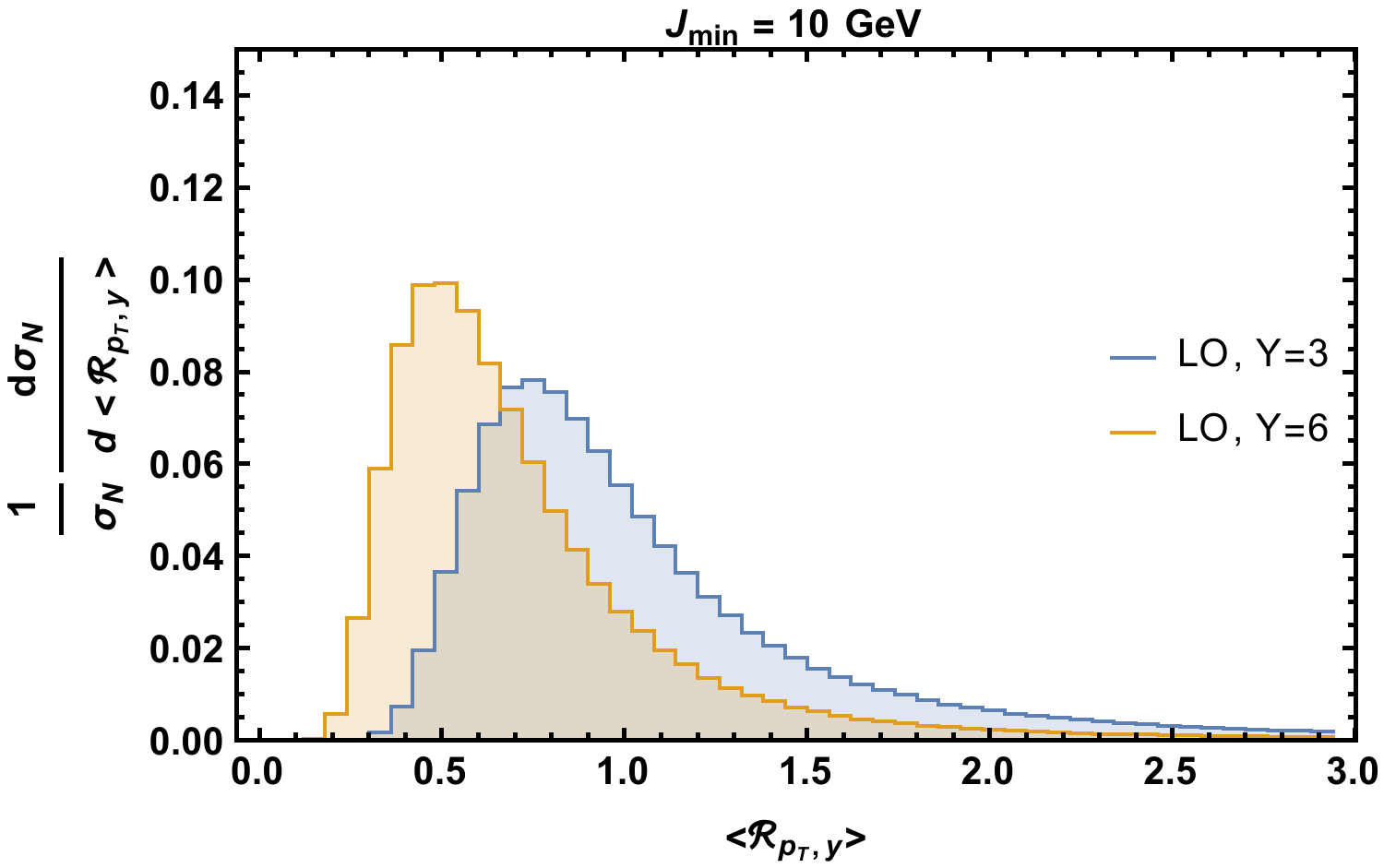}
  \caption{}
  \label{fig:sfig1}
\end{subfigure}%
\begin{subfigure}{.5\textwidth}
  \centering
  \includegraphics[width=.8\linewidth]{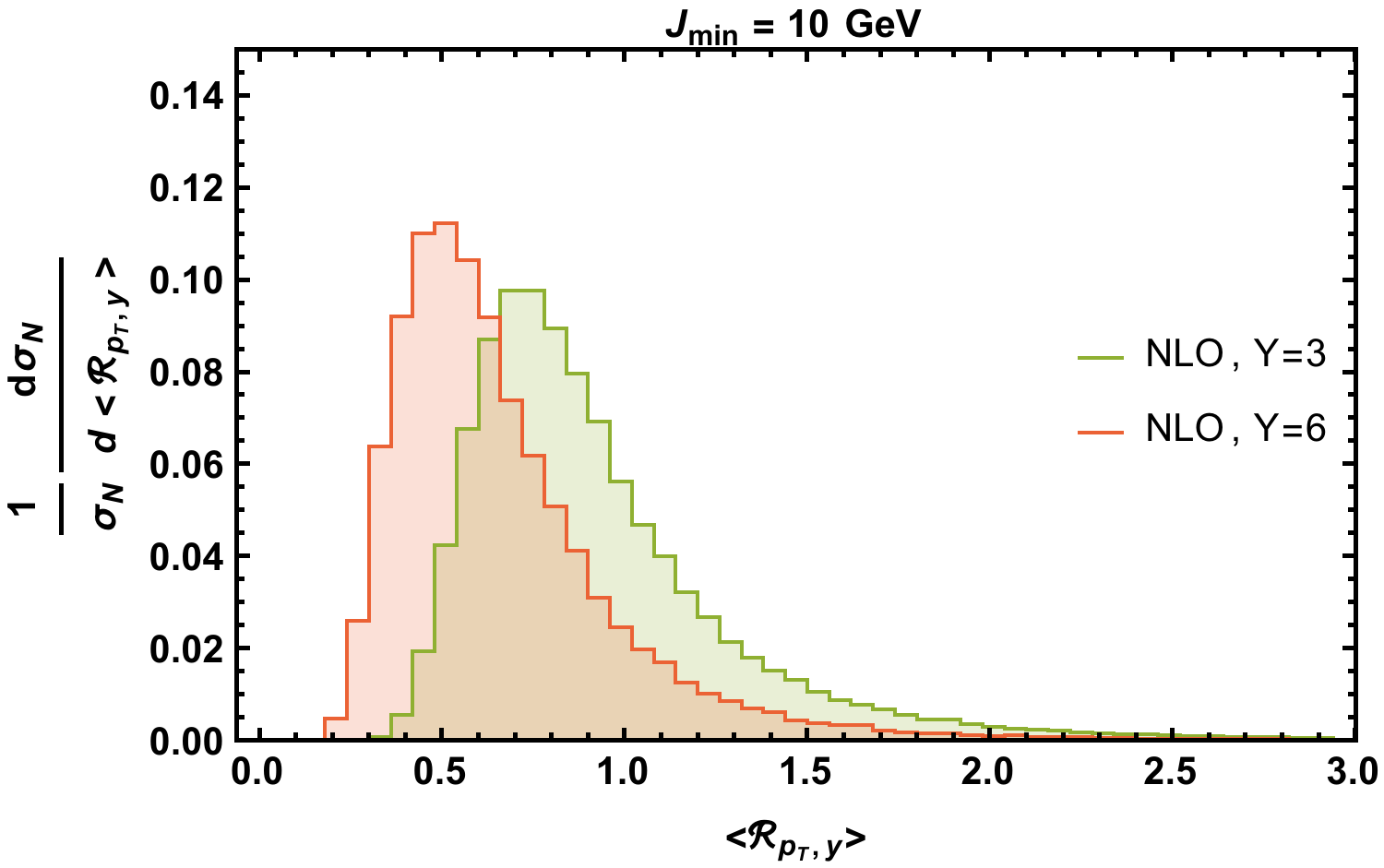}
  \caption{}
  \label{fig:sfig2}
\end{subfigure}
\\
\begin{subfigure}{.5\textwidth}
  \centering
  \includegraphics[width=.8\linewidth]{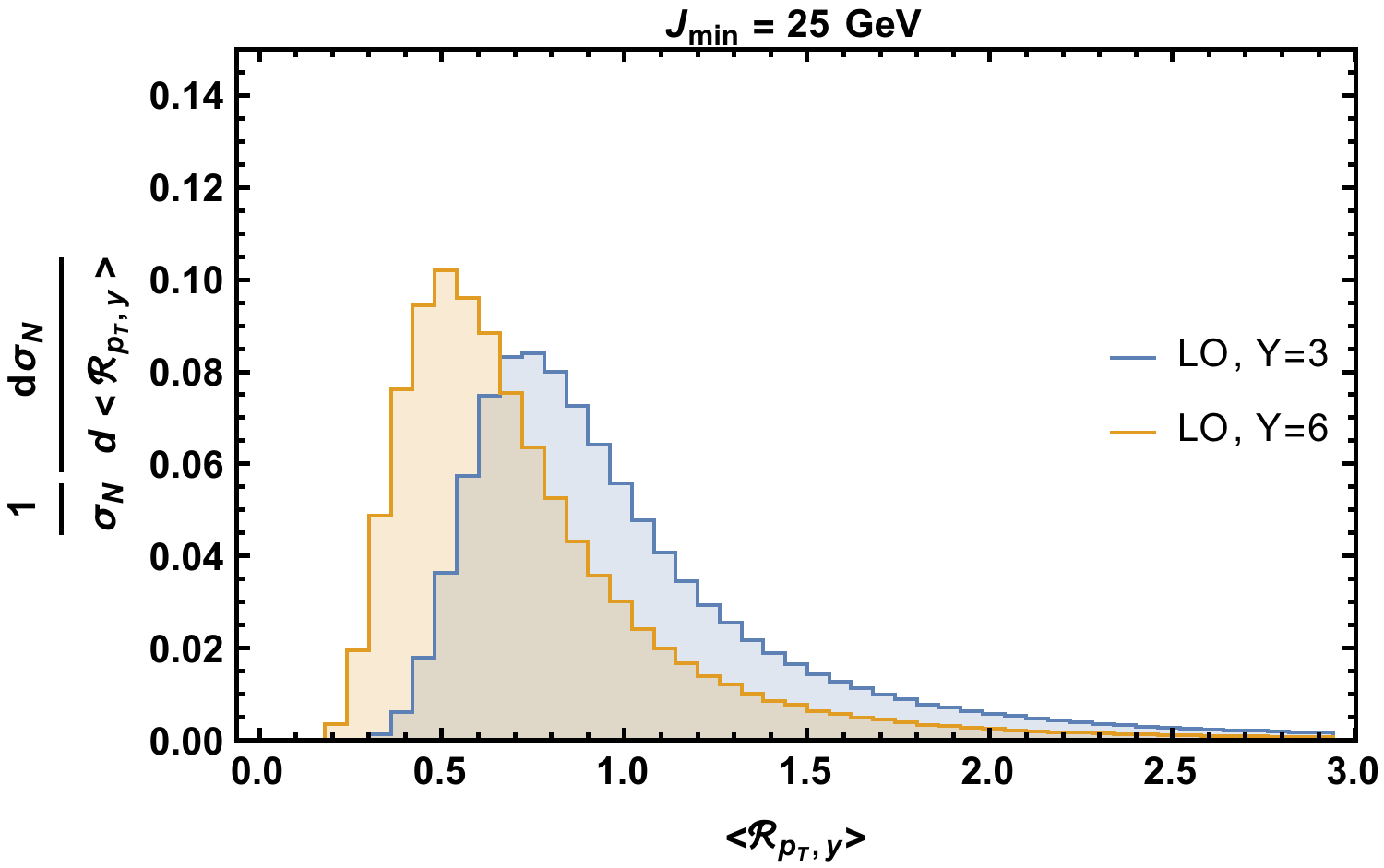}
  \caption{}
  \label{fig:sfig1}
\end{subfigure}%
\begin{subfigure}{.5\textwidth}
  \centering
  \includegraphics[width=.8\linewidth]{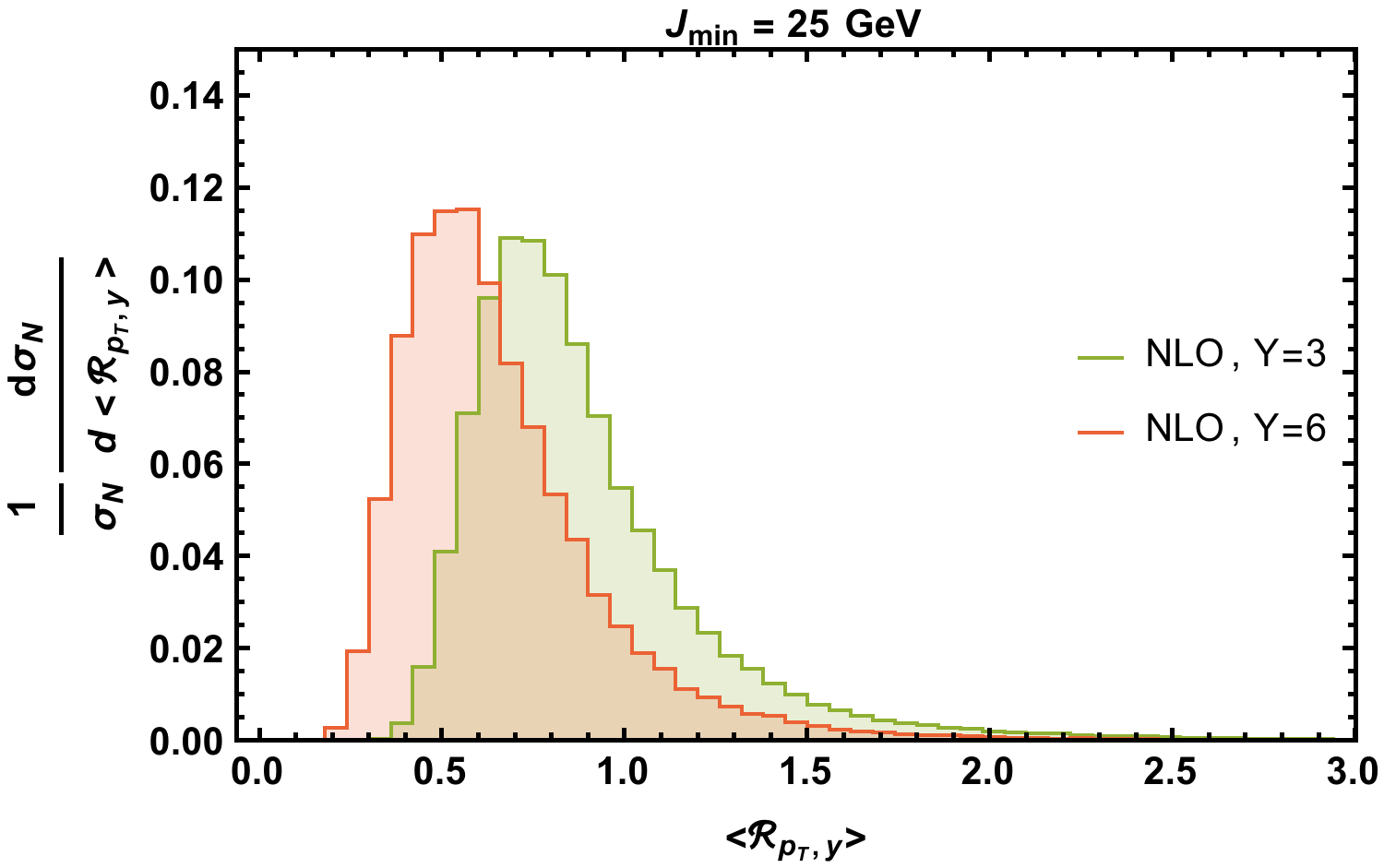}
  \caption{}
  \label{fig:sfig2}
\end{subfigure}
\caption{Normalized $\langle {\mathcal R}_{p_T, y} \rangle$ at LO (left) and NLO+DLs (right) for multiplicity $N = 5$, rapidity difference $Y = 3, 6$ and $J_{\text{min}} = 2$ (top), $J_{\text{min}} = 10$ (middle), $J_{\text{min}} = 25$ (bottom).}
\label{fig:rpty5}
\end{figure}

In Figs.~\ref{fig:rpty3},~\ref{fig:rpty4} and ~\ref{fig:rpty5} we plot $\frac{1}{\sigma_N} \frac{d \sigma_N}{d \langle {\mathcal R}_{p_T, y} \rangle}$ for $N=3, 4, 5$ multiplicities respectively. As in the previous subsection, in each subplot of all three figures, we show two distributions, one
curve is reserved for rapidity $Y=3$ and the other for $Y=6$ whereas, subplots $(a), (c)$ and $(e)$ are at LO and $(b), (d)$ 
and $(f)$ are at NLO+DLs. The two subplots on the top row have $J_{min} = 2$ GeV, the ones in the middle have $J_{min} = 10$ GeV
and the two subplots at the bottom have $J_{min} = 25$ GeV.
In all cases, the values of $\langle {\mathcal R}_{p_T, y} \rangle$ are bounded by 0 but do not have
an upper limit since we average over ratios
$\frac{k_i e^{y_i}}{k_{i-1} e^{y_{i-1}}}$. The factor $\frac{ e^{y_i}}{e^{y_{i-1}}}$ is always smaller
than 1 but the factor $\frac{k_i }{k_{i-1} }$ is practically unbounded (although it is expected to be typically of order one in multi-Regge kinematics).

For multiplicity $N=3$, in Fig.~\ref{fig:rpty3} we see that  we have clear peaks situated in the region between $0.2$ and
$0.5$ and that the distributions drop fast near $\langle {\mathcal R}_{p_T, y} \rangle \sim 0$ whereas they present a smoother drop for $\langle {\mathcal R}_{p_T, y} \rangle > 1$. 
The peaks appear at lower
values and the distributions become more narrow as $Y$ increases. An increase of the value $J_{min}$ also makes the distributions more narrow although this is more striking when we move from $J_{min} = 2$ GeV to $J_{min} = 10$ GeV and less so when we move from $J_{min} = 10$ GeV to $J_{min} = 25$ GeV.
The impact of increasing $J_{min}$ on the positions of the peaks is more complicated. For $Y=3$, the peaks are clearly shifting to larger values whereas for $Y=6$ they still seem to follow that tread but to a much lesser degree.

As we increase the multiplicity to $N=4$ and $N=5$ the above observations generally hold. However, the actual positions of the peaks are gradually found at larger values such that for $N=5$ they are in the region $\langle {\mathcal R}_{p_T, y} \rangle \sim 0.5$ ($Y=6$) and $\langle {\mathcal R}_{p_T, y} \rangle \sim 0.8$ ($Y=3$). We should also note that the distributions become more narrow as the multiplicity increases.

\subsection{Comparison between LO  and NLO+DLs}
\begin{figure}
\begin{subfigure}{.5\textwidth}
  \centering
  \includegraphics[width=.8\linewidth]{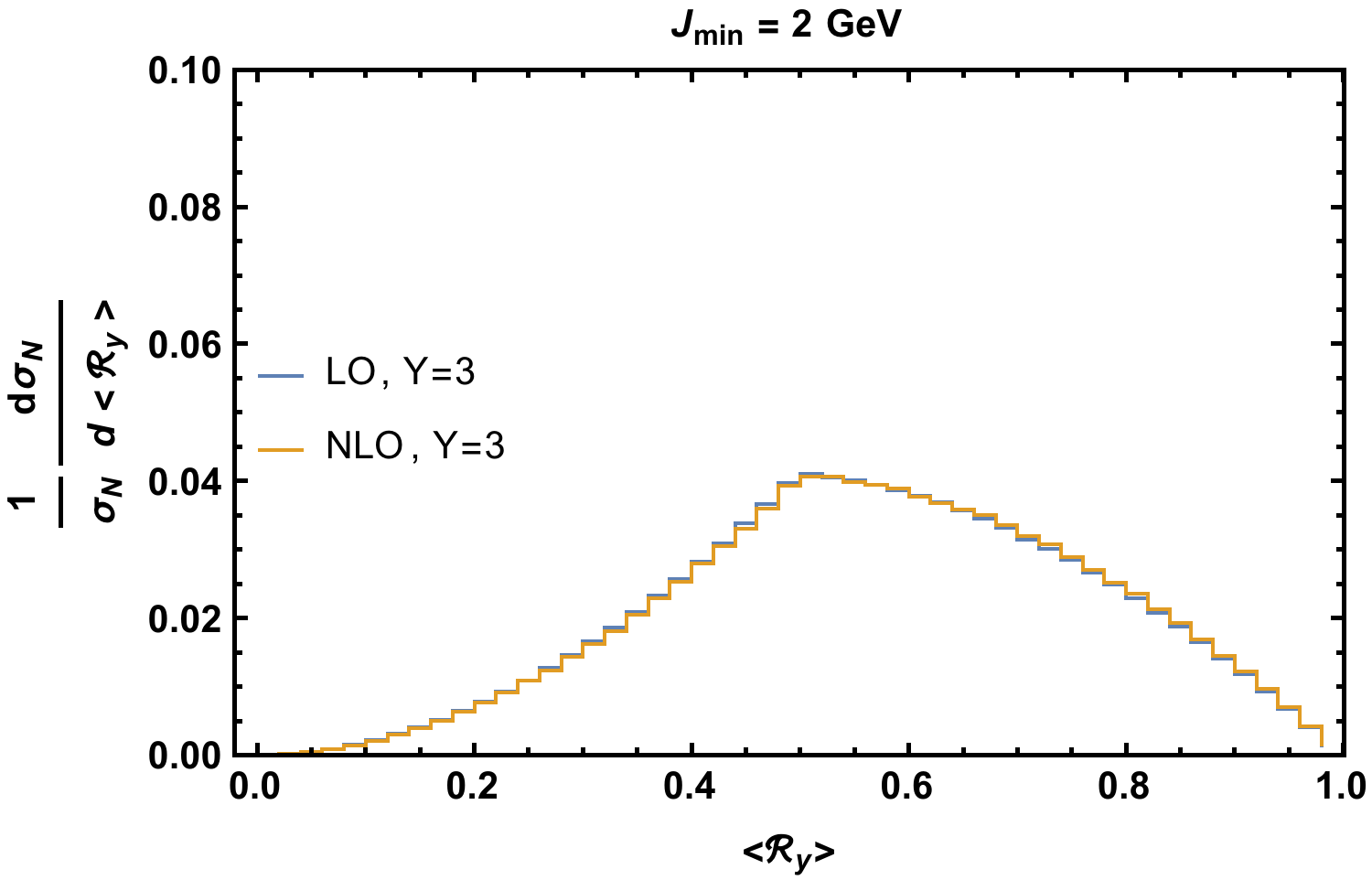}
  \caption{}
  \label{fig:sfig1}
\end{subfigure}%
\begin{subfigure}{.5\textwidth}
  \centering
  \includegraphics[width=.8\linewidth]{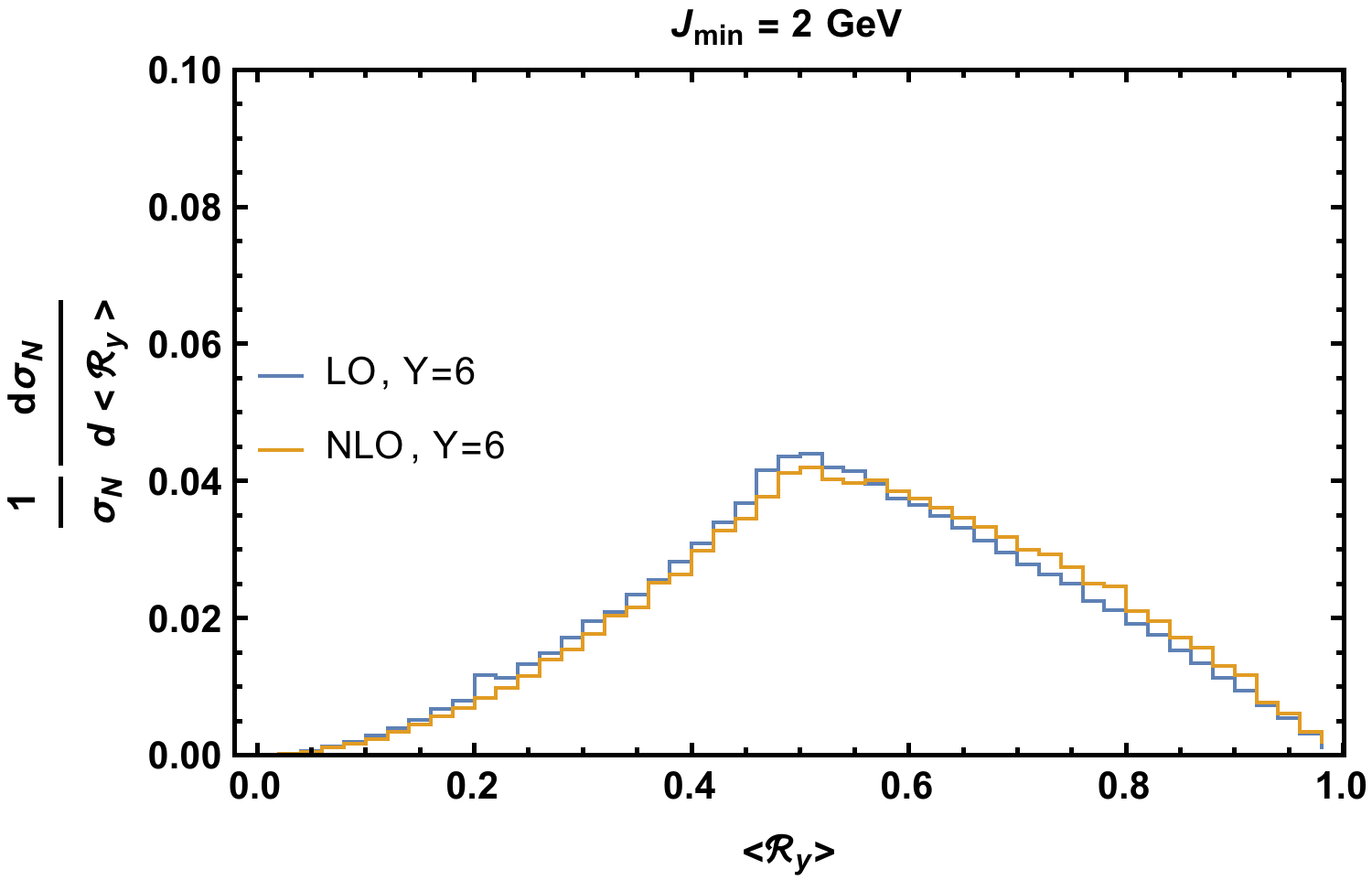}
  \caption{}
  \label{fig:sfig2}
\end{subfigure}
\\
\begin{subfigure}{.5\textwidth}
  \centering
  \includegraphics[width=.8\linewidth]{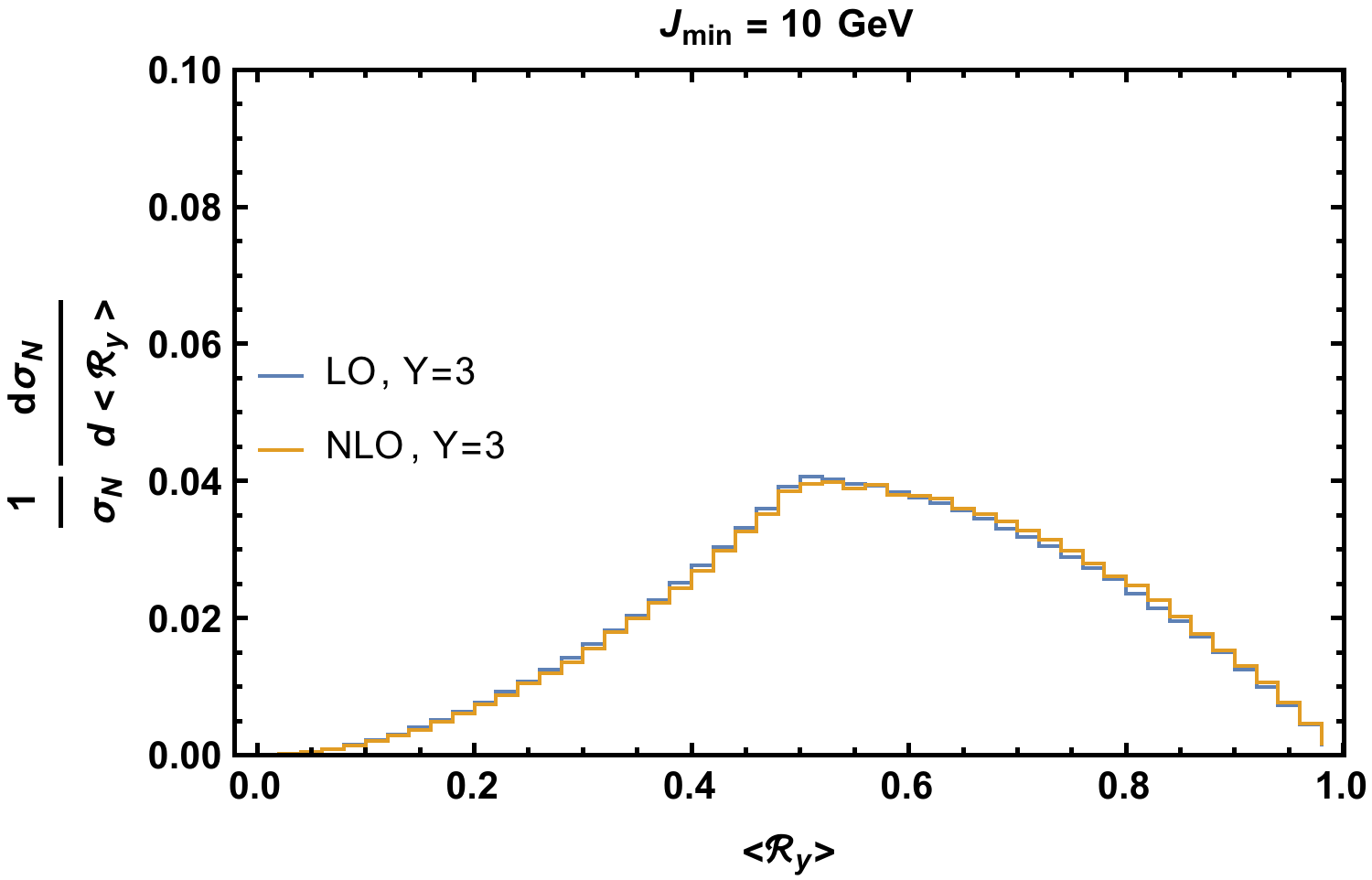}
  \caption{}
  \label{fig:sfig1}
\end{subfigure}%
\begin{subfigure}{.5\textwidth}
  \centering
  \includegraphics[width=.8\linewidth]{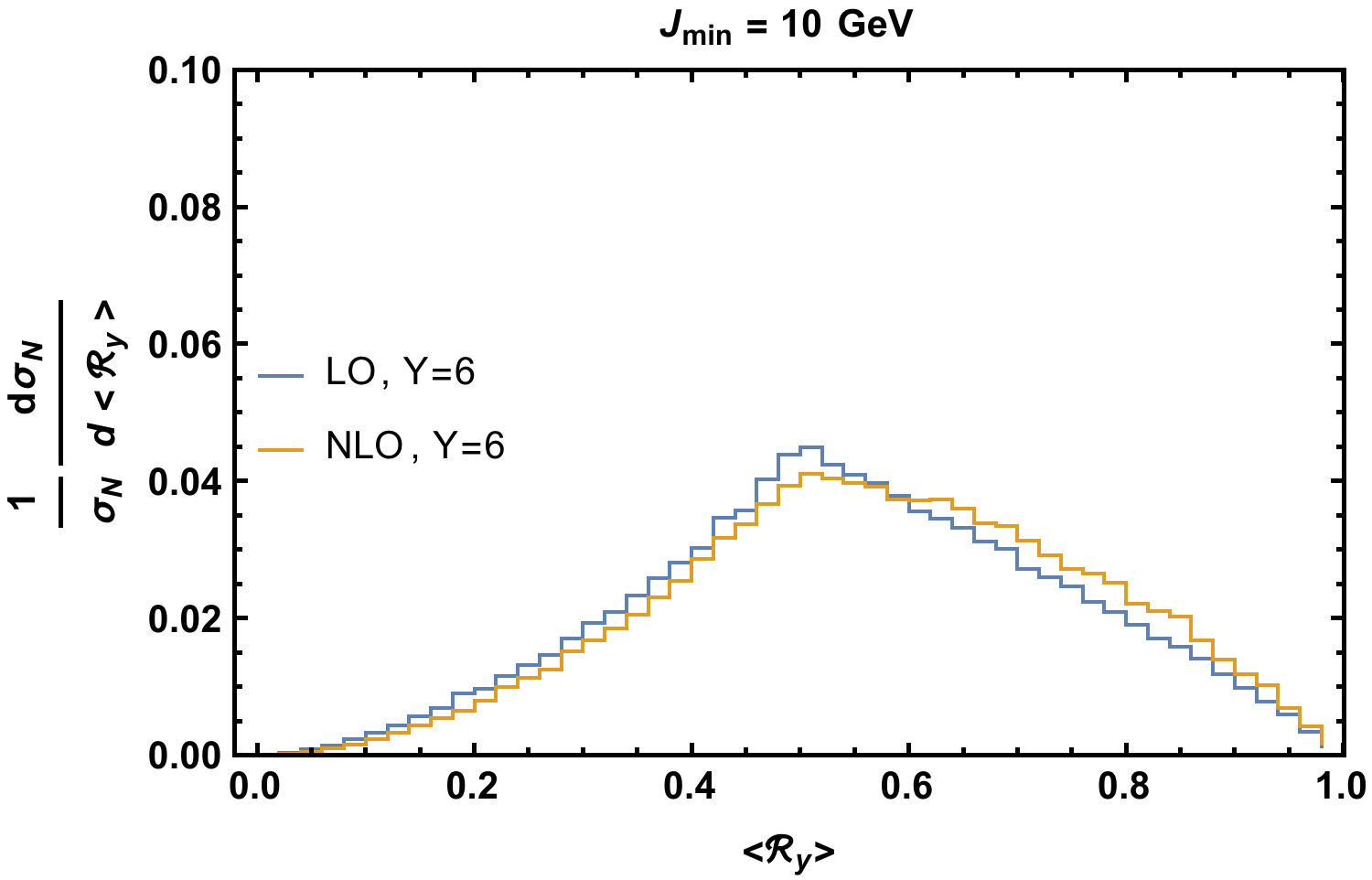}
  \caption{}
  \label{fig:sfig2}
\end{subfigure}
\\
\begin{subfigure}{.5\textwidth}
  \centering
  \includegraphics[width=.8\linewidth]{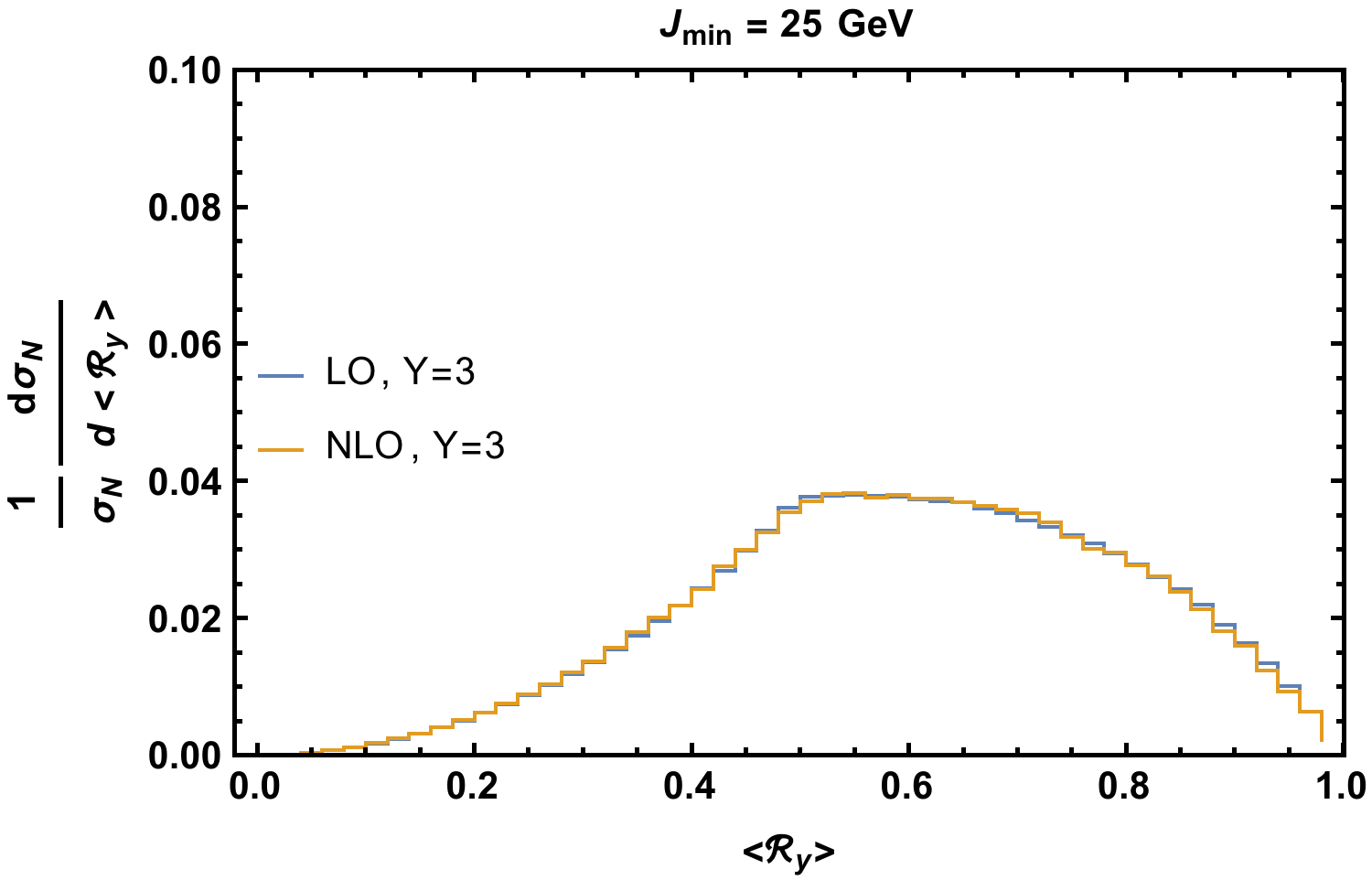}
  \caption{}
  \label{fig:sfig1}
\end{subfigure}%
\begin{subfigure}{.5\textwidth}
  \centering
  \includegraphics[width=.8\linewidth]{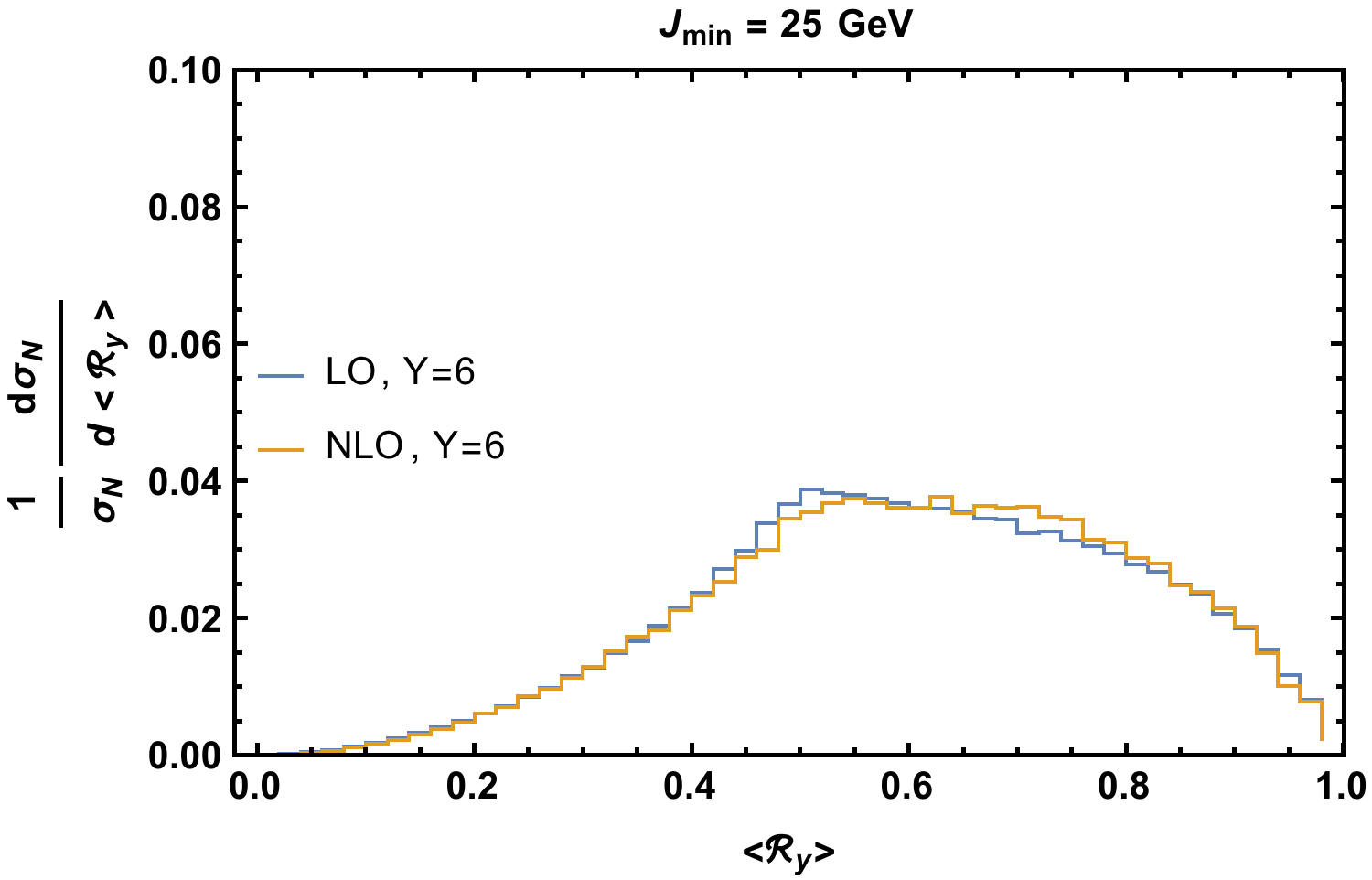}
  \caption{}
  \label{fig:sfig2}
\end{subfigure}
\caption{Comparison of LO and NLO+DLs normalized $\langle {\mathcal R}_{y} \rangle$ distributions for multiplicity $N = 3$, rapidity differences $Y = 3$ (left) and $Y = 6$ (right)  and $J_{\text{min}} = 2$ (top), $J_{\text{min}} = 10$ (middle), $J_{\text{min}} = 25$ (bottom).}
\label{fig:ry-LO-NLO-3}
\end{figure}

\begin{figure}
\begin{subfigure}{.5\textwidth}
  \centering
  \includegraphics[width=.8\linewidth]{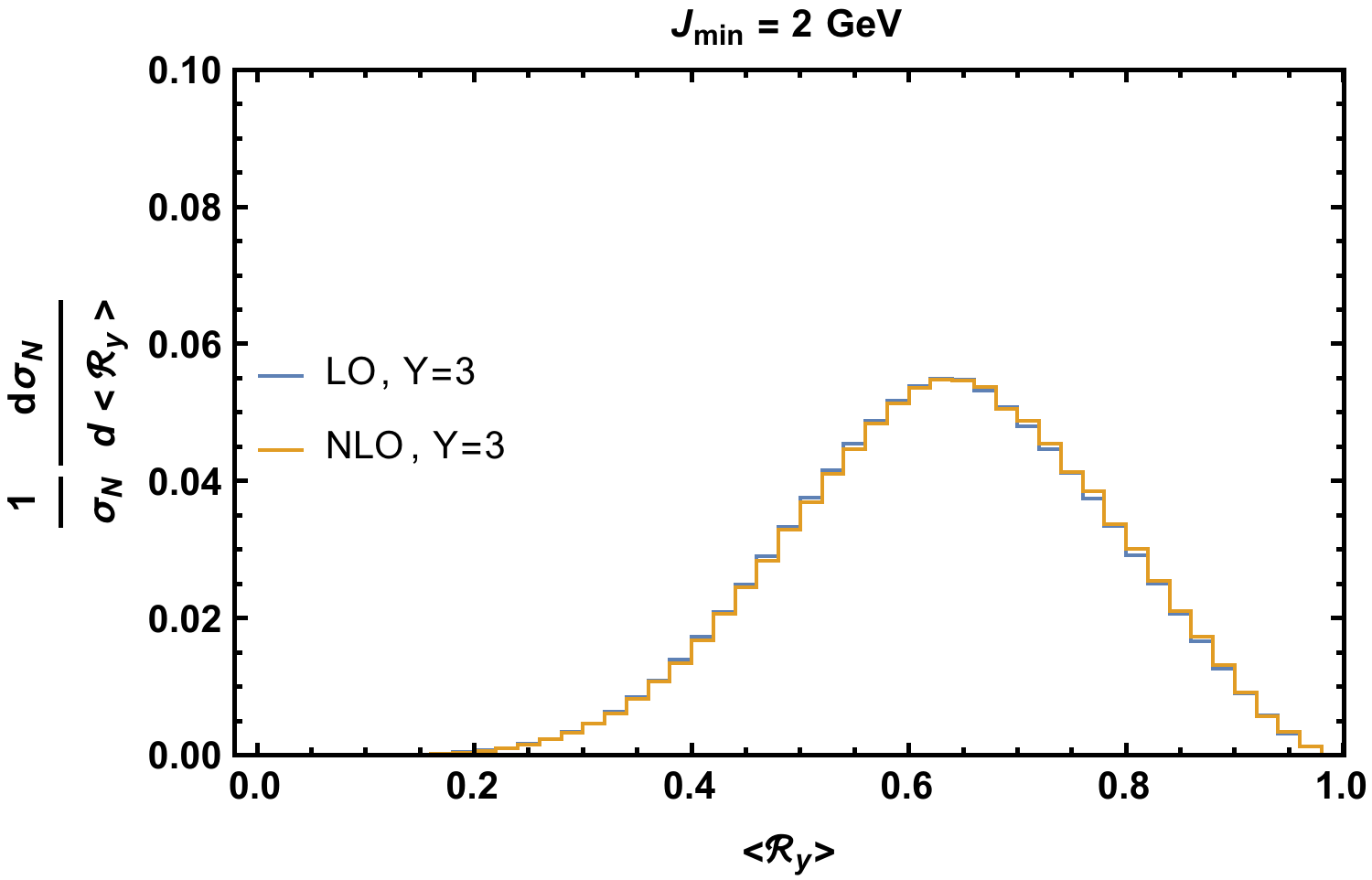}
  \caption{}
  \label{fig:sfig1}
\end{subfigure}%
\begin{subfigure}{.5\textwidth}
  \centering
  \includegraphics[width=.8\linewidth]{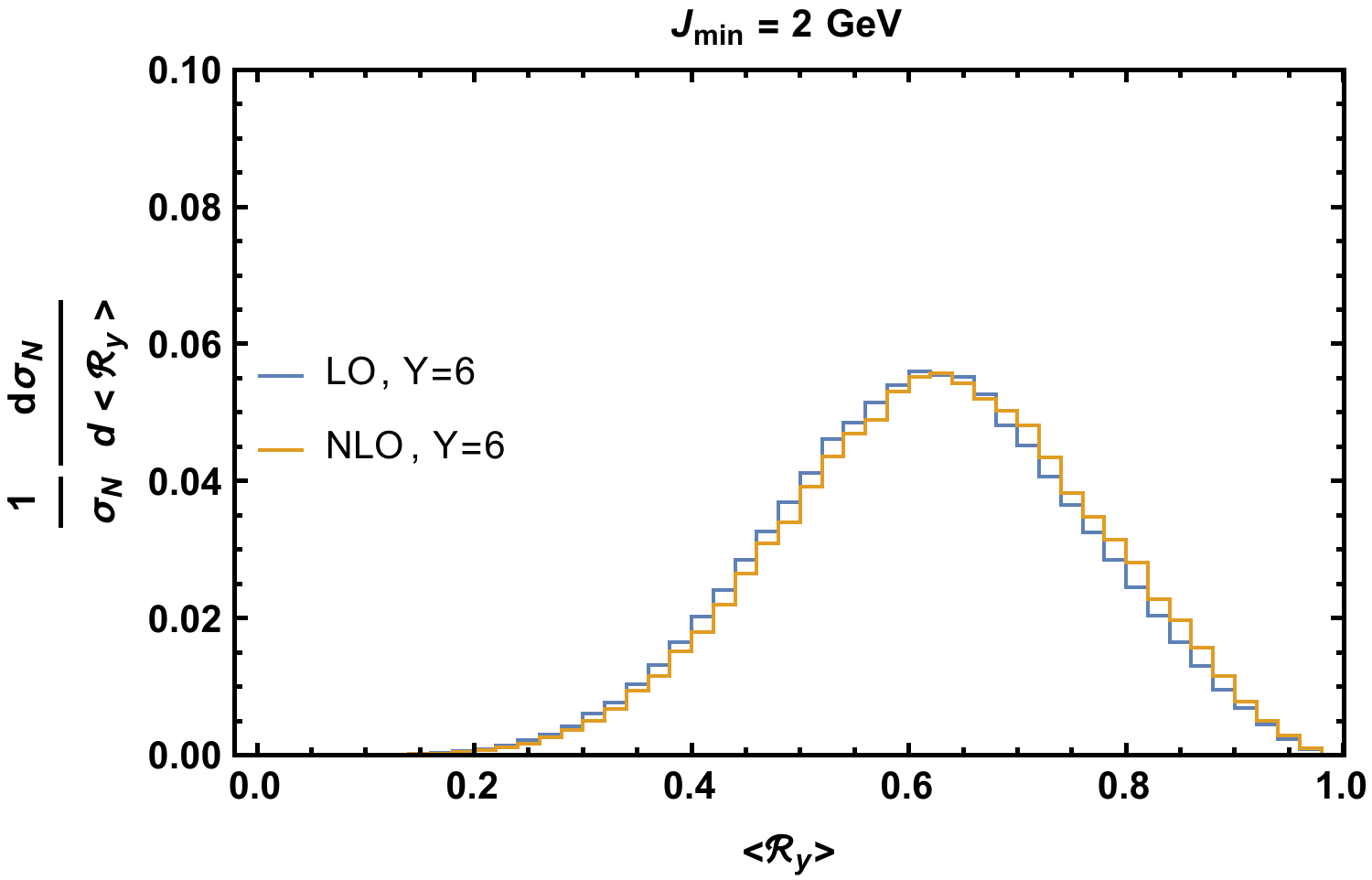}
  \caption{}
  \label{fig:sfig2}
\end{subfigure}
\\
\begin{subfigure}{.5\textwidth}
  \centering
  \includegraphics[width=.8\linewidth]{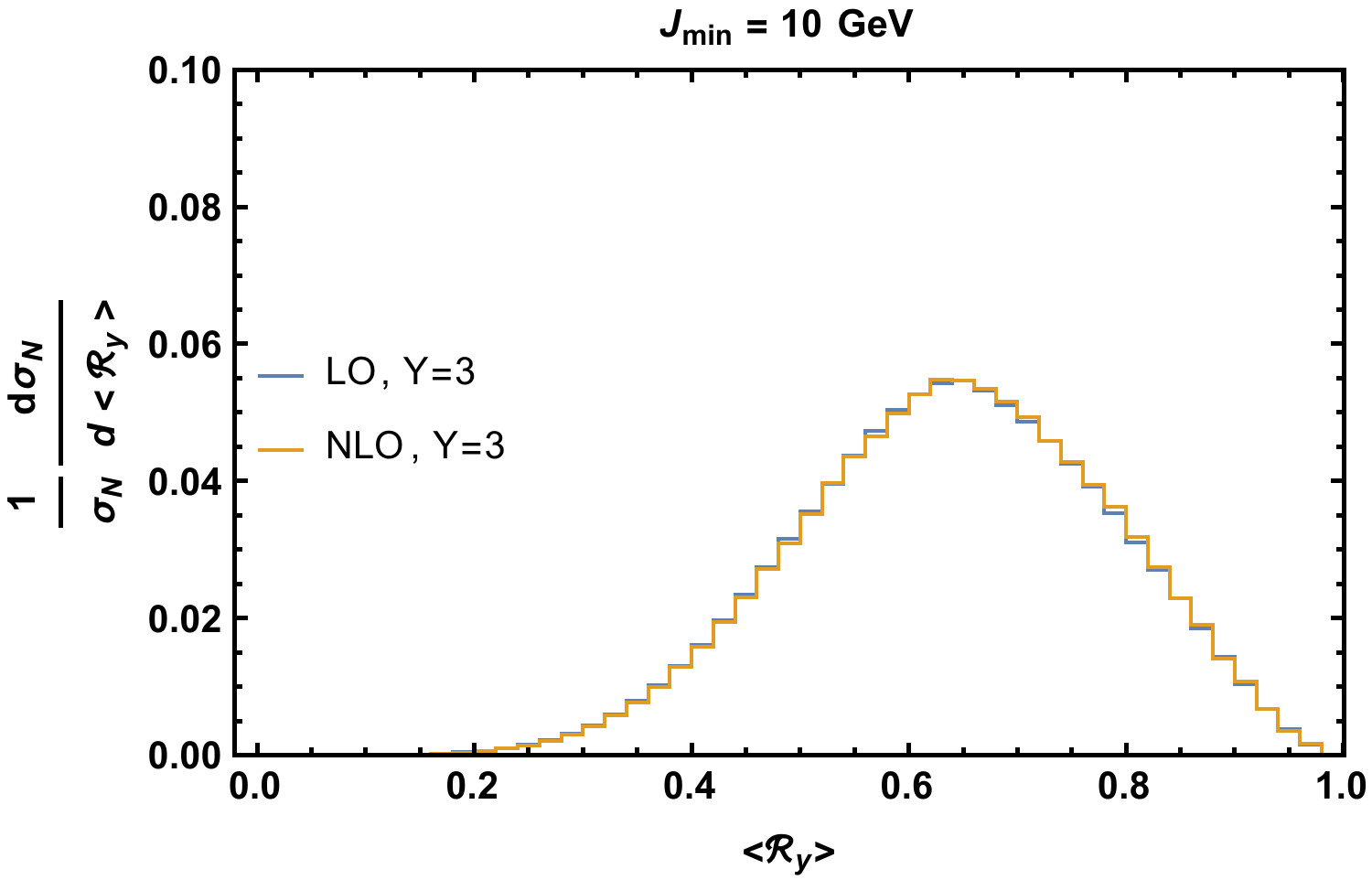}
  \caption{}
  \label{fig:sfig1}
\end{subfigure}%
\begin{subfigure}{.5\textwidth}
  \centering
  \includegraphics[width=.8\linewidth]{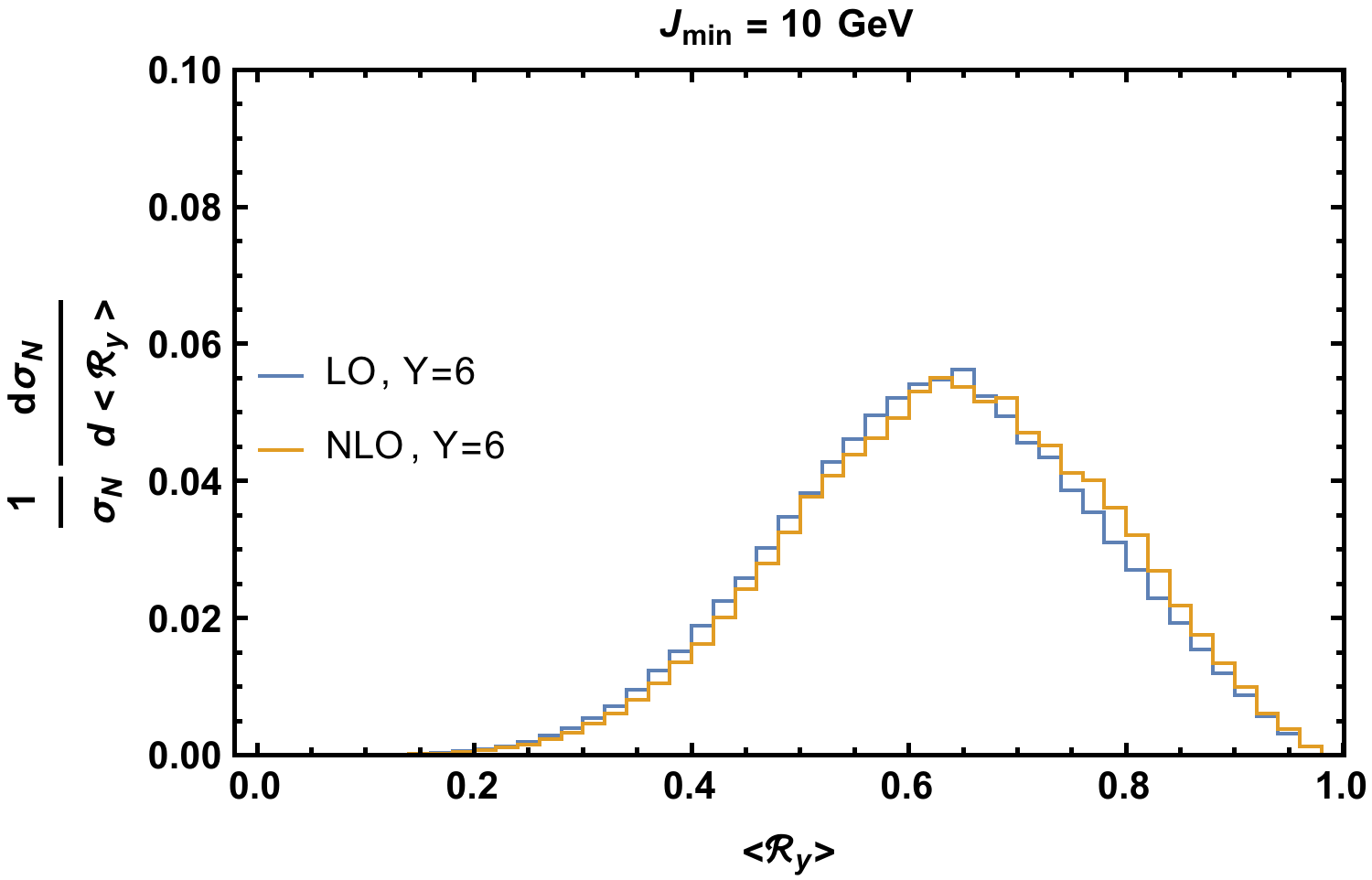}
  \caption{}
  \label{fig:sfig2}
\end{subfigure}
\\
\begin{subfigure}{.5\textwidth}
  \centering
  \includegraphics[width=.8\linewidth]{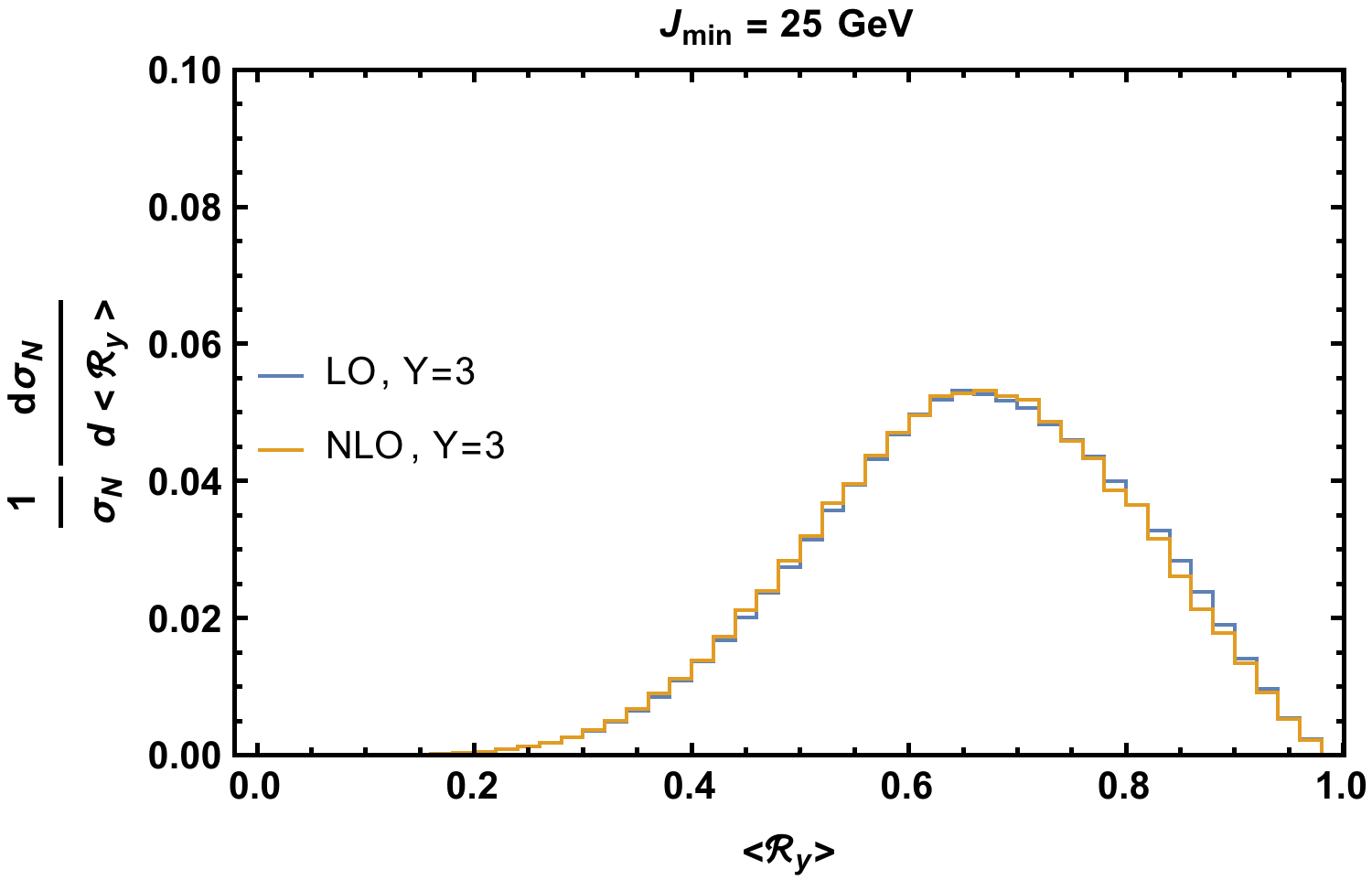}
  \caption{}
  \label{fig:sfig1}
\end{subfigure}%
\begin{subfigure}{.5\textwidth}
  \centering
  \includegraphics[width=.8\linewidth]{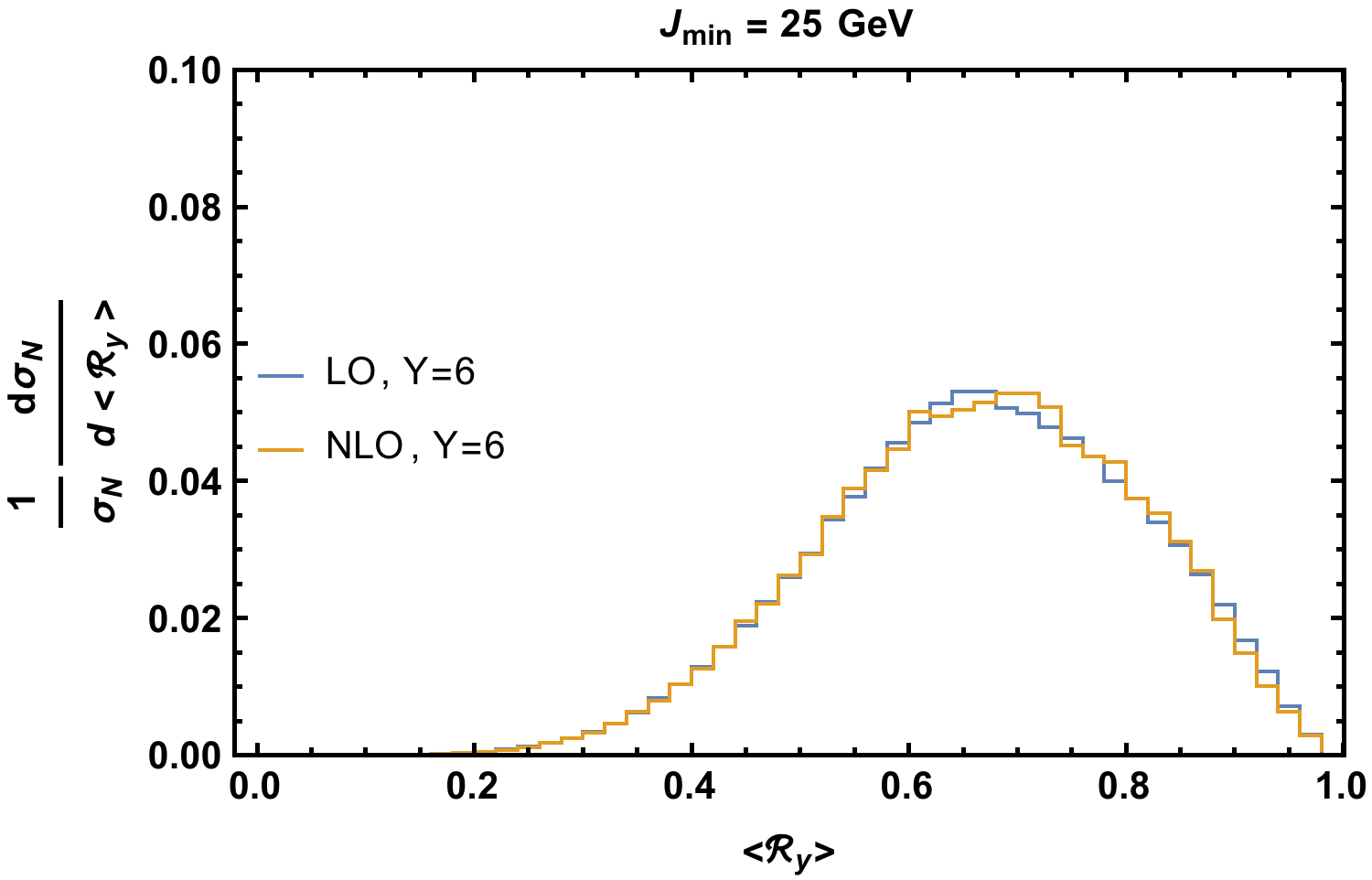}
  \caption{}
  \label{fig:sfig2}
\end{subfigure}
\caption{Comparison of LO and NLO+DLs normalized $\langle {\mathcal R}_{y} \rangle$ distributions for multiplicity $N = 4$, rapidity differences $Y = 3$ (left) and $Y = 6$ (right)  and $J_{\text{min}} = 2$ (top), $J_{\text{min}} = 10$ (middle), $J_{\text{min}} = 25$ (bottom).}
\label{fig:ry-LO-NLO-4}
\end{figure}

\begin{figure}
\begin{subfigure}{.5\textwidth}
  \centering
  \includegraphics[width=.8\linewidth]{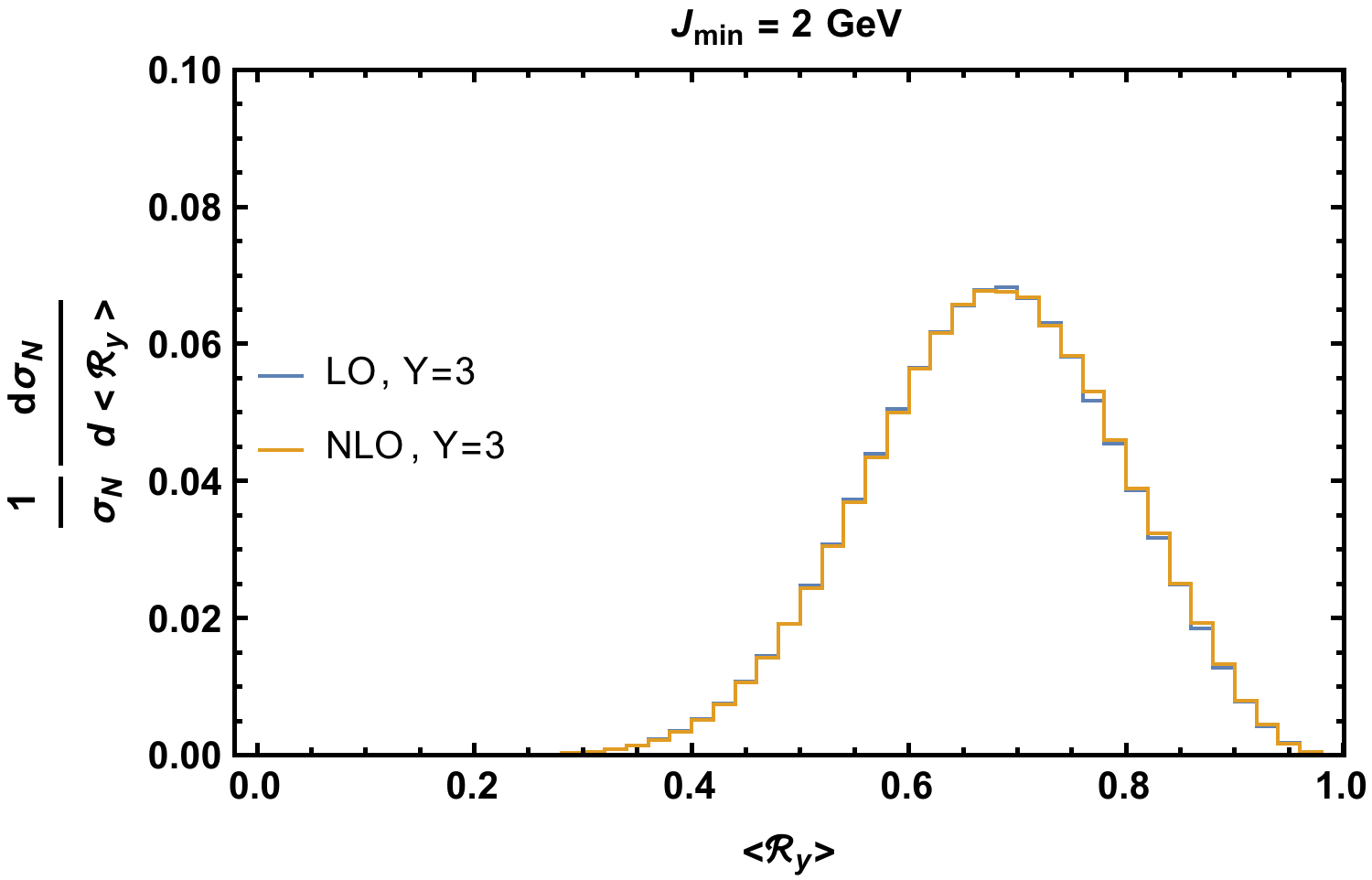}
  \caption{}
  \label{fig:sfig1}
\end{subfigure}%
\begin{subfigure}{.5\textwidth}
  \centering
  \includegraphics[width=.8\linewidth]{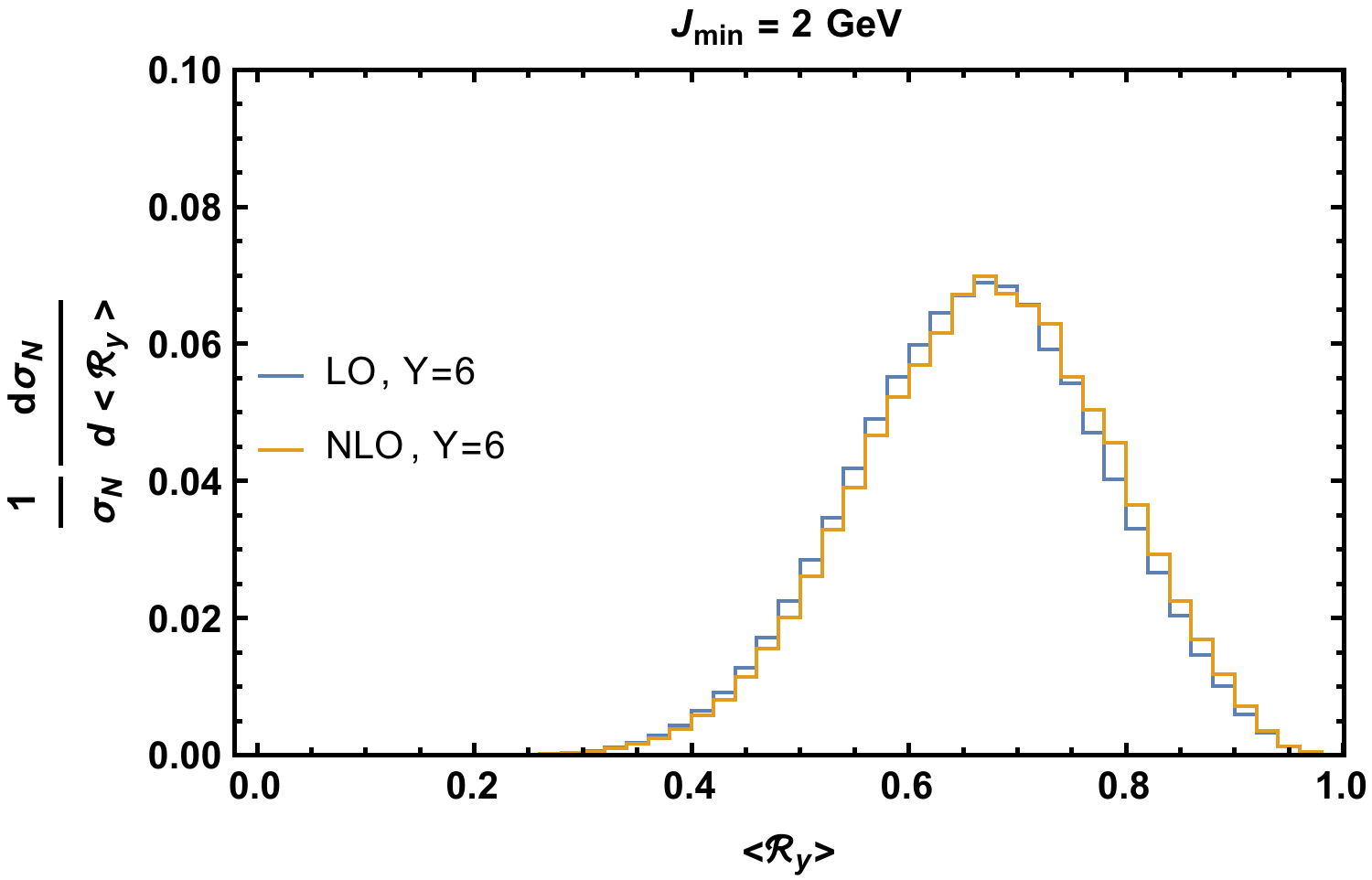}
  \caption{}
  \label{fig:sfig2}
\end{subfigure}
\\
\begin{subfigure}{.5\textwidth}
  \centering
  \includegraphics[width=.8\linewidth]{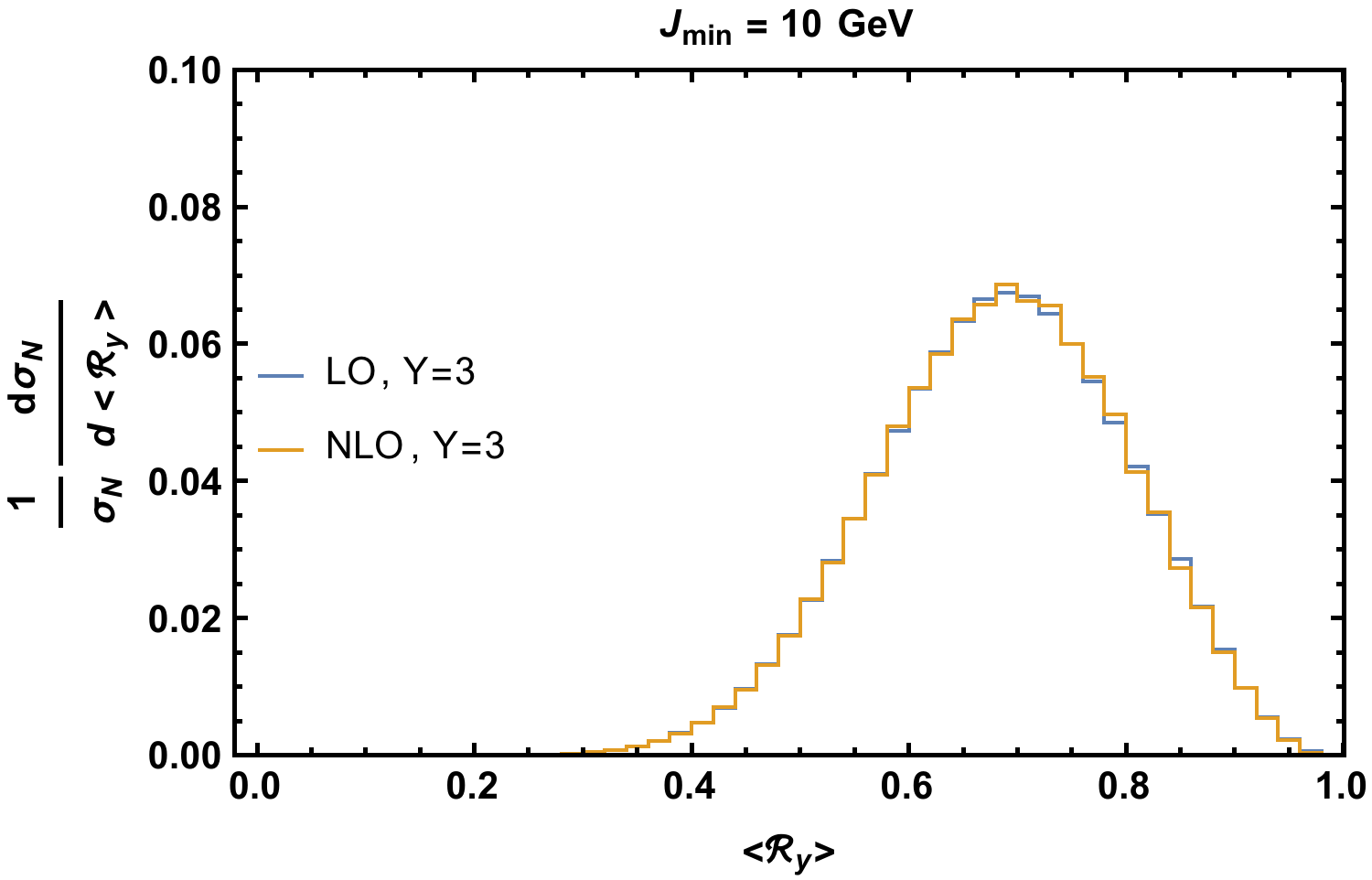}
  \caption{}
  \label{fig:sfig1}
\end{subfigure}%
\begin{subfigure}{.5\textwidth}
  \centering
  \includegraphics[width=.8\linewidth]{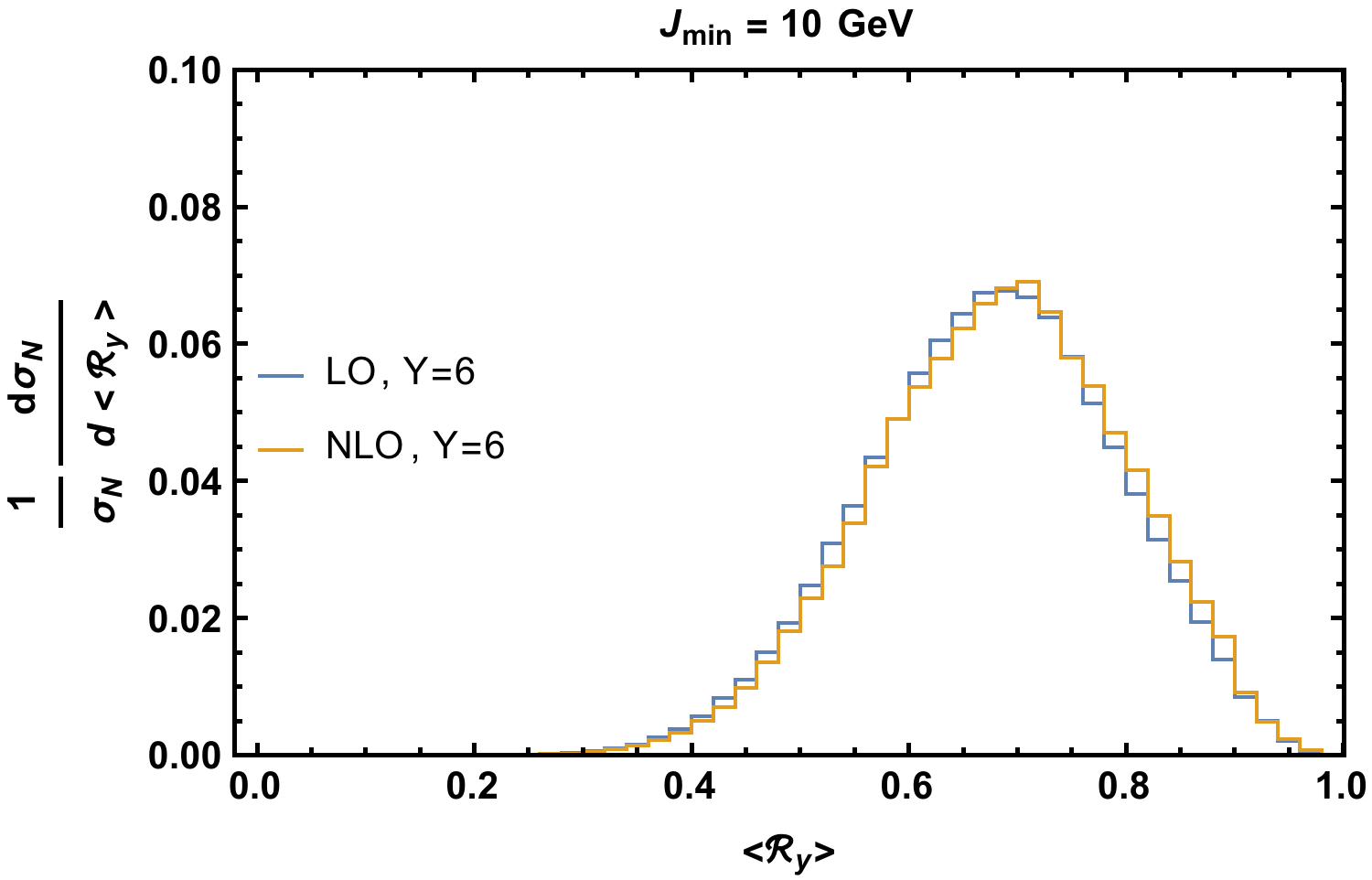}
  \caption{}
  \label{fig:sfig2}
\end{subfigure}
\\
\begin{subfigure}{.5\textwidth}
  \centering
  \includegraphics[width=.8\linewidth]{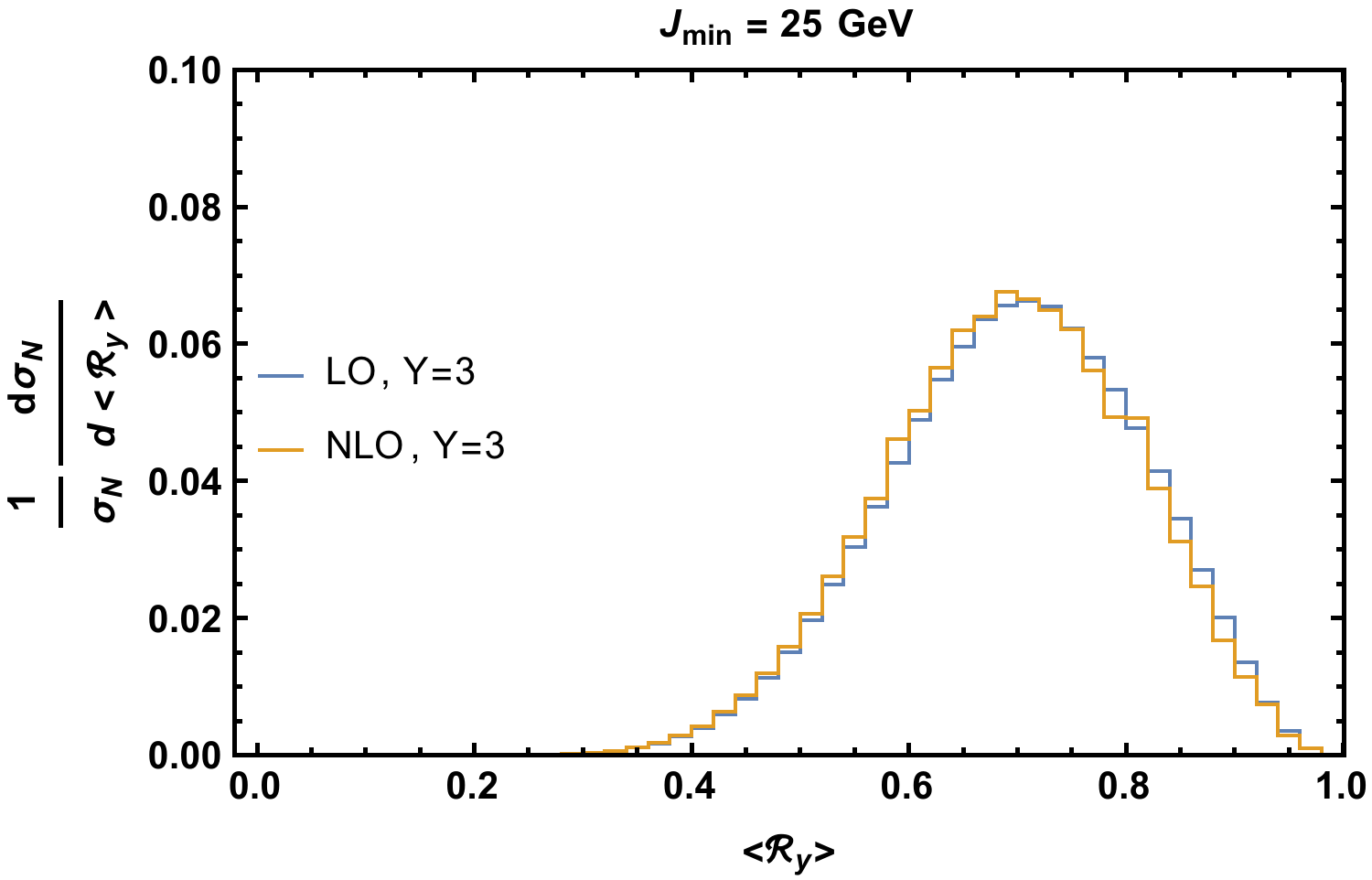}
  \caption{}
  \label{fig:sfig1}
\end{subfigure}%
\begin{subfigure}{.5\textwidth}
  \centering
  \includegraphics[width=.8\linewidth]{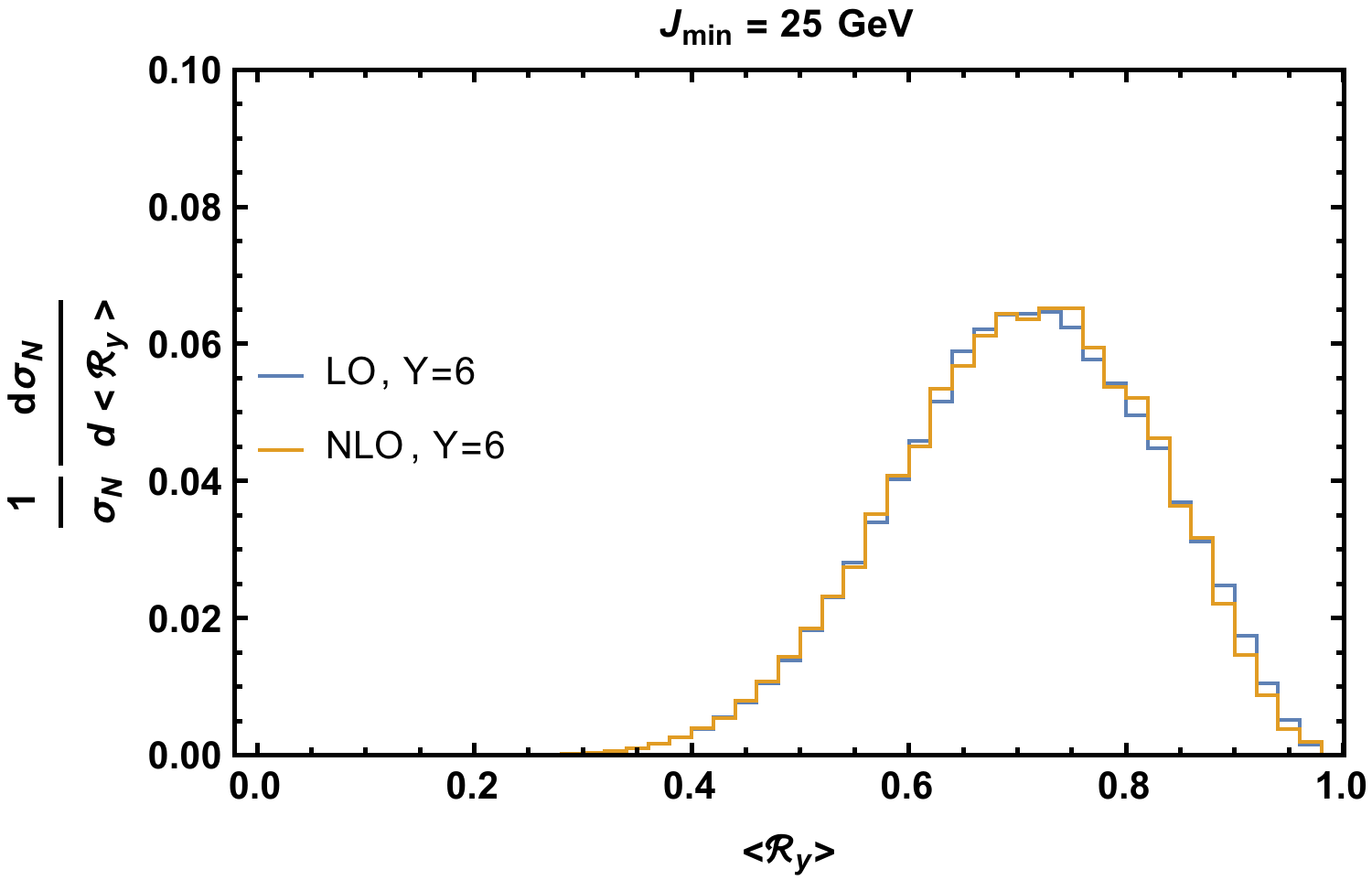}
  \caption{}
  \label{fig:sfig2}
\end{subfigure}
\caption{Comparison of LO and NLO+DLs normalized $\langle {\mathcal R}_{y} \rangle$ distributions for multiplicity $N = 5$, rapidity differences $Y = 3$ (left) and $Y = 6$ (right)  and $J_{\text{min}} = 2$ (top), $J_{\text{min}} = 10$ (middle), $J_{\text{min}} = 25$ (bottom).}
\label{fig:ry-LO-NLO-5}
\end{figure}

\begin{figure}
\begin{subfigure}{.5\textwidth}
  \centering
  \includegraphics[width=.8\linewidth]{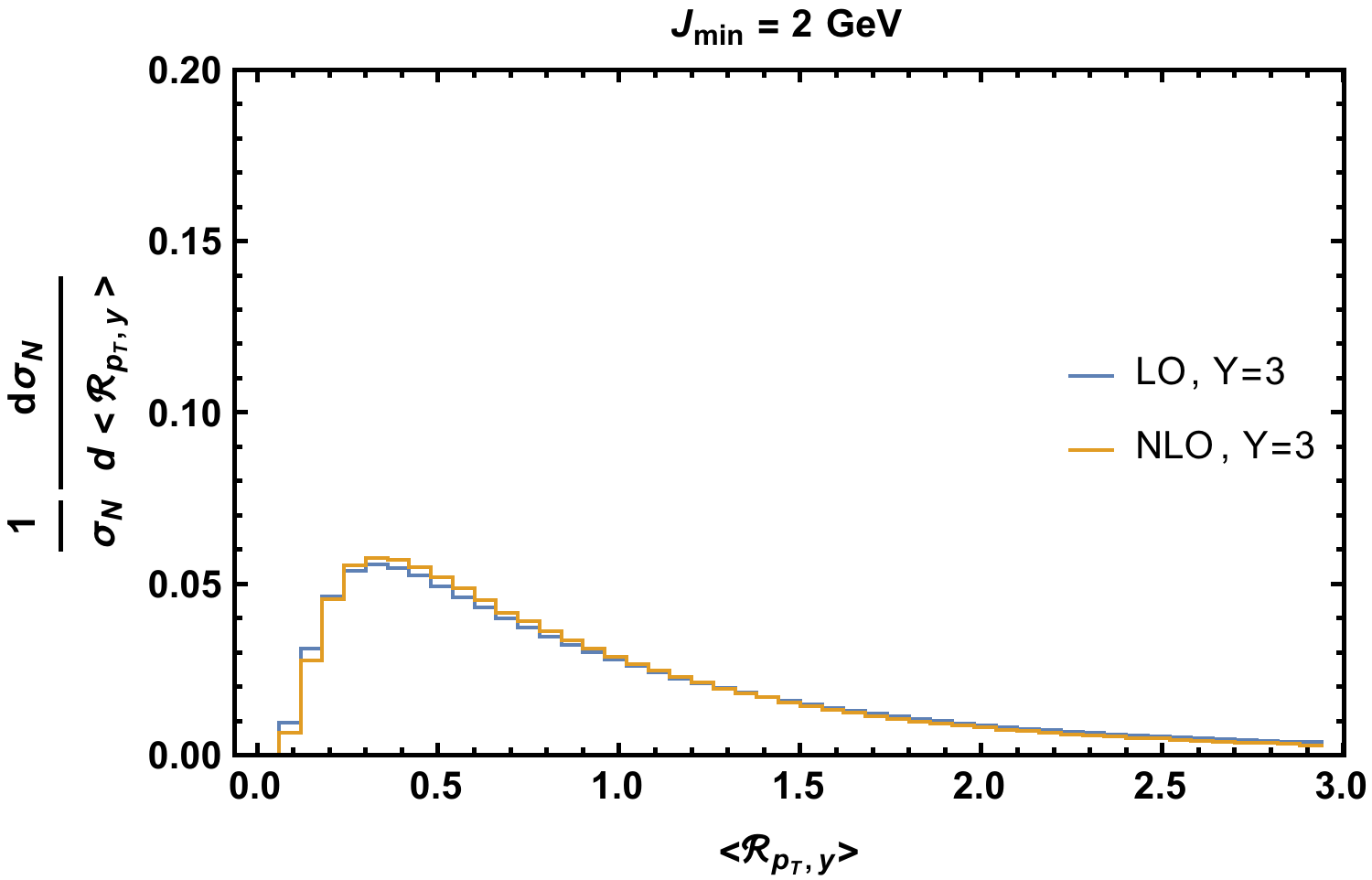}
  \caption{}
  \label{fig:sfig1}
\end{subfigure}%
\begin{subfigure}{.5\textwidth}
  \centering
  \includegraphics[width=.8\linewidth]{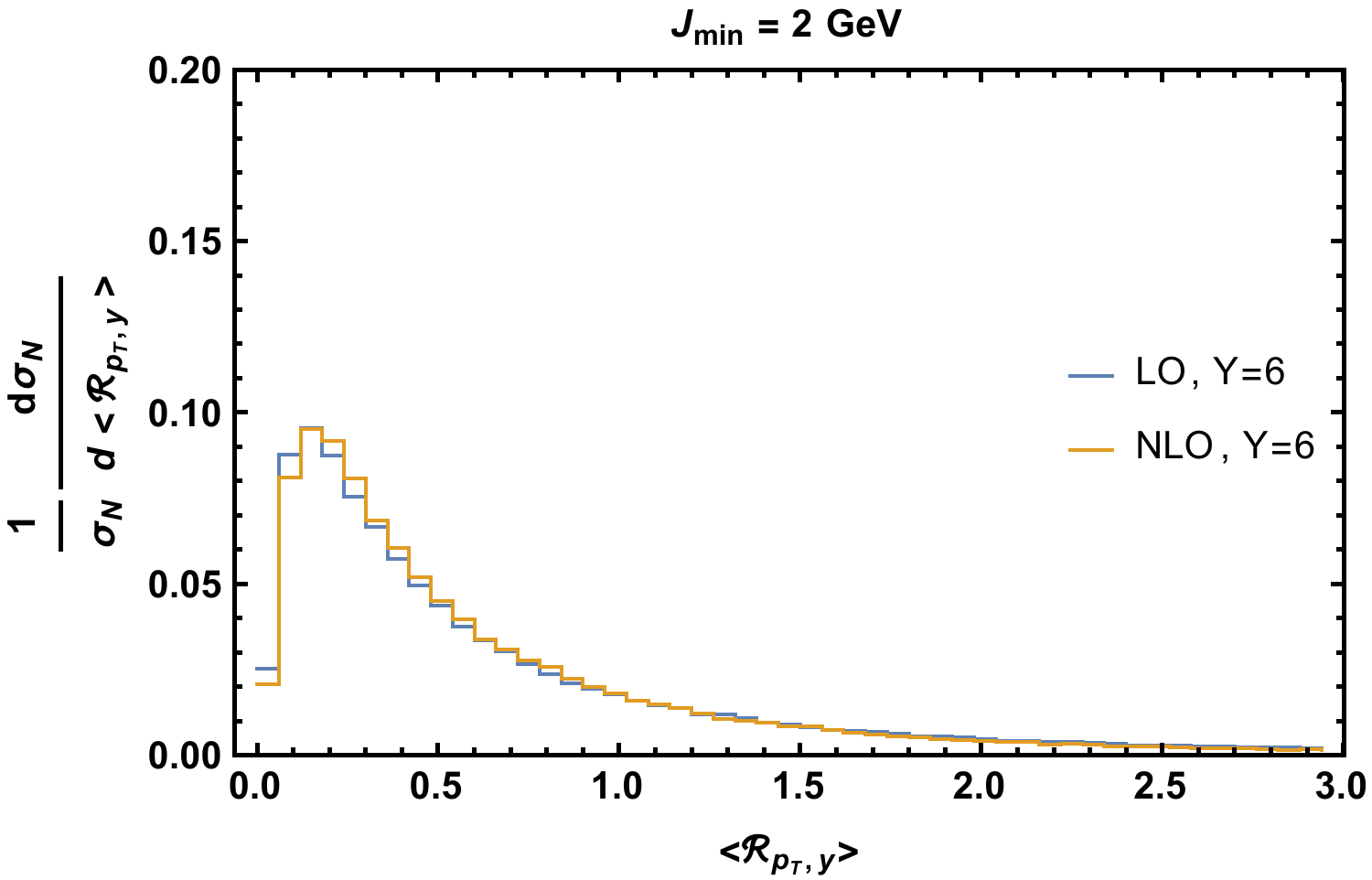}
  \caption{}
  \label{fig:sfig2}
\end{subfigure}
\\
\begin{subfigure}{.5\textwidth}
  \centering
  \includegraphics[width=.8\linewidth]{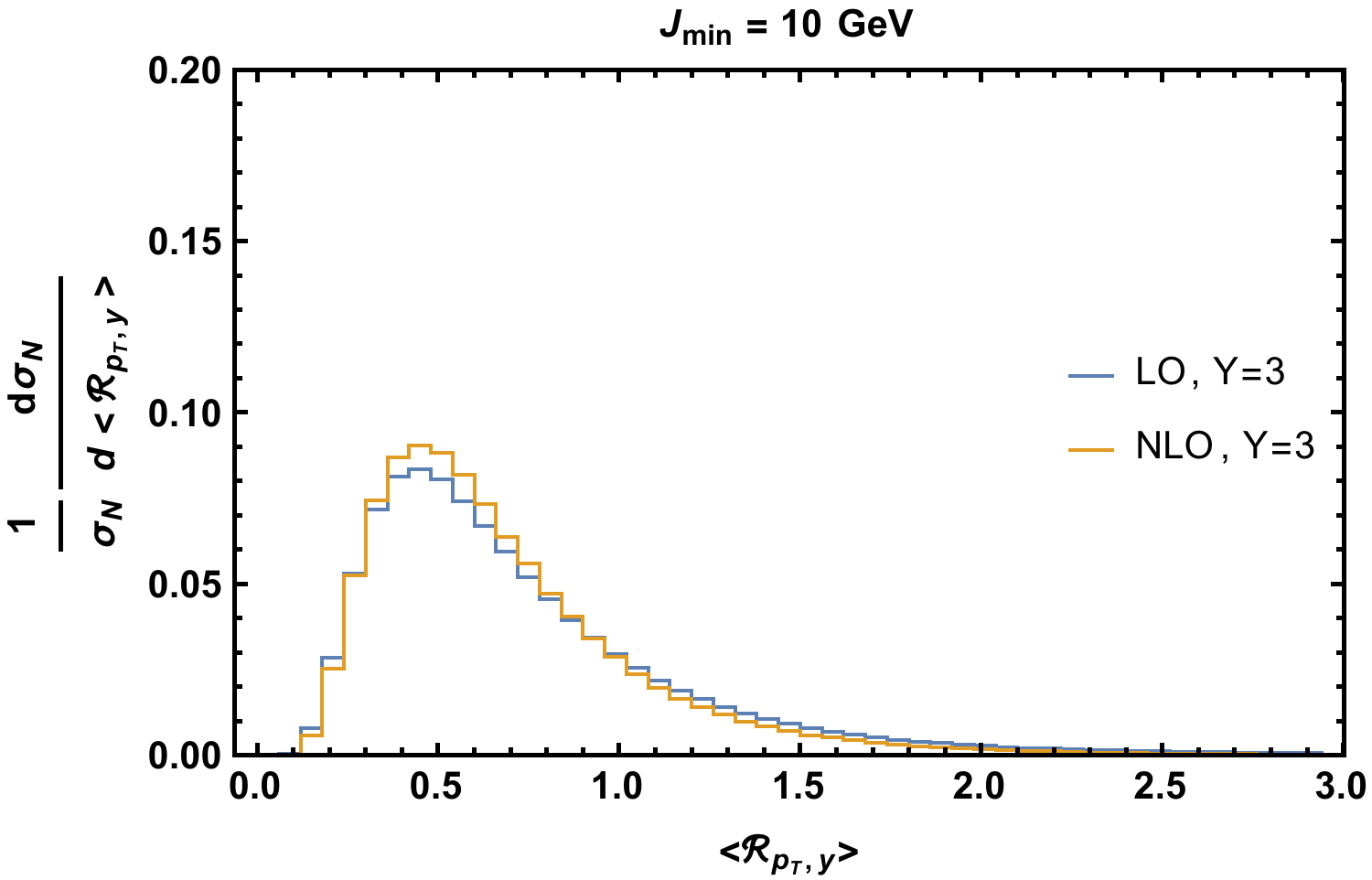}
  \caption{}
  \label{fig:sfig1}
\end{subfigure}%
\begin{subfigure}{.5\textwidth}
  \centering
  \includegraphics[width=.8\linewidth]{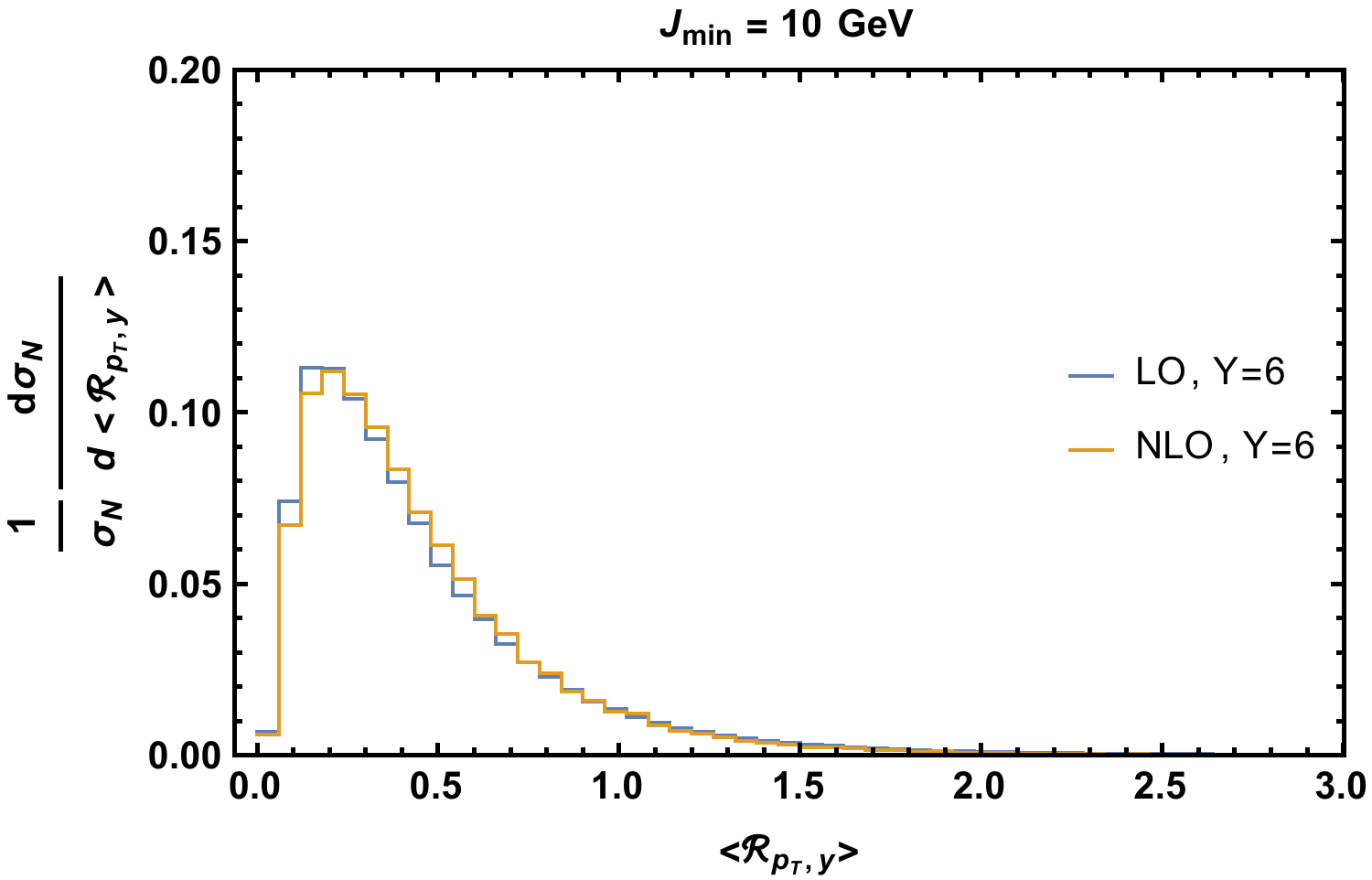}
  \caption{}
  \label{fig:sfig2}
\end{subfigure}
\\
\begin{subfigure}{.5\textwidth}
  \centering
  \includegraphics[width=.8\linewidth]{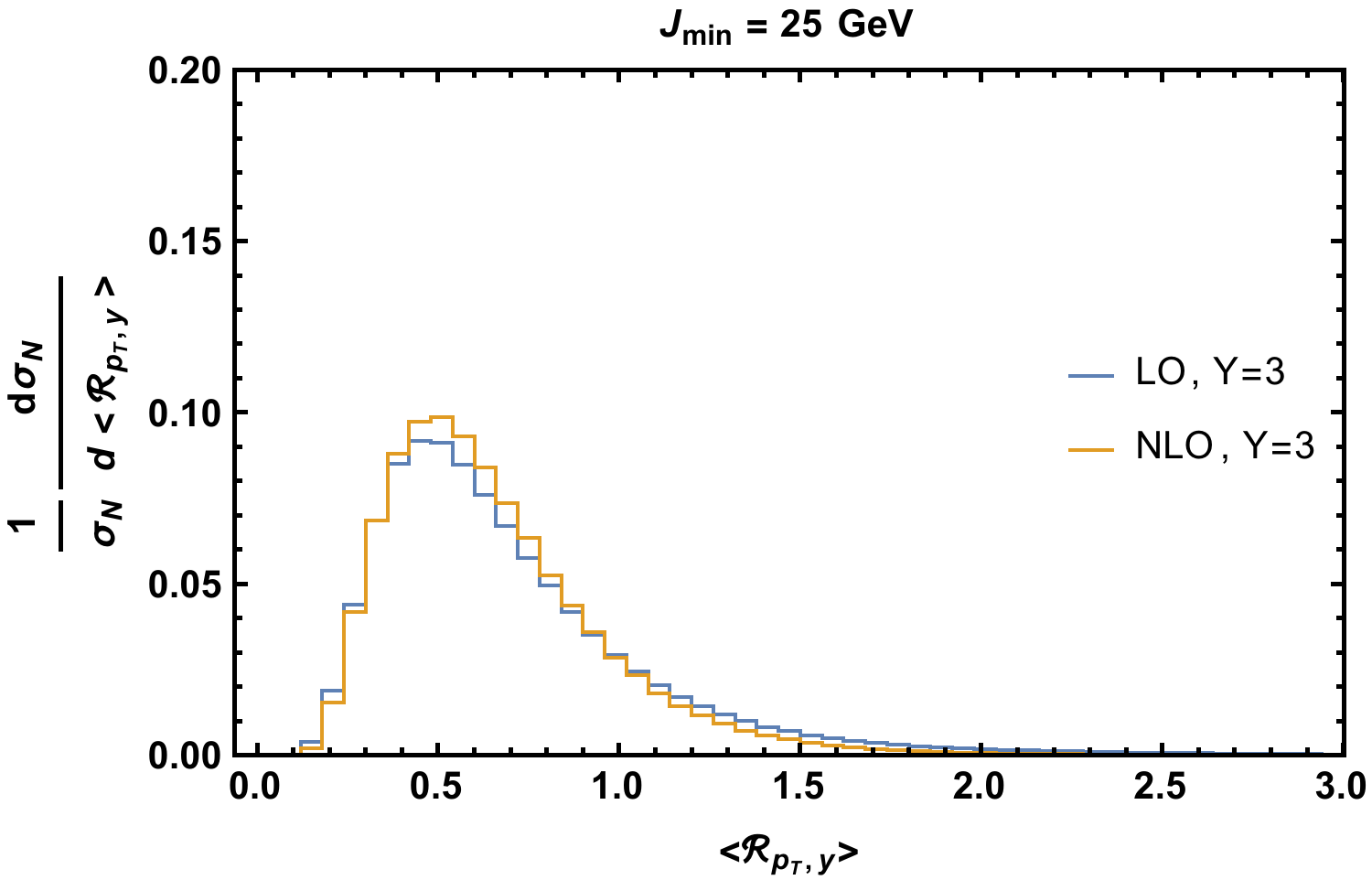}
  \caption{}
  \label{fig:sfig1}
\end{subfigure}%
\begin{subfigure}{.5\textwidth}
  \centering
  \includegraphics[width=.8\linewidth]{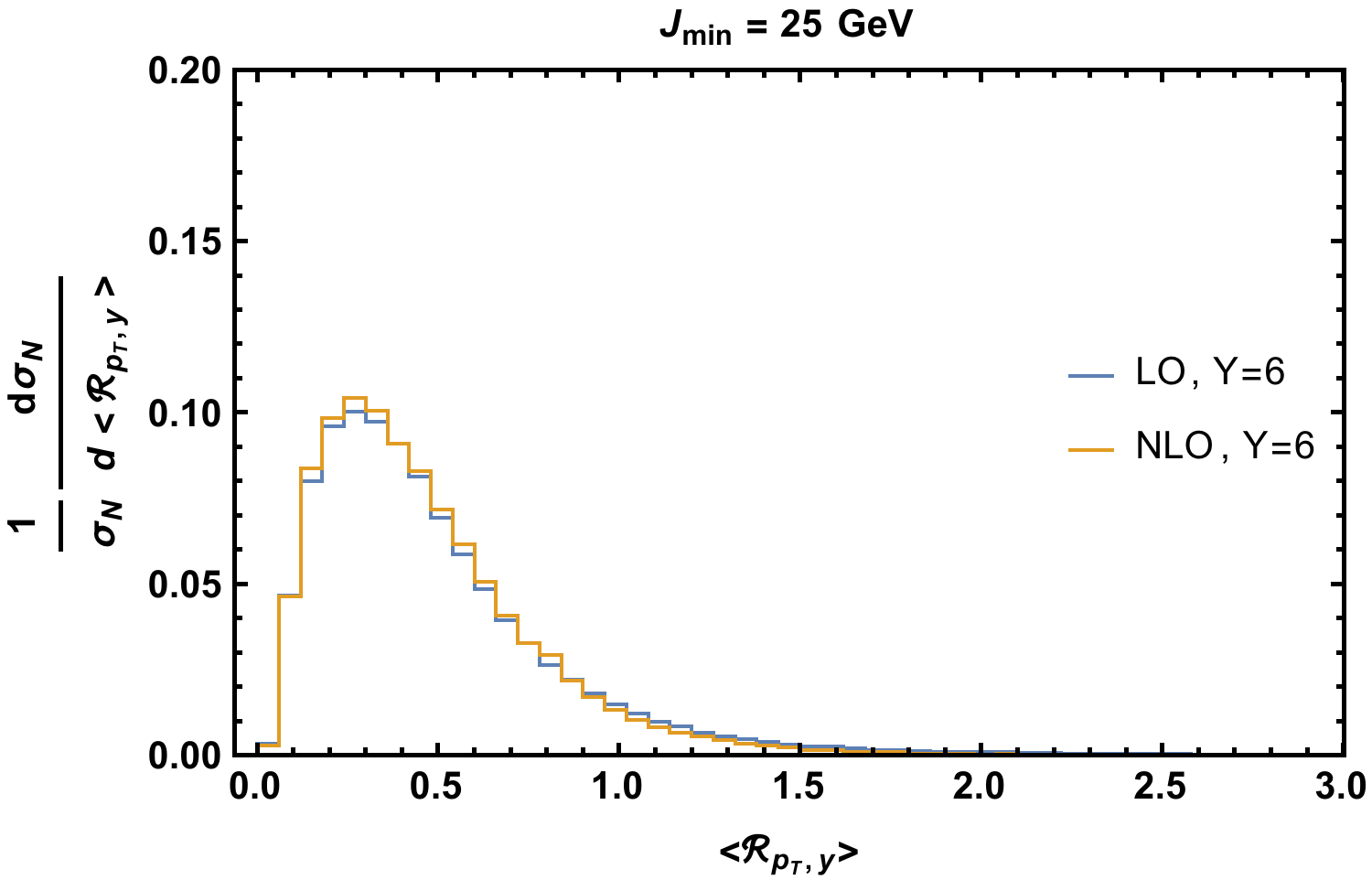}
  \caption{}
  \label{fig:sfig2}
\end{subfigure}
\caption{Comparison of LO and NLO+DLs normalized $\langle {\mathcal R}_{p_T, y} \rangle$ distributions for multiplicity $N = 3$, rapidity differences $Y = 3$ (left) and $Y = 6$ (right)  and $J_{\text{min}} = 2$ (top), $J_{\text{min}} = 10$ (middle), $J_{\text{min}} = 25$ (bottom).}
\label{fig:rpty-LO-NLO-3}
\end{figure}

\begin{figure}
\begin{subfigure}{.5\textwidth}
  \centering
  \includegraphics[width=.8\linewidth]{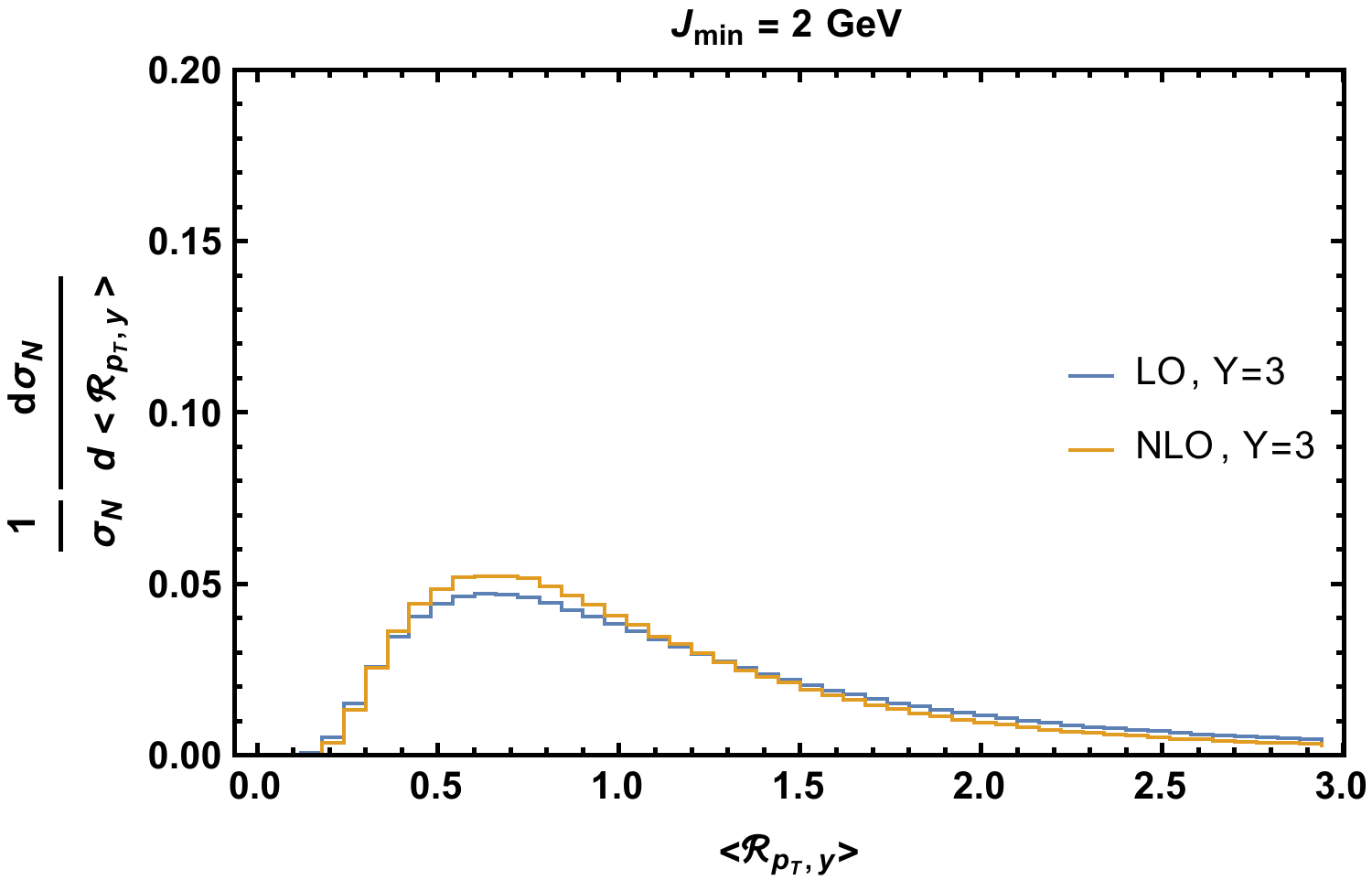}
  \caption{}
  \label{fig:sfig1}
\end{subfigure}%
\begin{subfigure}{.5\textwidth}
  \centering
  \includegraphics[width=.8\linewidth]{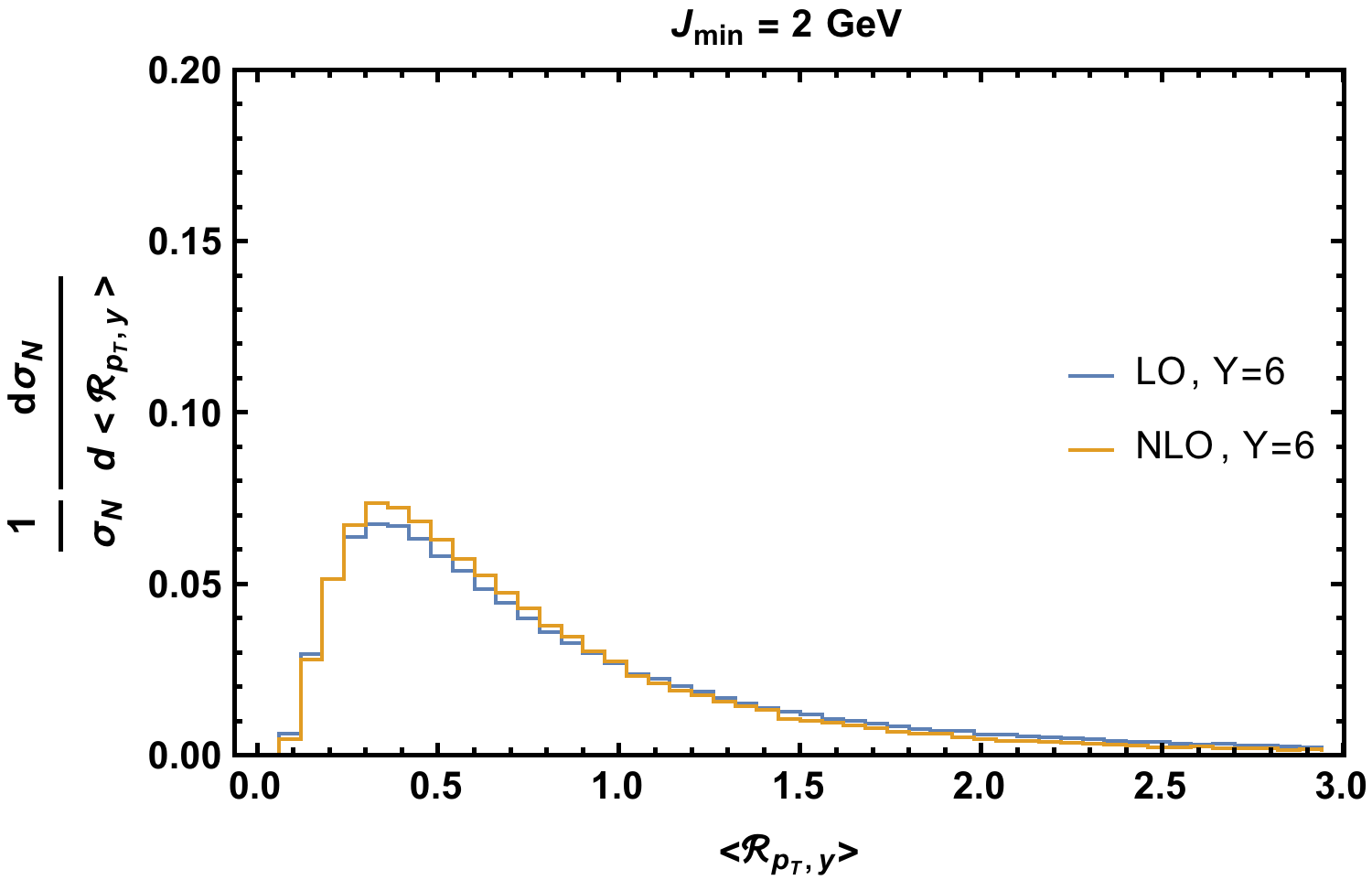}
  \caption{}
  \label{fig:sfig2}
\end{subfigure}
\\
\begin{subfigure}{.5\textwidth}
  \centering
  \includegraphics[width=.8\linewidth]{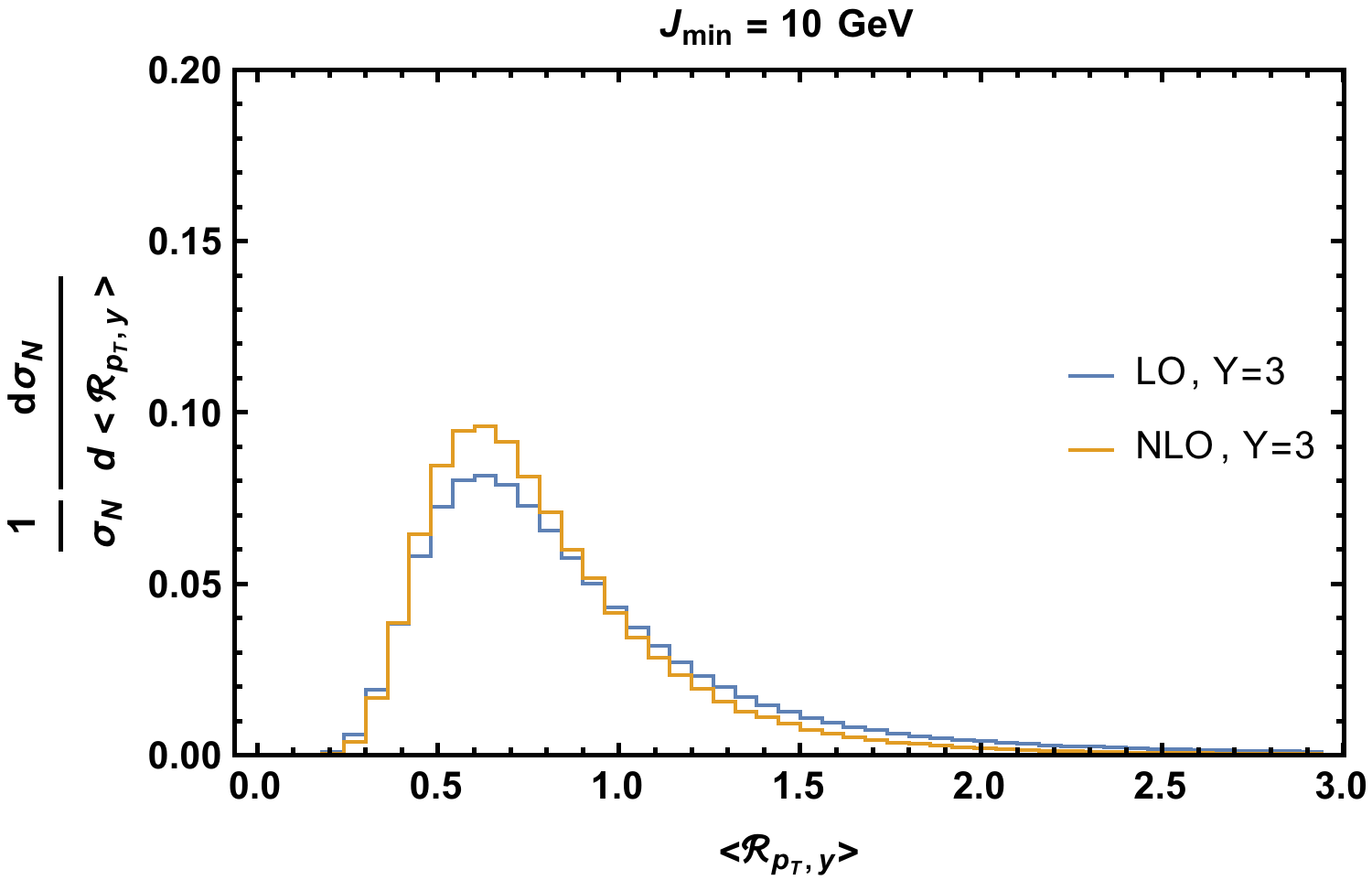}
  \caption{}
  \label{fig:sfig1}
\end{subfigure}%
\begin{subfigure}{.5\textwidth}
  \centering
  \includegraphics[width=.8\linewidth]{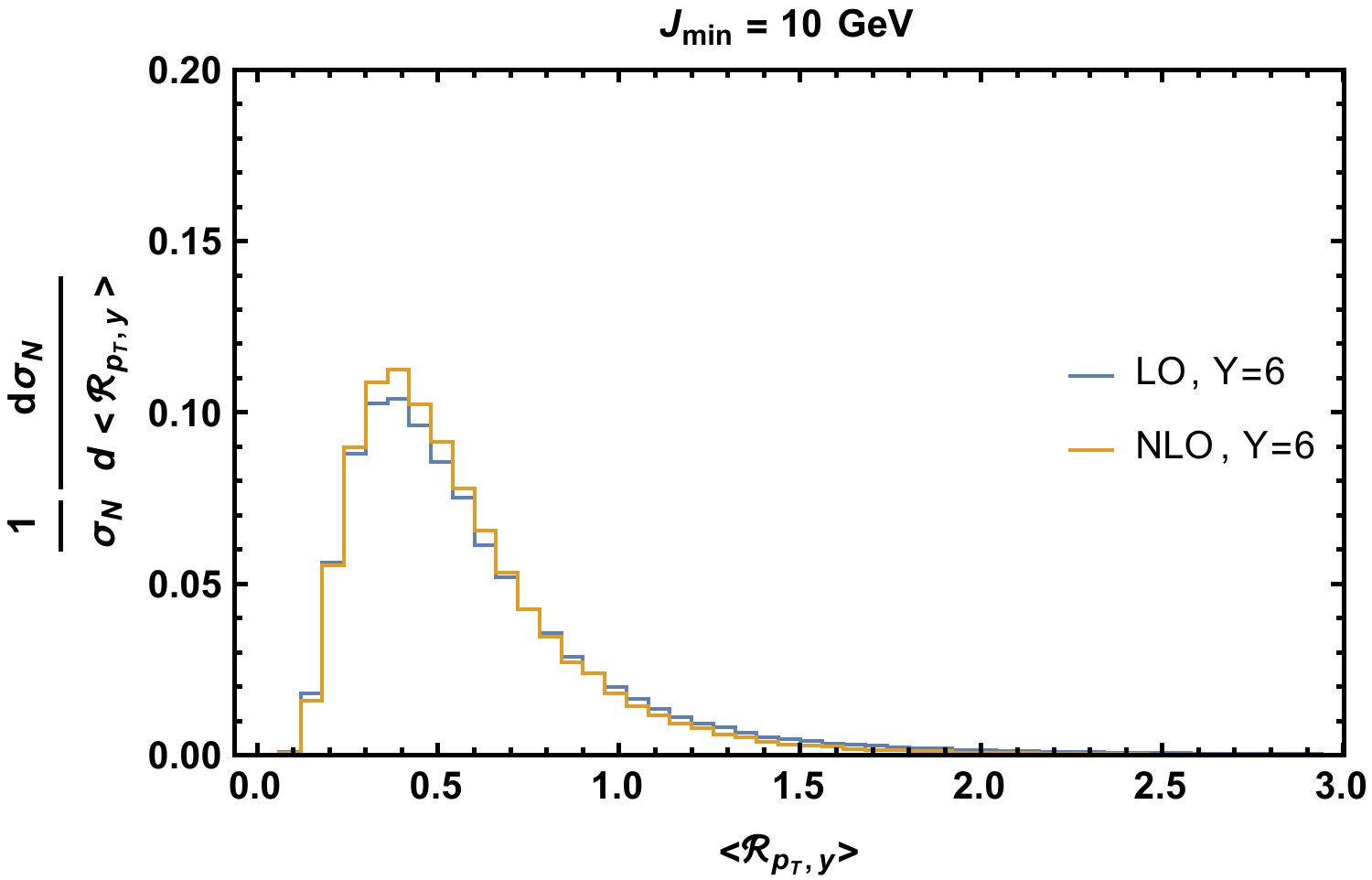}
  \caption{}
  \label{fig:sfig2}
\end{subfigure}
\\
\begin{subfigure}{.5\textwidth}
  \centering
  \includegraphics[width=.8\linewidth]{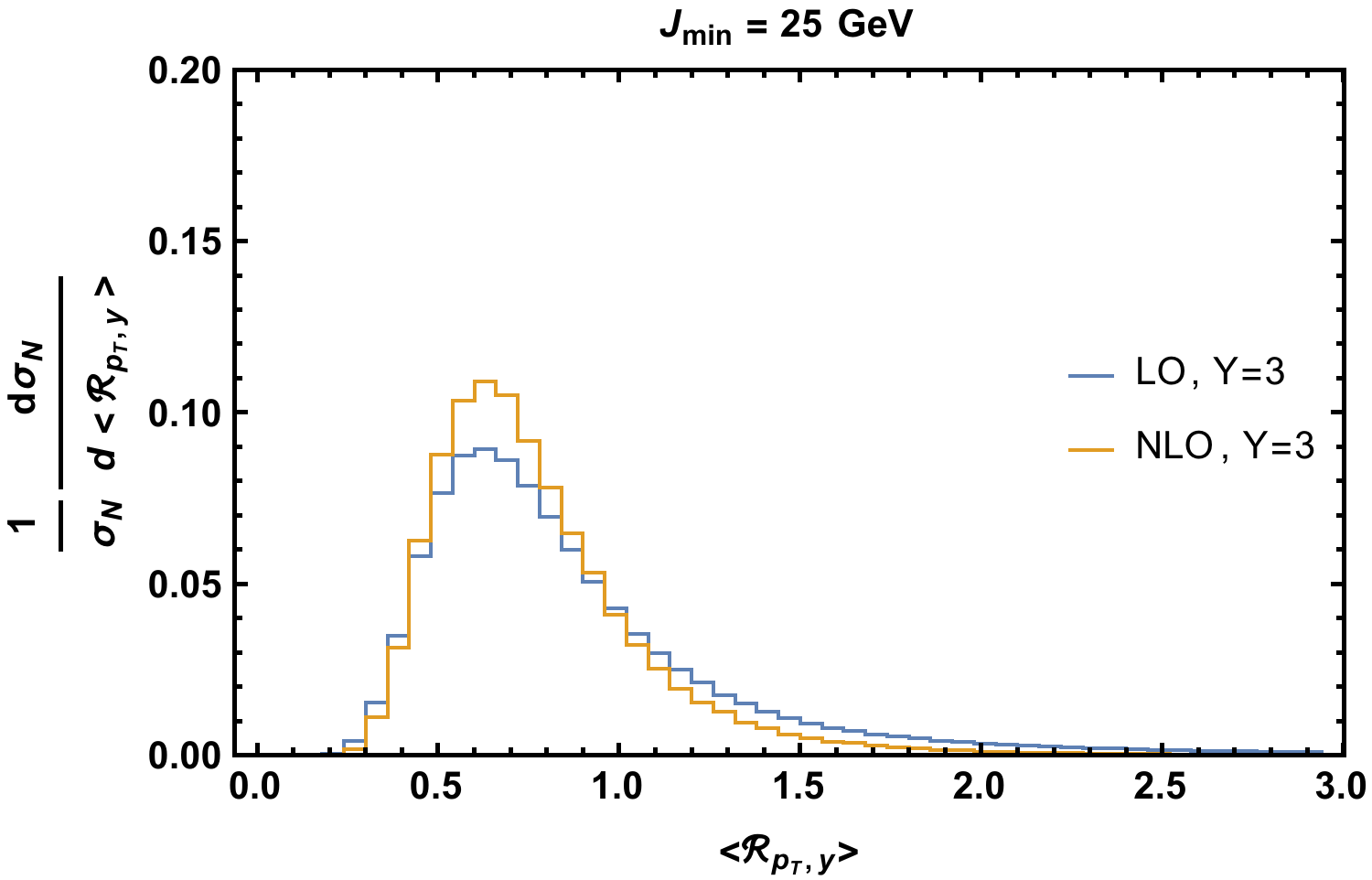}
  \caption{}
  \label{fig:sfig1}
\end{subfigure}%
\begin{subfigure}{.5\textwidth}
  \centering
  \includegraphics[width=.8\linewidth]{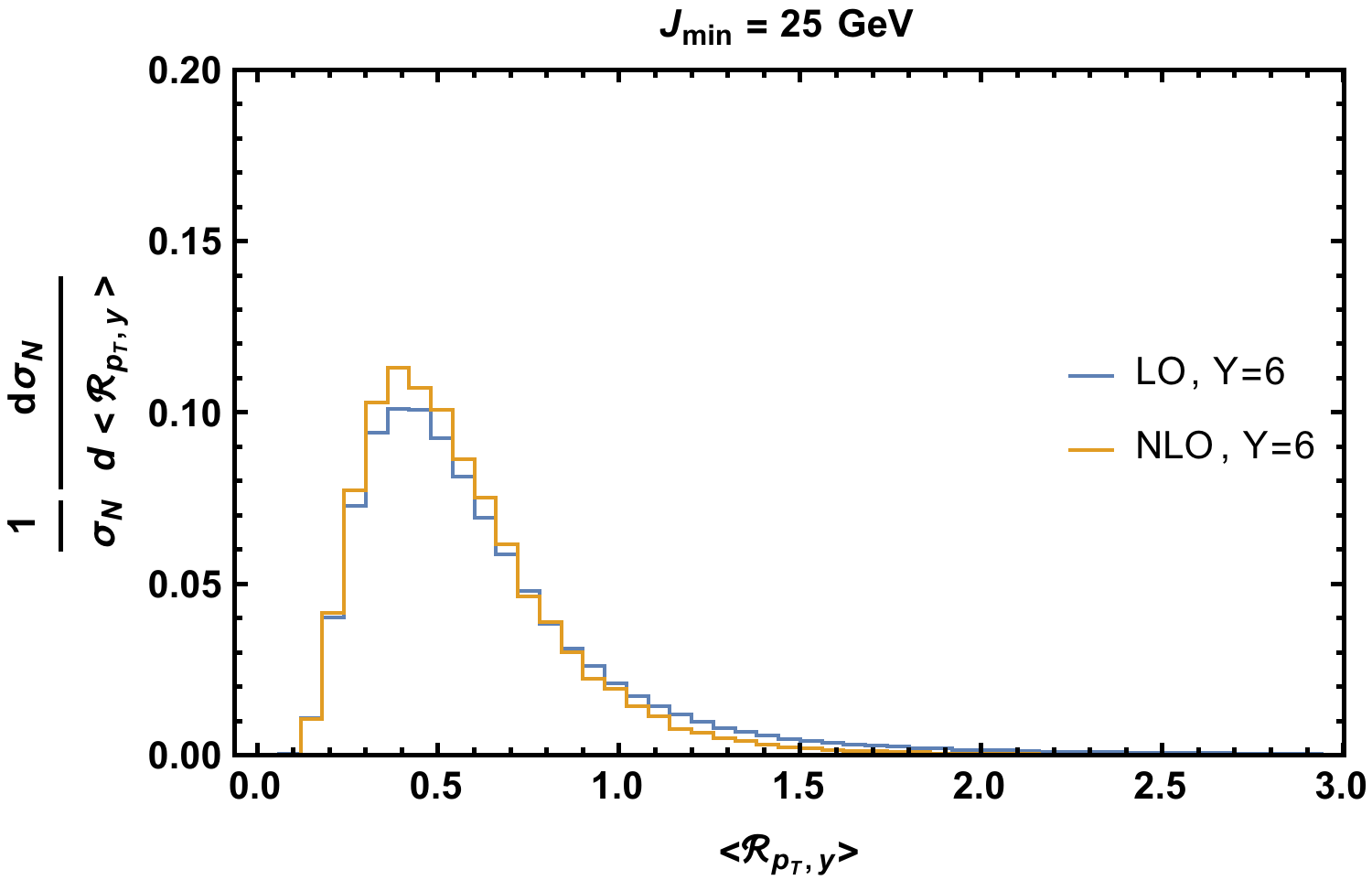}
  \caption{}
  \label{fig:sfig2}
\end{subfigure}
\caption{Comparison of LO and NLO+DLs normalized $\langle {\mathcal R}_{p_T, y} \rangle$ distributions for multiplicity $N = 4$, rapidity differences $Y = 3$ (left) and $Y = 6$ (right)  and $J_{\text{min}} = 2$ (top), $J_{\text{min}} = 10$ (middle), $J_{\text{min}} = 25$ (bottom).}
\label{fig:rpty-LO-NLO-4}
\end{figure}

\begin{figure}
\begin{subfigure}{.5\textwidth}
  \centering
  \includegraphics[width=.8\linewidth]{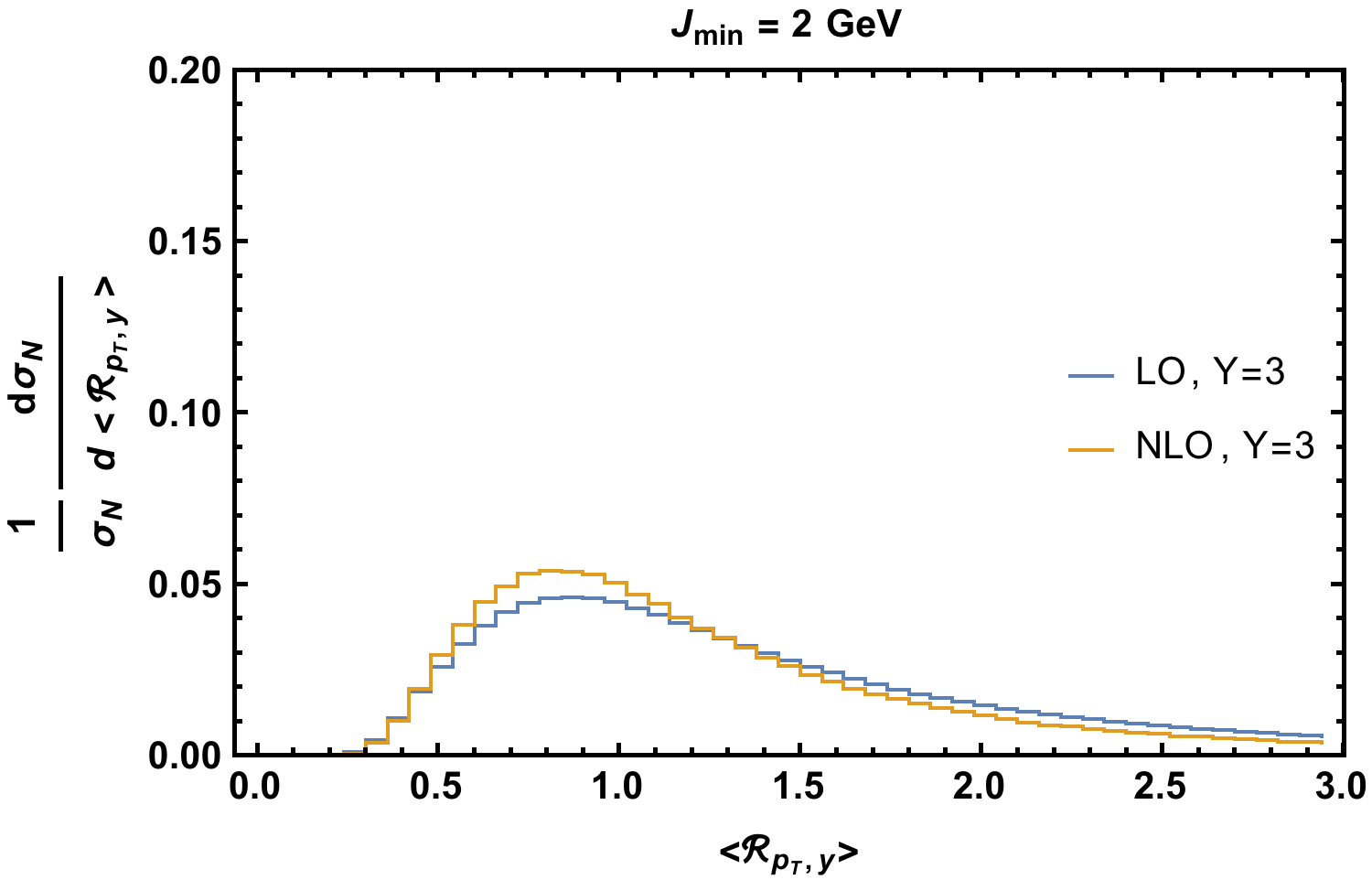}
  \caption{}
  \label{fig:sfig1}
\end{subfigure}%
\begin{subfigure}{.5\textwidth}
  \centering
  \includegraphics[width=.8\linewidth]{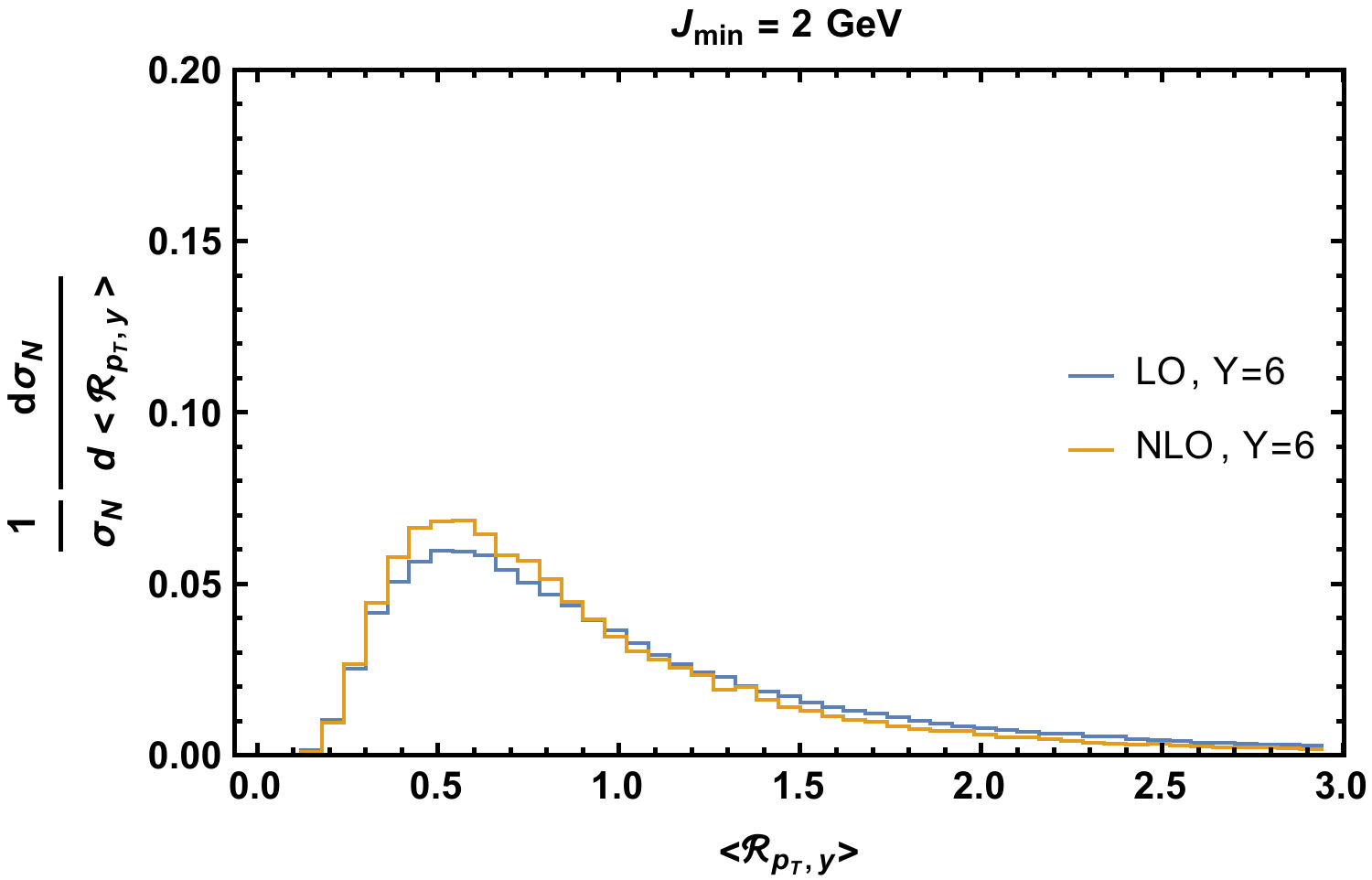}
  \caption{}
  \label{fig:sfig2}
\end{subfigure}
\\
\begin{subfigure}{.5\textwidth}
  \centering
  \includegraphics[width=.8\linewidth]{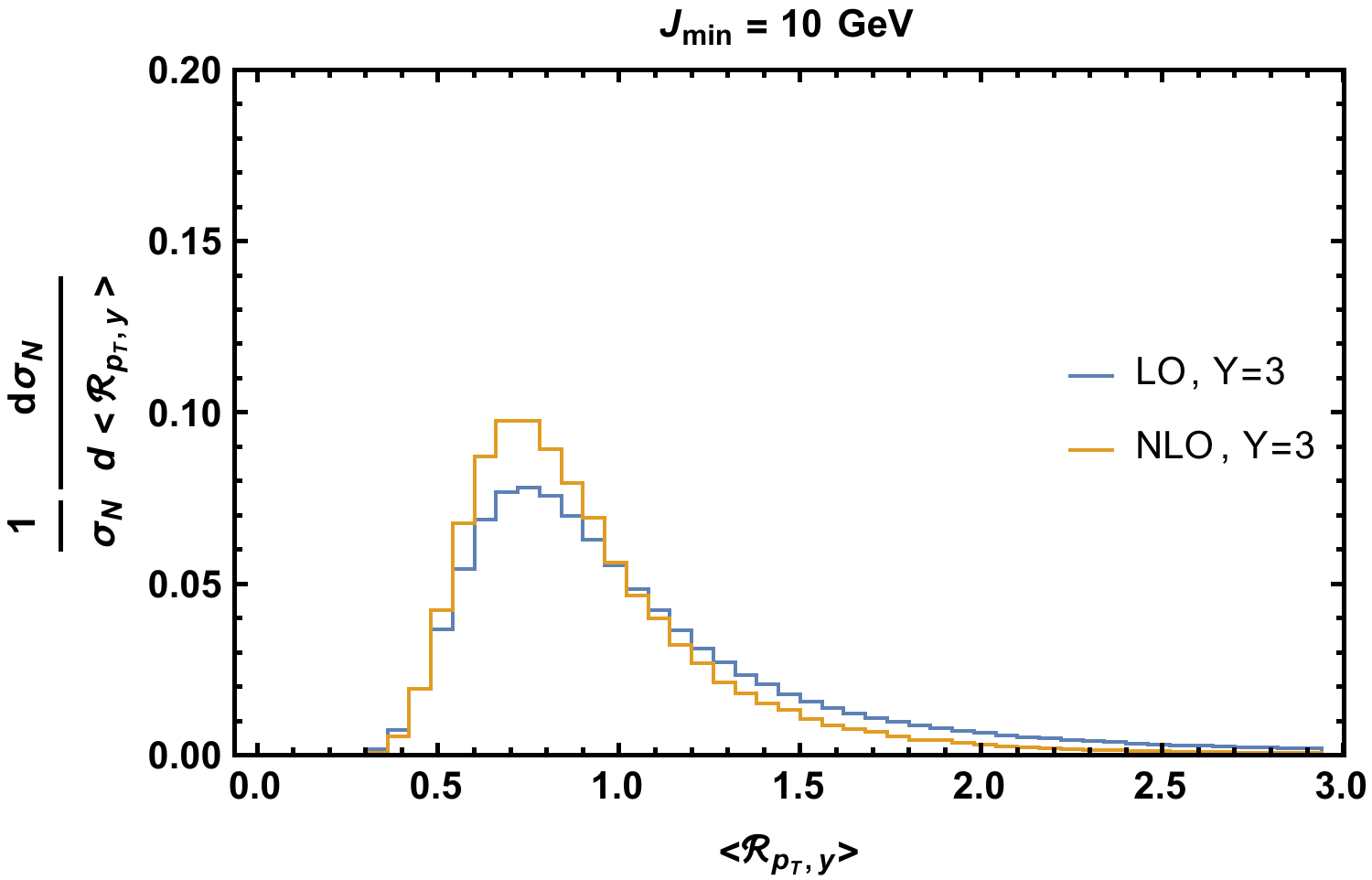}
  \caption{}
  \label{fig:sfig1}
\end{subfigure}%
\begin{subfigure}{.5\textwidth}
  \centering
  \includegraphics[width=.8\linewidth]{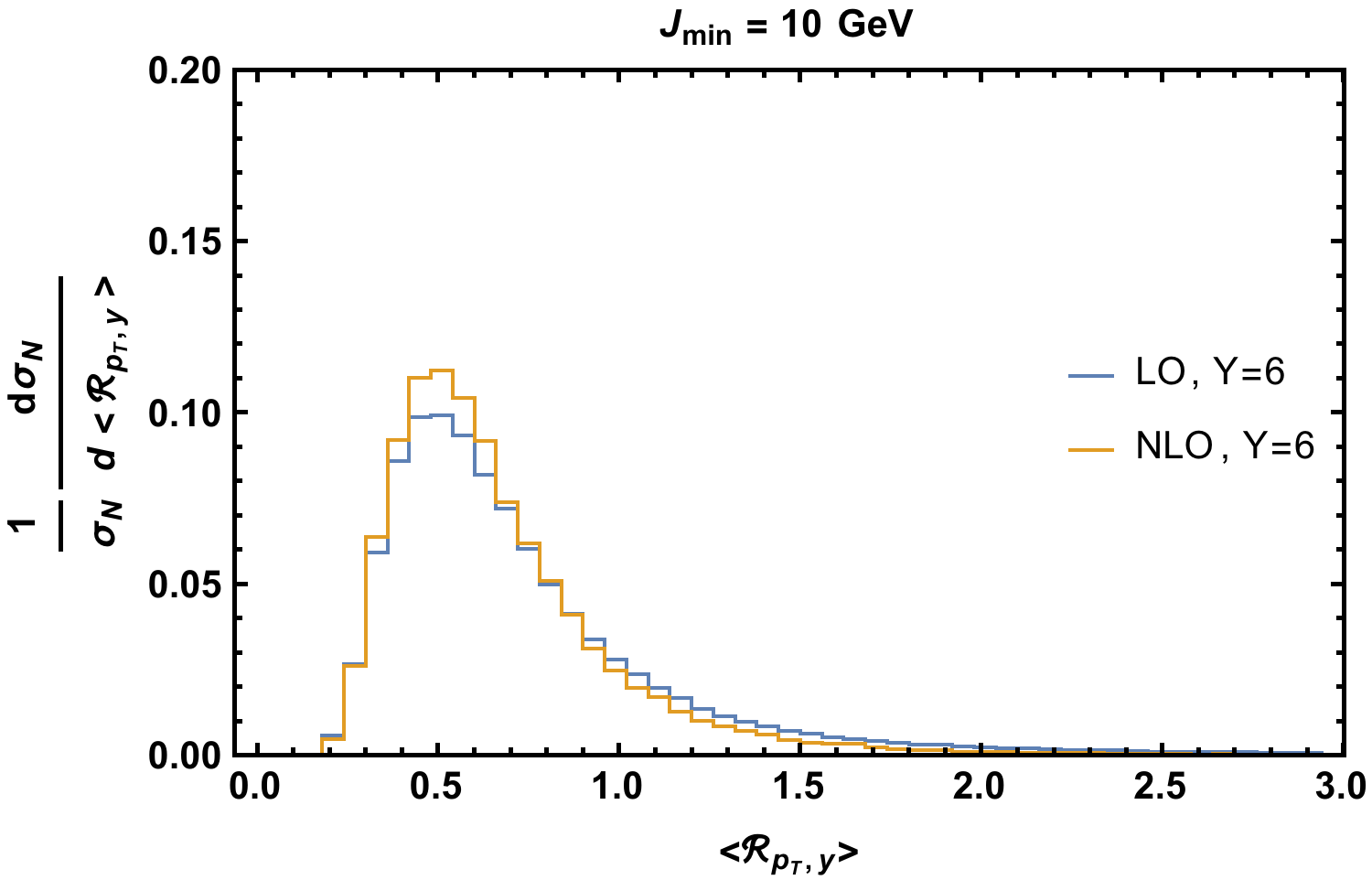}
  \caption{}
  \label{fig:sfig2}
\end{subfigure}
\\
\begin{subfigure}{.5\textwidth}
  \centering
  \includegraphics[width=.8\linewidth]{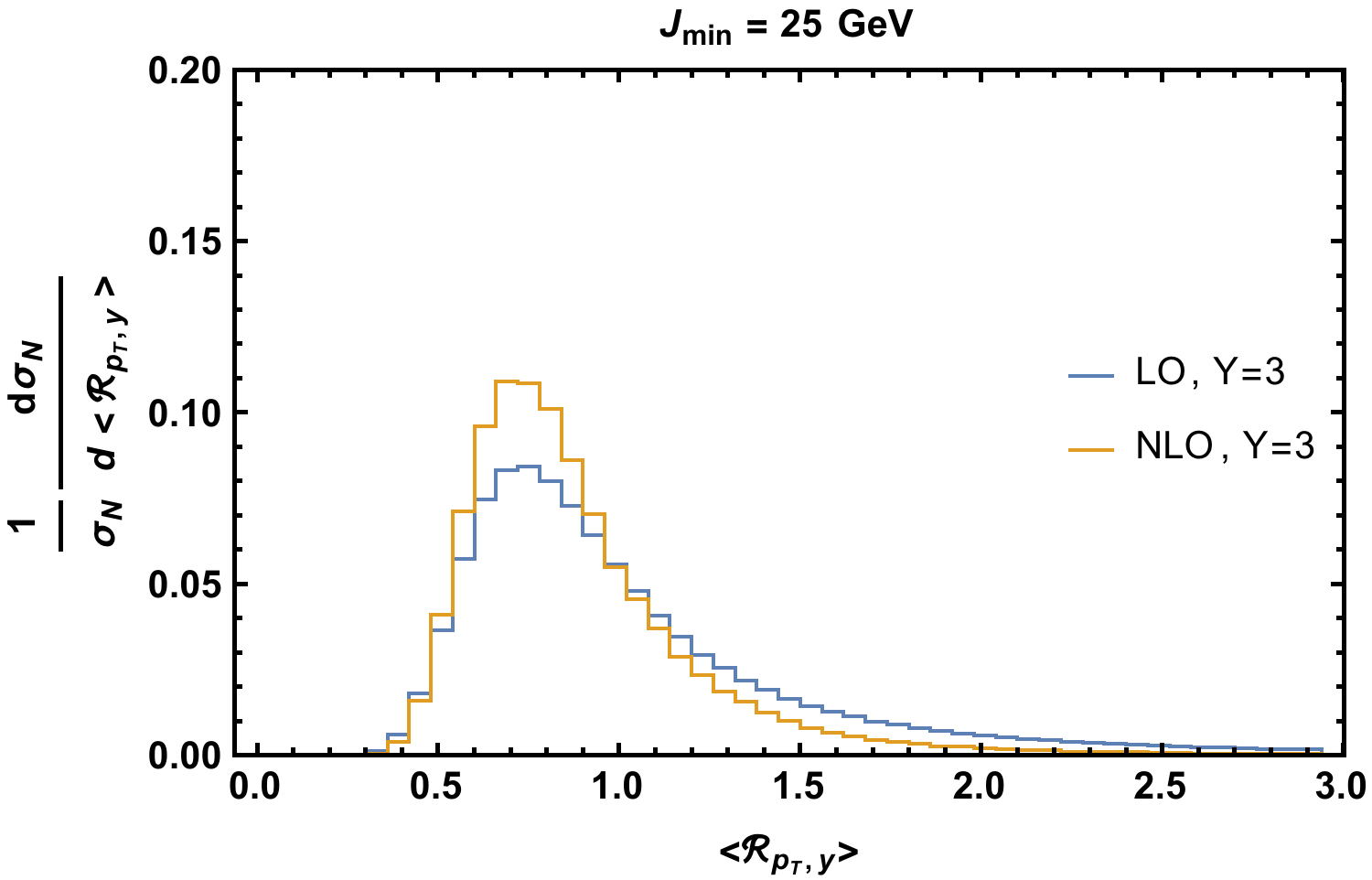}
  \caption{}
  \label{fig:sfig1}
\end{subfigure}%
\begin{subfigure}{.5\textwidth}
  \centering
  \includegraphics[width=.8\linewidth]{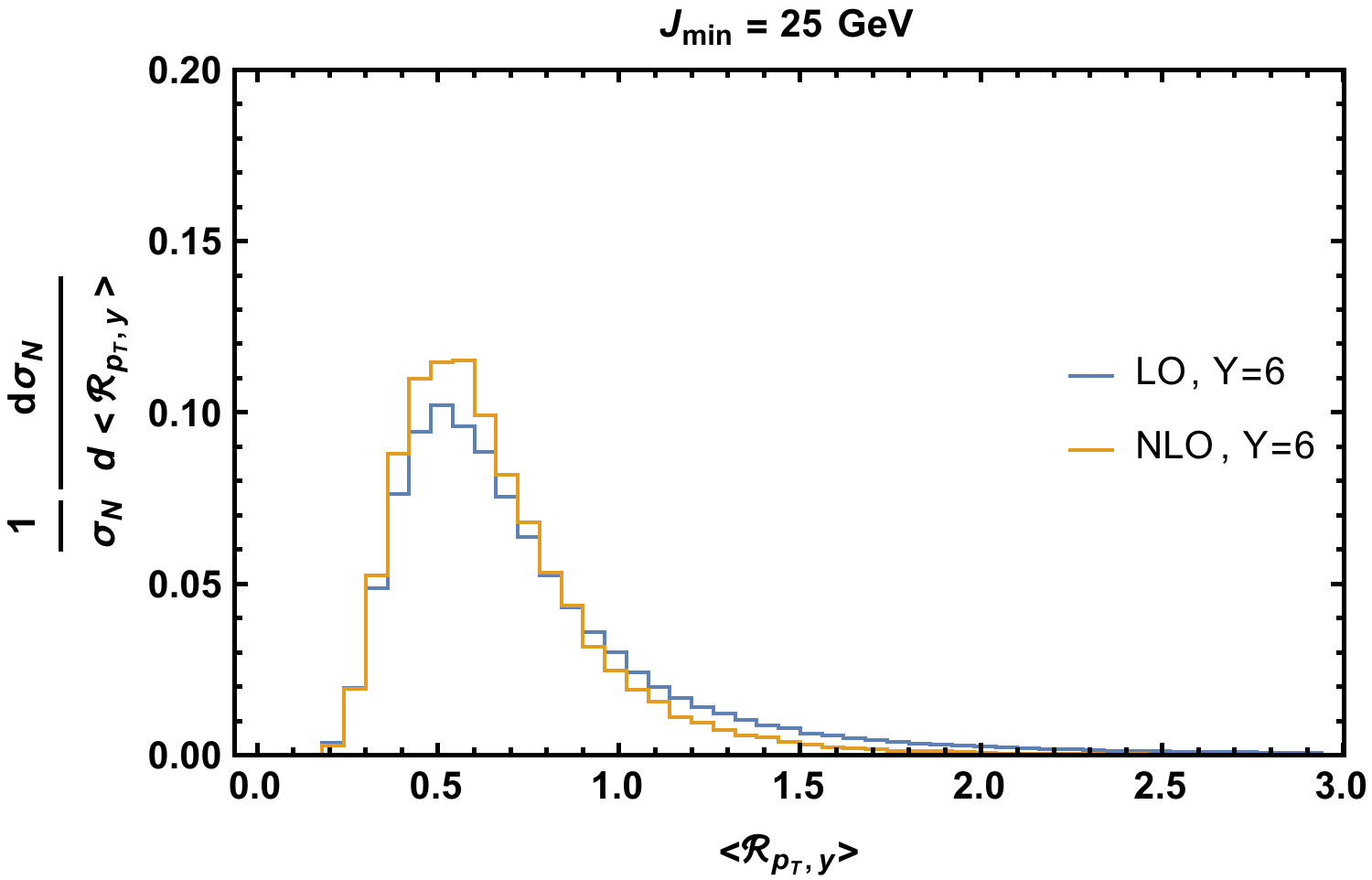}
  \caption{}
  \label{fig:sfig2}
\end{subfigure}
\caption{Comparison of LO and NLO+DLs normalized $\langle {\mathcal R}_{p_T, y} \rangle$ distributions for multiplicity $N = 5$, rapidity differences $Y = 3$ (left) and $Y = 6$ (right)  and $J_{\text{min}} = 2$ (top), $J_{\text{min}} = 10$ (middle), $J_{\text{min}} = 25$ (bottom).}
\label{fig:rpty-LO-NLO-5}
\end{figure}

Here,
in Figs.~\ref{fig:ry-LO-NLO-3},~\ref{fig:ry-LO-NLO-4} and
~\ref{fig:ry-LO-NLO-5} we compare the LO and the NLO+DLs distributions of 
$\langle {\mathcal R}_{ y} \rangle$, whereas in
Figs.~\ref{fig:rpty-LO-NLO-3},~\ref{fig:rpty-LO-NLO-4} and
~\ref{fig:rpty-LO-NLO-5} we compare the LO and the NLO+DLs distributions of
 $\langle {\mathcal R}_{p_T, y} \rangle$. The layout of subplots in all these figures
 is the same, namely, $J_{min} = 2$ GeV on the top,  $J_{min} = 10$ GeV in the middle
 and $J_{min} = 25$ GeV at the bottom. To the left, we place the plots with $Y=3$ and
 to the right the ones with $Y=6$. Each subplot contains two curves, one at LO and the other
 at NLO+DLs.

 Regarding the average ratio $\langle {\mathcal R}_{y} \rangle$ studied in Figs.~\ref{fig:ry-LO-NLO-3},~\ref{fig:ry-LO-NLO-4} and
~\ref{fig:ry-LO-NLO-5}, we notice that the stability upon the inclusion of the NLO+DLs corrections
is total for all three multiplicities. Even more so since it holds for all three different values of $J_{min}$ and for both values of $Y$. The stability 
is also very high for the average ratio $\langle {\mathcal R}_{p_T, y} \rangle$ as can be seen in
Figs.~\ref{fig:rpty-LO-NLO-3},~\ref{fig:rpty-LO-NLO-4} and \ref{fig:rpty-LO-NLO-5} but not as remarkable as for $\langle {\mathcal R}_{y} \rangle$. In particular,
it is better for $Y=6$ and for smaller $N$. The position of the peak of the distributions seems to be
extremely stable whereas the most noticeable difference between LO and NLO+DLs concerns the actual height of the peak.

\section{Conclusions}

We have revisited some observables  proposed some time ago in MN configurations at the LHC.
These were initially the average jet azimuthal angle $\phi$, the average jet
$p_T$ and the average ratio of rapidities. In the present study we were concerned with the rapidities and the $p_T$ of the minijets.
We have improved the initial observables in two ways. Firstly, we removed any direct influence of the MN jets by using as input  the rapidities and the transverse momenta of the minijets in between them to probe better the inner dynamics of the multi-Regge kinematics. Secondly, we combined the $p_T$ and the rapidities of the minijets into a new observable so that in the end we have two average
rapidity ratios, $\langle {\mathcal R}_{y} \rangle$ which is very similar to the original proposal and $\langle {\mathcal R}_{p_T, y} \rangle$ which is scaled by the transverse momenta of the minijets.

In addition,  since both ATLAS and CMS use a veto on the resolved minijets of the order of 20 GeV, we introduced in the present study three different values of a theoretical equivalent to the minijet $p_T$ veto, namely 2, 10 and 25 GeV.

We presented plots of the distributions of the two average rapidity ratios for fixed minijet multiplicities both at LO and at NLO+DLs. We found that the LO and NLO+DLs distributions are in remarkable agreements for $\langle {\mathcal R}_{y} \rangle$ for all $J_{min}$ values and for both reference rapidities whereas for $\langle {\mathcal R}_{p_T, y} \rangle$ they are still in very good agreement but more so for the large values of $Y$.

We foresee that these observables will contribute to isolate BFKL related effects in MN jets final states at the LHC. For that, we will need to perform a full phenomenological study including PDFs and jet vertices at NLO which we hope to do in the near future. We also hope that this work will draw the attention of the  community to the importance of taking into account in theoretical works the minijet $p_T$ veto used by the LHC experiments as well as restricting the multiplicity of resolved final state jets into fixed values.

\section*{Acknowledgements}

This work has been supported by the Spanish Research Agency (Agencia Estatal de Investigaci{\'o}n) through the grant IFT Centro de Excelencia Severo Ochoa SEV-2016-0597 and the Spanish Government grant FPA2016-78022-P.  It has also received funding from the European Union’s Horizon 2020 research
and innovation programme under grant agreement No. 824093. The work of GC was supported by the Funda\c{c}\~ao para a Ci\^encia e a Tecnologia (Portugal) under project CERN/FIS-PAR/0024/2019 and contract `Investigador auxiliar FCT - Individual Call CEECIND/03216/2017'.

\end{document}